\documentclass[12pt]{article}
\usepackage{amssymb}
\usepackage{amsmath}
\usepackage{hyperref}
\usepackage[dvips]{graphicx}
\usepackage{bm}
\usepackage{latexsym}
\usepackage{booktabs}
\usepackage{colortbl}
\usepackage{float}
\usepackage{xcolor}
\usepackage[margin=2cm]{geometry}
\usepackage{longtable}
\usepackage{booktabs}
\usepackage[table]{xcolor}
\usepackage{subcaption}
\begin{document}
\title{\bf  Topological Phase Transition of $ F_{\alpha }{}^{\gamma } F^{\alpha \beta } F_{\beta }{}^{\eta } \
	F^{\lambda \mu } R_{\gamma \eta \lambda \mu } $ AdS Black Holes}

\author{Mehdi Sadeghi\thanks{Email: mehdi.sadeghi@abru.ac.ir}\,\,\, and \,\,Faramaz Rahmani\thanks{Corresponding author: faramarz.rahmani@abru.ac.ir}\hspace{2mm}\\
{\small {\em Department of Physics, Faculty of Basic Sciences,}}\\
{\small {\em Ayatollah Boroujerdi University, Boroujerd, Iran}}
}
\date{\today}

\maketitle

\begin{abstract}
\noindent We investigate the thermodynamic criticality and topological properties of four-dimensional charged anti-de Sitter black holes in the presence of a nonminimal electromagnetic-curvature interaction of the form \(F_{\alpha}{}^{\gamma}F^{\alpha\beta}F_{\beta}{}^{\eta}F^{\lambda\mu}R_{\gamma\eta\lambda\mu}\). Treating the coupling perturbatively to first order, we construct the corrected black hole solutions and derive their thermodynamic quantities in the extended phase space. Conventional critical behavior is examined through the criticality conditions for the Hawking temperature in both canonical and grand canonical ensembles. We then use thermodynamic topology to investigate the global structure of equilibrium configurations and its dependence on the thermodynamic parameters. In the canonical ensemble, two distinct regimes emerge: a noncritical region characterized by a total winding number \(W=+1\), and a critical region with \(W=0\). In the grand canonical ensemble, three distinct regimes are identified according to the electric potential. Interestingly, the low- and intermediate-potential regions share the same topological number, \(W=+1\), although conventional critical points occur only in the intermediate-potential region. The high-potential region, by contrast, is characterized by \(W=0\) and exhibits no conventional critical behavior. Our findings show that thermodynamic topology provides a global characterization of equilibrium configurations that complements conventional temperature-based criticality analysis, and that the global winding number alone does not uniquely determine the existence of conventional critical points.
\end{abstract}
\section{Introduction}
\label{sec:1}

Black hole thermodynamics provides a framework for exploring the interplay between gravitation, quantum theory, and statistical physics. The identification of black hole entropy with horizon area, together with Hawking radiation, established that black holes possess genuine thermodynamic properties. In asymptotically anti-de Sitter (AdS) spacetimes, these properties exhibit a particularly rich phase structure. The AdS/CFT correspondence further motivates their investigation by relating gravitational thermodynamics to thermal phenomena in the dual conformal field theory \cite{Bekenstein:2020lub, Bardeen:1973gs,Hawking:1975vcx,Hawking:1982dh,Witten:1998qj}.

An important development in this direction is the extended phase-space formulation, in which the cosmological constant is treated as a thermodynamic pressure and the black hole mass is interpreted as enthalpy. This perspective, commonly referred to as black hole chemistry, has revealed a variety of phase phenomena, including van der Waals-like criticality in charged AdS black holes and Hawking--Page transitions. The resulting phase structure depends not only on the gravitational background but also on the conserved charges(such as the electric charge), thermodynamic ensemble, and interactions included in the underlying theory. These observations motivate the investigation of how modifications to the gravitational and electromagnetic sectors affect black hole thermodynamics \cite{Kastor:2009wy, Dolan2011, Kubiznak:2012wp, Belhaj:2012bg,Banerjee:2011cz,Fernando:2006gh,Cai:2001dz,Frassino:2014pha,Cai:2014znn,Hendi:2017fxp}.

A complementary perspective has recently emerged through the topological classification of black hole thermodynamics. In this approach, suitable thermodynamic functions are used to construct vector fields whose zeros characterize distinguished thermodynamic configurations. Assigning winding numbers to these zeros and summing their contributions yields a global topological number that characterizes the corresponding thermodynamic structure. This classification provides information that is not necessarily captured by conventional analyses based on equations of state or local stability criteria. In particular, the number and winding numbers of the zeros may change as external parameters vary, while different configurations can yield the same total topological number. Consequently, the topological description and the conventional analysis of phase transitions address related but distinct aspects of black hole thermodynamics \cite{Wei:2021vdx,WeiLiu2026,Ali:2023jox,Gogoi:2025ied,Yerra:2022eov,EslamPanah:2024fls,Hazarika:2024xar,WeiLiuMann2022,Wei:2024gfz,Wu:2022whe,ZhuWu2024,WuGu2024,Chen:2024sow,Liu:2022aqt,Ahmed:2022kyv,Chen:2024atr,Zhu:2024zcl,Chen:2025fse,Tian:2026qnt,Chen:2023ddv,Wu:2025xxo}.

Nonminimal couplings between electromagnetic fields and spacetime curvature provide a systematic extension of the minimally coupled Einstein--Maxwell theory and can arise in higher-order effective gravitational theories. Such interactions modify the coupled electromagnetic and gravitational field equations and can consequently affect black hole solutions and their thermodynamic properties. In particular, the \(F^4R\) interaction considered here introduces corrections that depend on both the electromagnetic field strength and spacetime curvature, potentially becoming more pronounced in the strong-field regime. While conventional thermodynamic analyses primarily focus on the existence and location of critical points, thermodynamic topology provides a complementary perspective by classifying thermodynamic critical points through their associated winding numbers. This raises the question of whether the nonminimal interaction can alter the global topological classification of black hole thermodynamic states, rather than merely modifying the quantitative features of their phase structure \cite{Drummond:1979pp,Balakin:2005fu,Balakin:2015gpq,Sadeghi:2024msd,Sadeghi:2026jic,Rahmani:2025iks}.

In this work, we investigate the thermodynamic phase structure and topological classification of four-dimensional charged AdS black holes in the presence of the nonminimal interaction
$F_{\alpha}{}^{\gamma}F^{\alpha\beta}
F_{\beta}{}^{\eta}F^{\lambda\mu}
R_{\gamma\eta\lambda\mu}$.
We treat the coupling perturbatively to first order and examine both the conventional critical behavior and the thermodynamic topology in the canonical and grand canonical ensembles. In addition to determining the critical behavior, we study how the electric charge and electric potential control the thermodynamic configurations and their associated winding numbers. Particular attention is paid to the relation between the number of thermodynamic zero points and the resulting global topological number, including configurations in which contributions with opposite winding numbers cancel. By considering both thermodynamic descriptions within the same model, we aim to clarify their respective roles in characterizing the phase structure of nonminimally coupled AdS black holes.

The paper is organized as follows. In Sec.~\ref{sec:2}, we obtain the perturbative black hole solutions associated with the nonminimal interaction. Section~\ref{sec:3} develops the thermodynamic framework and examines conventional criticality in the canonical and grand canonical ensembles, including a discussion of the validity of the perturbative approximation. In Sec.~\ref{sec:4}, we introduce the topological approach and determine the thermodynamic winding numbers in both ensembles, discussing their dependence on the relevant thermodynamic parameters. Finally, Sec.~\ref{sec:5} summarizes our results and conclusions.

\section{Perturbative Solutions of $F_{\alpha}^{\gamma} F^{\alpha\beta} F_{\beta}^{\eta} F^{\lambda\mu} R_{\gamma\eta\lambda\mu}$--AdS black holes}
\label{sec:2}

We consider four-dimensional Einstein--Maxwell gravity with a negative cosmological constant and a nonminimal coupling between the electromagnetic field and spacetime curvature. The corresponding action is given by 
\begin{equation}\label{action}
	S=\frac{1}{\kappa}\int d^4x\sqrt{-g}\left[
	(R-2\Lambda)-\frac{\alpha}{4}F_{\mu\alpha}F^{\mu\alpha}
	+\epsilon F_{\alpha}{}^{\gamma}F^{\alpha\beta}
	F_{\beta}{}^{\eta}F^{\lambda\mu}R_{\gamma\eta\lambda\mu}
	\right],
\end{equation}
where $\kappa$ is the gravitational coupling, $R$ is the Ricci scalar, and $\Lambda=-3/\ell^2$ is the cosmological constant, with $\ell$ denoting the AdS radius. The parameter $\alpha$ is dimensionless, and $\epsilon$ is the coupling constant associated with the nonminimal interaction. In units where $\hbar=c=1$, the dimension of $\epsilon$ is $L^4$. We also introduce the Maxwell invariant $\mathcal{F}=F_{\mu\nu}F^{\mu\nu}$, where the electromagnetic field strength is defined as
\begin{equation}\label{YM}
	F_{\mu\nu}=\partial_\mu A_\nu-\partial_\nu A_\mu,
\end{equation}
with $A_\mu$ denoting the electromagnetic gauge potential.

Variation of the action with respect to the spacetime metric yields the modified Einstein equations,
\begin{equation}\label{EOM1}
	R_{\mu\nu}-\frac{1}{2}g_{\mu\nu}R+\Lambda g_{\mu\nu}
	= T_{\mu\nu}^{\text{(eff)}},
\end{equation}
where
\begin{equation}
	T_{\mu\nu}^{\text{(eff)}}=
	\alpha T_{\mu\nu}^{\text{(M)}}+\epsilon T_{\mu\nu}^{\text{(I)}}.
\end{equation}
Here, $T_{\mu\nu}^{\text{(M)}}$ is the Maxwell energy--momentum tensor,
\begin{equation}
	T_{\mu\nu}^{\text{(M)}}=
	\frac{1}{2}F_{\mu}{}^{\alpha}F_{\nu\alpha}
	-\frac{1}{8}g_{\mu\nu}F_{\alpha\beta}F^{\alpha\beta},
\end{equation}
while $T_{\mu\nu}^{\text{(I)}}$ represents the contribution arising from the nonminimal interaction. See Appendix \ref{app:A}. 

Variation with respect to the gauge potential $A_\mu$ gives the modified Maxwell equations:
\begin{eqnarray}\label{EOM-Maxwell}
	\nabla_{\mu}\Big(&-\frac{\alpha}{2}F^{\mu\nu}
	+\epsilon F^{\gamma\eta}F^{\mu\alpha}F^{\nu\beta}R_{\alpha\beta\gamma\eta}
	+\epsilon F_{\alpha}{}^{\beta}F^{\gamma\eta}F^{\nu\alpha}
	R_{\beta}{}^{\mu}{}_{\gamma\eta}\nonumber\\
	&-\epsilon F_{\alpha}{}^{\beta}F^{\gamma\eta}F^{\mu\alpha}
	R_{\beta}{}^{\nu}{}_{\gamma\eta}
	+\epsilon F_{\alpha}{}^{\gamma}F^{\alpha\beta}
	F_{\beta}{}^{\eta}R_{\gamma\eta}{}^{\mu\nu}
	\Big)=0.
\end{eqnarray}

To construct static, electrically charged solutions with spherical symmetry, we adopt the metric ansatz
\begin{equation}\label{metric}
	ds^2=-e^{-2H(r)}f(r)dt^2+\frac{dr^2}{f(r)}
	+r^2d\theta^2+r^2\sin^2\theta\,d\varphi^2,
\end{equation}
where $f(r)$ is the metric function and $H(r)$ is the redshift function. For the electromagnetic field, we consider the purely electric ansatz
\begin{equation}\label{background}
	A_\mu=(h(r),0,0,0),
\end{equation}
where $h(r)$ is the radial electric potential.

The coupled Einstein and Maxwell equations are generally difficult to solve analytically in the presence of the nonminimal interaction. We therefore employ a perturbative expansion in the coupling parameter \(\epsilon\) and write
\begin{align}
	f(r)&=f_0(r)+\epsilon f_1(r),\label{f}\\
	h(r)&=h_0(r)+\epsilon h_1(r),\label{h}\\
	H(r)&=H_0(r)+\epsilon H_1(r).\label{H}
\end{align}
Here, the functions with subscript \(0\) represent the leading-order solutions, whereas those with subscript \(1\) describe the first-order corrections. The perturbative treatment is valid provided that the nonminimal corrections remain sufficiently small compared with the corresponding leading-order contributions. In particular, the dimensionless coupling \(\epsilon/\ell^4\) should be small, although the validity of the expansion may also depend on the black hole parameters. See  Appendix \ref{app:B}.

At zeroth order in $\epsilon$, the field equations reduce to those of the minimally coupled Einstein--Maxwell theory with a negative cosmological constant. The resulting Reissner--Nordström--AdS solution is given by
\begin{equation}\label{f0}
	f_0(r)=1-\frac{2m_0}{r}-\frac{\Lambda r^2}{3}
	+\frac{\kappa\alpha Q^2}{4r^2},
\end{equation}
where the integration constant $m_0$ is determined by the horizon condition $f_0(r_h)=0$:
\begin{equation}\label{m0}
	m_0=\frac{r_h}{2}-\frac{\Lambda r_h^3}{6}
	+\frac{\kappa\alpha Q^2}{8r_h}.
\end{equation}

The corresponding electric potential, in a gauge in which it vanishes at the horizon, is given by
\begin{equation}\label{h0}
	h_0(r)=Q\left(\frac{1}{r}-\frac{1}{r_h}\right).
\end{equation}

Furthermore, subtracting the $tt$ and $rr$ components of the zeroth-order Einstein equations gives
\begin{equation}
	-2f_0(r)rH_0'(r)=0.
\end{equation}
Since $f_0(r)\neq0$ outside the horizon, this implies that $H_0(r)$ is constant. The remaining constant can be absorbed into a rescaling of the time coordinate, and we consequently set
\begin{equation}\label{H0}
	H_0(r)=0.
\end{equation}

The first-order field equations yield differential equations for the correction functions. The explicit forms of the \(tt\) and \(rr\) components of the modified Einstein equations, together with the Maxwell equations, are presented in Appendix~\ref{app:A}. Solving this system, we obtain
\begin{equation}\label{H1}
	H_1(r)=\frac{21\kappa Q^4}{4r^8},
\end{equation}
\begin{equation}\label{h1}
	h_1(r)=-\frac{7\kappa Q^5}{12r^9}
	+\frac{2\kappa Q^3}{7r^7}
	-\frac{4\kappa Q^3\Lambda}{5r^5}
	-\frac{8Q\Lambda}{3\alpha r^3}
	+\frac{16Q\Lambda^2}{\alpha r}+A,
\end{equation}
and
\begin{align}\label{f1}
	f_1(r)={}&\frac{2\alpha\kappa^2Q^6}{r^{10}}
	-\frac{41m_0\kappa Q^4}{2r^9}
	+\frac{12\kappa Q^4}{r^8}
	+\frac{\alpha\kappa^2Q^4}{7r^8}\nonumber\\
	&-\frac{64\kappa Q^4\Lambda}{15r^6}
	-\frac{2\alpha\kappa^2Q^4\Lambda}{5r^6}
	-\frac{4\kappa Q^2\Lambda}{3r^4}
	+\frac{8\kappa Q^2\Lambda^2}{r^2}
	+\frac{B}{r},
\end{align}
where \(A\) and \(B\) are integration constants.

To determine these constants, we impose the horizon
conditions on the perturbed solution. In particular, the
corrected horizon radius is written as
\begin{equation}
	r_h'=r_h+\epsilon r_h^{(1)},
\end{equation}
where $r_h$ denotes the horizon radius of the zeroth-order
geometry and $r_h^{(1)}$ represents its first-order
correction. We first require the electromagnetic potential
to vanish at the corrected horizon, namely
$h(r_h')=0$. Expanding this condition to first order in
$\epsilon$, we obtain
\begin{align}
	h(r_h+\epsilon r_h^{(1)})
	={}&h_0(r_h)
	+\epsilon\left[
	h_0'(r_h)r_h^{(1)}+h_1(r_h)
	\right]\nonumber\\
	&+\mathcal{O}(\epsilon^2)=0.
\end{align}
Since $h_0(r_h)=0$, the first-order correction to the
horizon radius is given by
\begin{equation}
	r_h^{(1)}=-\frac{h_1(r_h)}{h_0'(r_h)}.
\end{equation}
Consequently, the corrected horizon can be expressed as
\begin{equation}\label{ph}
	r_h'=r_h-\epsilon
	\frac{h_1(r_h)}{h_0'(r_h)}.
\end{equation}

The condition $h(r_h')=0$ determines the integration
constant $A$. Substituting the resulting constant into
the first-order gauge potential, we obtain
\begin{align}\label{hfinal}
	h(r)={}&Q\left(\frac{1}{r}-\frac{1}{r_h}\right)\nonumber\\
	&+\epsilon\Bigg[
	-\frac{7\kappa Q^5}{12}
	\left(\frac{1}{r^9}-\frac{1}{r_h^9}\right)
	+\frac{2\kappa Q^3}{7}
	\left(\frac{1}{r^7}-\frac{1}{r_h^7}\right)\nonumber\\
	&\qquad
	-\frac{4\kappa Q^3\Lambda}{5}
	\left(\frac{1}{r^5}-\frac{1}{r_h^5}\right)
	-\frac{8Q\Lambda}{3\alpha}
	\left(\frac{1}{r^3}-\frac{1}{r_h^3}\right)\nonumber\\
	&\qquad
	+\frac{16Q\Lambda^2}{\alpha}
	\left(\frac{1}{r}-\frac{1}{r_h}\right)
	\Bigg].
\end{align}
It is then evident that $h_1(r_h)=0$, and therefore
Eq.~(\ref{ph}) gives $r_h^{(1)}=0$.

We now impose the metric horizon condition
$f(r_h')=0$. Its expansion to first order in $\epsilon$
reads
\begin{align}
	f(r_h+\epsilon r_h^{(1)})
	={}&f_0(r_h)
	+\epsilon\left[
	f_0'(r_h)r_h^{(1)}+f_1(r_h)
	\right]\nonumber\\
	&+\mathcal{O}(\epsilon^2)=0.
\end{align}
Using $f_0(r_h)=0$ and $r_h^{(1)}=0$, we find
\begin{equation}
	f_1(r_h)=0.
\end{equation}
This condition determines the remaining integration
constant $B$ and ensures that the corrected metric
function vanishes at the same horizon radius to the
considered perturbative order.

The resulting metric function can be written as
\begin{align}\label{fr}
	f(r)={}&1-\frac{\Lambda r^2}{3}-\frac{2M}{r}
	+\frac{8\epsilon\kappa Q^2\Lambda^2
		+\frac{1}{4}\kappa\alpha Q^2}{r^2}
	-\frac{4\epsilon\kappa Q^2\Lambda}{3r^4}\nonumber\\
	&+\frac{\epsilon\left(-\frac{64}{15}\kappa Q^4\Lambda
		-\frac{2}{5}\alpha\kappa^2Q^4\Lambda\right)}{r^6}
	+\frac{\epsilon\left(12\kappa Q^4
		+\frac{1}{7}\alpha\kappa^2Q^4\right)}{r^8}\nonumber\\
	&-\frac{41\epsilon\left(
		-\frac{r_h^3\Lambda}{6}
		+\frac{\kappa\alpha Q^2}{8r_h}
		+\frac{r_h}{2}\right)\kappa Q^4}{2r^9}
	+\frac{2\epsilon\alpha\kappa^2Q^6}{r^{10}}.
\end{align}
The corresponding mass parameter is
\begin{align}\label{Mass}
	M={}&\epsilon\left(
	\frac{4\kappa Q^2\Lambda^2}{r_h}
	-\frac{\alpha\kappa^2Q^4\Lambda}{5r_h^5}
	-\frac{17\kappa Q^4\Lambda}{40r_h^5}
	-\frac{2\kappa Q^2\Lambda}{3r_h^3}\right.\nonumber\\
	&\left.\qquad
	-\frac{9\alpha\kappa^2Q^6}{32r_h^9}
	+\frac{\alpha\kappa^2Q^4}{14r_h^7}
	+\frac{7\kappa Q^4}{8r_h^7}
	\right)\nonumber\\
	&-\frac{r_h^3\Lambda}{6}
	+\frac{\kappa\alpha Q^2}{8r_h}
	+\frac{r_h}{2}
	=m_0+\epsilon m_1,
\end{align}
where $m_1$ denotes the first-order correction to the
mass parameter.

Finally, the asymptotic behavior of the temporal metric
component, $g_{tt}=-f(r)e^{-2H(r)}$, is given by
\begin{equation}
	g_{tt}=-\left[
	1-\frac{\Lambda r^2}{3}
	-\frac{2(m_0+\epsilon m_1)}{r}
	+\mathcal{O}\left(\frac{1}{r^2}\right)
	\right].
\end{equation}
The nonminimal corrections do not modify the leading
asymptotic mass term. Thus, the mass parameter is
identified as $M=m_0+\epsilon m_1$. 

As an additional check of the mass obtained from the asymptotic behavior of the metric, we consider the \(tt\)-component of the renormalized boundary stress-energy tensor using the counterterm prescription of Ref.~\cite{Balasubramanian:1999re}. This gives
\begin{equation}
	8\pi G T_{tt}^{\mathrm{ren}}
	= \frac{2Gm}{\ell r}+\mathcal{O}(r^{-2}).
\end{equation}
This result agrees with the corresponding expression
\begin{equation}
	8\pi G T_{tt}^{\mathrm{ren}} = \frac{r_0}{\ell r} + \mathcal{O}(r^{-2})
\end{equation}
given in Ref.~\cite{Balasubramanian:1999re}, where \(r_0=2Gm\). Thus, the holographic mass is identified as \(M_{\mathrm{hol}}=m\), where \(m=m_0+\epsilon m_1\) is determined by imposing the horizon condition on the perturbed metric function.


\section{Thermodynamic ensembles and criticality}\label{sec:3}

In this section, we derive the thermodynamic relations required to investigate the critical behavior of the system in both the canonical and grand canonical ensembles. By identifying the cosmological constant as $\Lambda=-8\pi P$, the mass relation (\ref{Mass}) can be expressed as the enthalpy in the extended phase space, which we denote by $\mathcal{H}$ to avoid confusion with the redshift function $H$. Throughout this section, we set $\kappa=\alpha=1$. The resulting enthalpy is

\begin{equation}\label{ent}
	\mathcal{H}=\left( \frac{256 Q^{2} \pi^{2} P^{2}}{r_{h}} + \frac{5 Q^{4} \pi P}{r_{h}^{5}} + \frac{16 Q^{2} \pi P}{3 r_{h}^{3}} - \frac{9 Q^{6}}{32 r_{h}^{9}} + \frac{53 Q^{4}}{56 r_{h}^{7}} \right) \epsilon + \frac{4 r_{h}^{3} \pi P}{3} + \frac{Q^{2}}{8 r_{h}} + \frac{r_{h}}{2}.
\end{equation}

The Hawking temperature is given by

\begin{equation}\label{Temp1}
	\begin{split}
		T_{H} &= \frac{1}{2\pi} \left[ \frac{1}{\sqrt{g_{rr}}} \frac{d}{dr} \sqrt{-g_{tt}} \right] \Bigg|_{r = r_h}
		= 2 P r_{h} - \frac{Q^{2}}{16 r_{h}^{3} \pi} + \frac{1}{4 \pi r_{h}}\\
		&+ \left( -\frac{128 \pi P^{2} Q^{2}}{r_{h}^{3}} - \frac{5 Q^{4} P}{2 r_{h}^{7}} - \frac{8 P Q^{2}}{r_{h}^{5}} + \frac{61 Q^{6}}{64 \pi r_{h}^{11}} - \frac{33 Q^{4}}{16 \pi r_{h}^{9}} \right) \epsilon.
	\end{split}
\end{equation}

Requiring that the first law of black hole thermodynamics holds, and treating the involved quantities perturbatively around \(\epsilon=0\), one obtains the entropy
\begin{equation}\label{entropy1}
	S=\int\frac{\partial \mathcal{H}/\partial r_h}{\partial S/\partial r_h}\,dr_h
	=\pi r_h^2+\frac{5\pi Q^4\epsilon}{3r_h^6}.
\end{equation}

Substituting the entropy and enthalpy into
\[
T=
\frac{
	\left(\dfrac{\partial \mathcal{H}}{\partial r_h}\right)_{P,Q,\epsilon}
}{
	\left(\dfrac{\partial S}{\partial r_h}\right)_{P,Q,\epsilon}
},
\]
we find that the resulting temperature exactly reproduces the Hawking temperature obtained from the surface gravity.

The first law of thermodynamics in the canonical ensemble takes the form
\begin{equation}
	d\mathcal{H}=T\,dS+V\,dP+\psi\,dQ+\Xi\,d\epsilon,
\end{equation}
where the coupling parameter \(\epsilon\) and its conjugate quantity \(\Xi\) are included to ensure consistency with the Smarr relation to first order in \(\epsilon\).

The equation of state is obtained by expanding the pressure as \(P=P_0+P_1\epsilon\) and substituting this expansion into the temperature expression~(\ref{Temp1}). This yields
\begin{equation}\label{pre}
	\begin{split}
		P ={}& \frac{T}{2r_h} + \frac{Q^2}{32\pi r_h^4} - \frac{1}{8\pi r_h^2} \\
		&+ \epsilon\left(
		\frac{16\pi Q^2 T^2}{r_h^6}
		- \frac{6Q^2 T}{r_h^7}
		+ \frac{Q^2}{2\pi r_h^8}
		+ \frac{21Q^4 T}{8r_h^9}
		\right. \\
		&\qquad\qquad\left.
		+ \frac{Q^4}{2\pi r_h^{10}}
		- \frac{3Q^6}{8\pi r_h^{12}}
		\right).
	\end{split}
\end{equation}

The thermodynamic volume is defined as
\begin{equation}\label{volume}
	\begin{aligned}
		V
		&=
		\left(
		\frac{\partial\mathcal{H}}{\partial P}
		\right)_{Q,S,\epsilon}=\left(
		\frac{\partial\mathcal{H}}{\partial P}
		\right)_{Q,r_h,\epsilon}
		-
		\frac{
			\left(
			\frac{\partial\mathcal{H}}{\partial r_h}
			\right)_{P,Q,\epsilon}
			\left(
			\frac{\partial S}{\partial P}
			\right)_{Q,r_h,\epsilon}
		}{
			\left(
			\frac{\partial S}{\partial r_h}
			\right)_{P,Q,\epsilon}
		}
		\\
		&=
		\frac{4\pi r_h^3}{3} + \epsilon\left(
		\frac{512\pi^2 Q^2 P}{r_h}
		+ \frac{5\pi Q^4}{r_h^5}
		+ \frac{16\pi Q^2}{3r_h^3}
		\right).
	\end{aligned}
\end{equation}

Similarly, the thermodynamic quantity \(\Xi\), conjugate to the coupling parameter \(\epsilon\), is obtained from the extended first law. Its expression is
\begin{equation}\label{Xi}
	\begin{aligned}
		\Xi
		&=
		\left(
		\frac{\partial\mathcal{H}}{\partial\epsilon}
		\right)_{S,P,Q}=\left(
		\frac{\partial\mathcal{H}}{\partial\epsilon}
		\right)_{P,Q,r_h}
		-
		\frac{
			\left(
			\frac{\partial\mathcal{H}}{\partial r_h}
			\right)_{P,Q,\epsilon}
			\left(
			\frac{\partial S}{\partial\epsilon}
			\right)_{P,Q,r_h}
		}{
			\left(
			\frac{\partial S}{\partial r_h}
			\right)_{P,Q,\epsilon}
		}
		\\
		&=
		\frac{256\pi^2 Q^2 P^2}{r_h}
		+ \frac{5\pi Q^4 P}{3r_h^5}
		+ \frac{16\pi Q^2 P}{3r_h^3}
		- \frac{17Q^6}{96r_h^9}
		+ \frac{89Q^4}{168r_h^7}
		\\
		&\quad
		+ \epsilon\left(
		\frac{640\pi^2 Q^6 P^2}{3r_h^9}
		+ \frac{25\pi Q^8 P}{6r_h^{13}}
		+ \frac{40\pi Q^6 P}{3r_h^{11}}
		- \frac{305Q^{10}}{192r_h^{17}}
		+ \frac{55Q^8}{16r_h^{15}}
		\right)
		\\
		&=
		\Xi_0+\epsilon\Xi_1.
	\end{aligned}
\end{equation}

The electric potential conjugate to the charge is given by
Equation~(\ref{potential}) This expression provides the relation needed to express the thermodynamic quantities in terms of the electric potential rather than the charge, thereby allowing us to formulate the thermodynamics in the grand canonical ensemble.
\begin{equation}\label{potential}
	\begin{aligned}
		\psi
		&=
		\left(
		\frac{\partial\mathcal{H}}{\partial Q}
		\right)_{P,S}=
		\left(
		\frac{\partial\mathcal{H}}{\partial Q}
		\right)_{P,r_h}
		-
		\frac{
			\left(
			\frac{\partial\mathcal{H}}{\partial r_h}
			\right)_{P,Q}
			\left(
			\frac{\partial S}{\partial Q}
			\right)_{P,r_h}
		}{
			\left(
			\frac{\partial S}{\partial r_h}
			\right)_{P,Q}
		}
		\\
		&=\frac{Q}{4 r_h} + \left( \frac{512 Q \pi^2 P^2}{r_h} + \frac{20 Q^3 \pi P}{3 r_h^5} + \frac{32 Q \pi P}{3 r_h^3} - \frac{61 Q^5}{48 r_h^9} + \frac{89 Q^3}{42 r_h^7} \right) \epsilon.
	\end{aligned}
\end{equation}
This relation allows us to express the thermodynamic quantities in terms of the electric potential rather than the charge, thereby transforming the description from the canonical to the grand canonical ensemble.

To first order in \(\epsilon\), the Smarr relation takes the form
\begin{equation}
	\mathcal{H}
	=2TS-2PV+\psi Q+4\Xi_0\epsilon.
\end{equation}

To obtain the thermodynamic relations in the **grand canonical ensemble**, we solve Eq.~(\ref{potential}) perturbatively for \(Q\) in terms of \(\psi\), retaining the physically admissible real solution. Substituting this solution into the canonical thermodynamic expressions, we express the relevant quantities in terms of the electric potential.

The enthalpy in the grand canonical ensemble is then given by
\begin{equation}\label{MG}
	\begin{split}
		\mathcal{M} = {} & \frac{4 r_{h}^{3} \pi P}{3} + 2 \psi^{2} r_{h} + \frac{r_{h}}{2} \\
		& + \left( -4096 \psi^{2} r_{h} \pi^{2} P^{2} - \frac{1280 \psi^{4} \pi P}{3 r_{h}} - \frac{256 \psi^{2} \pi P}{3 r_{h}} \right. \\
		& \qquad \left. + \frac{12160 \psi^{6}}{3 r_{h}^{3}} - \frac{6304 \psi^{4}}{21 r_{h}^{3}} \right) \epsilon.
	\end{split}
\end{equation}

The corresponding Hawking temperature is obtained by substituting the perturbative solution for \(Q\) into Eq.~(\ref{Temp1}), yielding
\begin{equation}\label{TG}
	T=2 r_{h} P - \frac{\psi^{2}}{r_{h} \pi} + \frac{1}{4 r_{h} \pi} + \left( \frac{2048 \pi \psi^{2} P^{2}}{r_{h}} + \frac{640 \psi^{4} P}{3 r_{h}^{3}} - \frac{128 \psi^{2} P}{3 r_{h}^{3}} + \frac{3904 \psi^{6}}{3 r_{h}^{5} \pi} - \frac{5392 \psi^{4}}{21 r_{h}^{5} \pi} \right) \epsilon.
\end{equation}

The corresponding entropy is
\begin{equation}\label{SG}
	S=\pi r_{h}^{2}+\frac{1280 \psi^{4} \pi \epsilon}{3 r_{h}^{2}}.
\end{equation}

To verify the consistency of these expressions with the first law, we note that, at fixed \(P\), \(\psi\), and \(\epsilon\), the thermodynamic quantities can be expressed as functions of the horizon radius \(r_h\), with \(Q=Q(r_h)\). Therefore, the first law yields
\begin{equation}\label{TGfirstlaw}
	T=
	\frac{
		\left(\frac{\partial\mathcal{M}}{\partial r_h}\right)_{P,\psi,\epsilon}
		-\psi\left(\frac{\partial Q}{\partial r_h}\right)_{P,\psi,\epsilon}
	}{
		\left(\frac{\partial S}{\partial r_h}\right)_{P,\psi,\epsilon}
	}.
\end{equation}
Substituting the corresponding expressions for \(\mathcal{M}\), \(Q\), and \(S\) into this relation reproduces the Hawking temperature given in Eq.~(\ref{TG}).

We first investigate the critical behavior of the system in both  thermodynamic ensembles using the corresponding temperature relations. We then turn to the topological analysis of the black hole solutions. We do not pursue the conventional thermodynamic analysis further, as the winding numbers and the \(r_h\)--\(\tau\) diagrams provide complementary information about the local stability and global structure of the thermodynamic states.

\subsection{Criticality in the canonical ensemble}

We now turn to the conventional thermodynamic analysis of the corrected solutions. The critical behavior of the system can be investigated using the local critical-point conditions
\begin{equation}\label{crit_t}
	\left(\frac{\partial T}{\partial r_h}\right)_{P,Q}
	=
	\left(\frac{\partial^2 T}{\partial r_h^2}\right)_{P,Q}
	=0,
\end{equation}
which identify the inflection points of the isobars in the \(T\)--\(r_h\) plane.
 In the canonical ensemble, the black hole charge
\(Q\) is held fixed, and we apply this condition along isobars, i.e.
at constant \(P\) and \(Q\). For \(\epsilon=10^{-4}\), the resulting
critical-point coordinates are collected in
Table~\ref{table1}.

\begin{table}[htbp]
	\centering
	\caption{Critical point coordinates for $\epsilon=10^{-4}$.
		The calculations show no critical behavior for $Q=0.01$--$0.14$,
		whereas critical points exist for $Q\geq 0.15$.
		The numerical scan was extended up to $Q=2.00$.}
	\label{table1}
	
	\begingroup
	\scriptsize
	\setlength{\tabcolsep}{2.7pt}
	\renewcommand{\arraystretch}{0.88}
	
	\begin{minipage}[t]{0.49\textwidth}
		\vspace{0pt}
		\centering
		\begin{tabular}{ccccc}
			\toprule
			$Q$ & $r_c$ & $P_c$ & $T_c$ & Status \\
			\midrule
			\rowcolor{gray!25} 0.01 & -- & -- & -- & no critical \\
			\rowcolor{gray!25} 0.02 & -- & -- & -- & no critical \\
			\rowcolor{gray!25} 0.03 & -- & -- & -- & no critical \\
			\rowcolor{gray!25} 0.04 & -- & -- & -- & no critical \\
			\rowcolor{gray!25} 0.05 & -- & -- & -- & no critical \\
			\rowcolor{gray!25} 0.06 & -- & -- & -- & no critical \\
			\rowcolor{gray!25} 0.07 & -- & -- & -- & no critical \\
			\rowcolor{gray!25} 0.08 & -- & -- & -- & no critical \\
			\rowcolor{gray!25} 0.09 & -- & -- & -- & no critical \\
			\rowcolor{gray!25} 0.10 & -- & -- & -- & no critical \\
			\rowcolor{gray!25} 0.11 & -- & -- & -- & no critical \\
			\rowcolor{gray!25} 0.12 & -- & -- & -- & no critical \\
			\rowcolor{gray!25} 0.13 & -- & -- & -- & no critical \\
			\rowcolor{gray!25} 0.14 & -- & -- & -- & no critical \\
			\midrule
			0.15 & 0.286729 & 0.304695 & 0.424425 & critical \\
			0.16 & 0.296480 & 0.282915 & 0.408597 & critical \\
			0.17 & 0.306103 & 0.263417 & 0.393900 & critical \\
			0.18 & 0.315622 & 0.245865 & 0.380198 & critical \\
			0.19 & 0.325060 & 0.229986 & 0.367378 & critical \\
			0.20 & 0.334436 & 0.215558 & 0.355346 & critical \\
			0.21 & 0.343770 & 0.202399 & 0.344022 & critical \\
			0.22 & 0.353076 & 0.190354 & 0.333338 & critical \\
			0.23 & 0.362371 & 0.179296 & 0.323236 & critical \\
			0.24 & 0.371668 & 0.169115 & 0.313667 & critical \\
			0.25 & 0.380979 & 0.159718 & 0.304586 & critical \\
			0.26 & 0.390316 & 0.151025 & 0.295955 & critical \\
			0.27 & 0.399688 & 0.142967 & 0.287740 & critical \\
			0.28 & 0.409106 & 0.135484 & 0.279911 & critical \\
			0.29 & 0.418577 & 0.128523 & 0.272442 & critical \\
			0.30 & 0.428109 & 0.122037 & 0.265309 & critical \\
			0.31 & 0.437708 & 0.115985 & 0.258490 & critical \\
			0.32 & 0.447379 & 0.110331 & 0.251967 & critical \\
			0.33 & 0.457127 & 0.105043 & 0.245720 & critical \\
			0.34 & 0.466955 & 0.100090 & 0.239735 & critical \\
			0.35 & 0.476867 & 0.095448 & 0.233996 & critical \\
			0.36 & 0.486863 & 0.091091 & 0.228491 & critical \\
			0.37 & 0.496946 & 0.087000 & 0.223206 & critical \\
			0.38 & 0.507117 & 0.083154 & 0.218129 & critical \\
			0.39 & 0.517374 & 0.079535 & 0.213251 & critical \\
			0.40 & 0.527718 & 0.076127 & 0.208560 & critical \\
			0.41 & 0.538147 & 0.072916 & 0.204048 & critical \\
			0.42 & 0.548660 & 0.069888 & 0.199706 & critical \\
			0.43 & 0.559256 & 0.067031 & 0.195526 & critical \\
			0.44 & 0.569931 & 0.064332 & 0.191499 & critical \\
			0.45 & 0.580684 & 0.061781 & 0.187619 & critical \\
			0.46 & 0.591512 & 0.059369 & 0.183878 & critical \\
			0.47 & 0.602412 & 0.057086 & 0.180270 & critical \\
			0.48 & 0.613381 & 0.054924 & 0.176789 & critical \\
			0.49 & 0.624417 & 0.052875 & 0.173429 & critical \\
			0.50 & 0.635516 & 0.050932 & 0.170184 & critical \\
			\bottomrule
		\end{tabular}
	\end{minipage}
	\hfill
	\begin{minipage}[t]{0.49\textwidth}
		\vspace{0pt}
		\centering
		\begin{tabular}{ccccc}
			\toprule
			$Q$ & $r_c$ & $P_c$ & $T_c$ & Status \\
			\midrule
			0.51 & 0.646676 & 0.049088 & 0.167049 & critical \\
			0.52 & 0.657893 & 0.047338 & 0.164020 & critical \\
			0.53 & 0.669166 & 0.045675 & 0.161091 & critical \\
			0.54 & 0.680490 & 0.044094 & 0.158259 & critical \\
			0.55 & 0.691864 & 0.042589 & 0.155518 & critical \\
			0.56 & 0.703284 & 0.041158 & 0.152865 & critical \\
			0.57 & 0.714748 & 0.039794 & 0.150296 & critical \\
			0.58 & 0.726255 & 0.038494 & 0.147808 & critical \\
			0.59 & 0.737800 & 0.037255 & 0.145396 & critical \\
			0.60 & 0.749383 & 0.036072 & 0.143058 & critical \\
			0.61 & 0.761001 & 0.034943 & 0.140791 & critical \\
			0.62 & 0.772652 & 0.033865 & 0.138592 & critical \\
			0.63 & 0.784334 & 0.032834 & 0.136457 & critical \\
			0.64 & 0.796045 & 0.031848 & 0.134385 & critical \\
			0.65 & 0.807785 & 0.030905 & 0.132372 & critical \\
			0.66 & 0.819550 & 0.030002 & 0.130417 & critical \\
			0.67 & 0.831340 & 0.029137 & 0.128517 & critical \\
			0.68 & 0.843153 & 0.028308 & 0.126669 & critical \\
			0.69 & 0.854988 & 0.027513 & 0.124873 & critical \\
			0.70 & 0.866844 & 0.026750 & 0.123125 & critical \\
			0.71 & 0.878719 & 0.026018 & 0.121424 & critical \\
			0.72 & 0.890612 & 0.025315 & 0.119768 & critical \\
			0.73 & 0.902523 & 0.024640 & 0.118156 & critical \\
			0.74 & 0.914450 & 0.023991 & 0.116586 & critical \\
			0.75 & 0.926392 & 0.023367 & 0.115055 & critical \\
			0.76 & 0.938349 & 0.022766 & 0.113564 & critical \\
			0.77 & 0.950319 & 0.022188 & 0.112110 & critical \\
			0.78 & 0.962303 & 0.021631 & 0.110692 & critical \\
			0.79 & 0.974299 & 0.021095 & 0.109309 & critical \\
			0.80 & 0.986306 & 0.020578 & 0.107959 & critical \\
			0.81 & 0.998324 & 0.020080 & 0.106642 & critical \\
			0.82 & 1.010353 & 0.019599 & 0.105356 & critical \\
			0.83 & 1.022391 & 0.019136 & 0.104100 & critical \\
			0.84 & 1.034439 & 0.018688 & 0.102873 & critical \\
			0.85 & 1.046495 & 0.018256 & 0.101675 & critical \\
			0.86 & 1.058560 & 0.017838 & 0.100503 & critical \\
			0.87 & 1.070633 & 0.017434 & 0.099358 & critical \\
			0.88 & 1.082713 & 0.017044 & 0.098239 & critical \\
			0.89 & 1.094800 & 0.016667 & 0.097144 & critical \\
			0.90 & 1.106894 & 0.016302 & 0.096073 & critical \\
			0.91 & 1.118994 & 0.015948 & 0.095025 & critical \\
			0.92 & 1.131100 & 0.015606 & 0.093999 & critical \\
			0.93 & 1.143213 & 0.015275 & 0.092995 & critical \\
			0.94 & 1.155330 & 0.014954 & 0.092013 & critical \\
			0.95 & 1.167453 & 0.014643 & 0.091050 & critical \\
			0.96 & 1.179581 & 0.014342 & 0.090107 & critical \\
			0.97 & 1.191713 & 0.014050 & 0.089184 & critical \\
			0.98 & 1.203851 & 0.013766 & 0.088279 & critical \\
			0.99 & 1.215992 & 0.013491 & 0.087392 & critical \\
			1.00 & 1.228137 & 0.013224 & 0.086522 & critical \\
			\bottomrule
		\end{tabular}
	\end{minipage}
	\endgroup
\end{table}

Repeating this analysis for \(\epsilon=10^{-3}\) and \(\epsilon=10^{-5}\) yields the same qualitative behavior as that found for \(\epsilon=10^{-4}\). However, the non-critical (NC) region shrinks as \(\epsilon\) decreases, as expected: in the limit \(\epsilon\to0\), the solution reduces to the Reissner--Nordström--AdS black hole, for which the canonical ensemble exhibits a critical point for every nonzero value of the electric charge.

We emphasize that the boundary of the NC region reported here is estimated by scanning discrete values of the electric charge, as listed in Table~\ref{table1}, rather than by determining the critical charge \(Q_c\) analytically. Thus, \(Q\simeq0.14\) should be understood as the largest tabulated charge for which no real critical point is found, with the actual boundary lying between the last NC entry and the first entry exhibiting criticality.

For (Q=0), the solution reduces to Schwarzschild--AdS, which does not exhibit van der Waals-type criticality; instead, it undergoes a Hawking--Page phase transition. The NC region is characterized by the absence of a critical point: for \(0<Q\lesssim 0.14\) at \(\epsilon=10^{-4}\), the criticality conditions admit no real solution. The presence of charge is therefore not sufficient to restore criticality in this range. As shown by the topological analysis in Sec.~\ref{sec:4}, however, the total winding number in this region is \(W=+1\), and a first-order transition between small and large charged black hole branches is nevertheless present, although the associated isobars do not exhibit the van der Waals form.

\begin{figure}[H]
	\centering
	\begin{subfigure}[b]{0.4\textwidth}
		\centering
		\includegraphics[width=\textwidth]{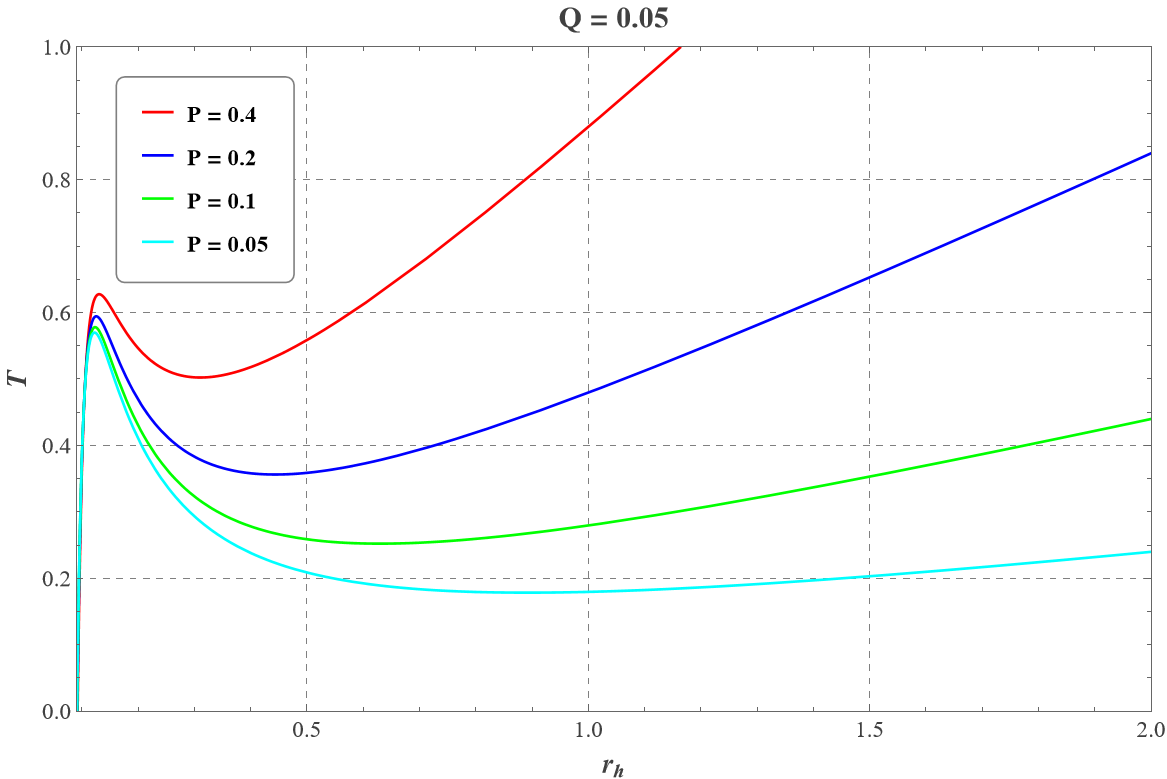}
		\caption{$Q=0.05$}
		\label{fig:temp_a}
	\end{subfigure}
	\quad
	\begin{subfigure}[b]{0.4\textwidth}
		\centering
		\includegraphics[width=\textwidth]{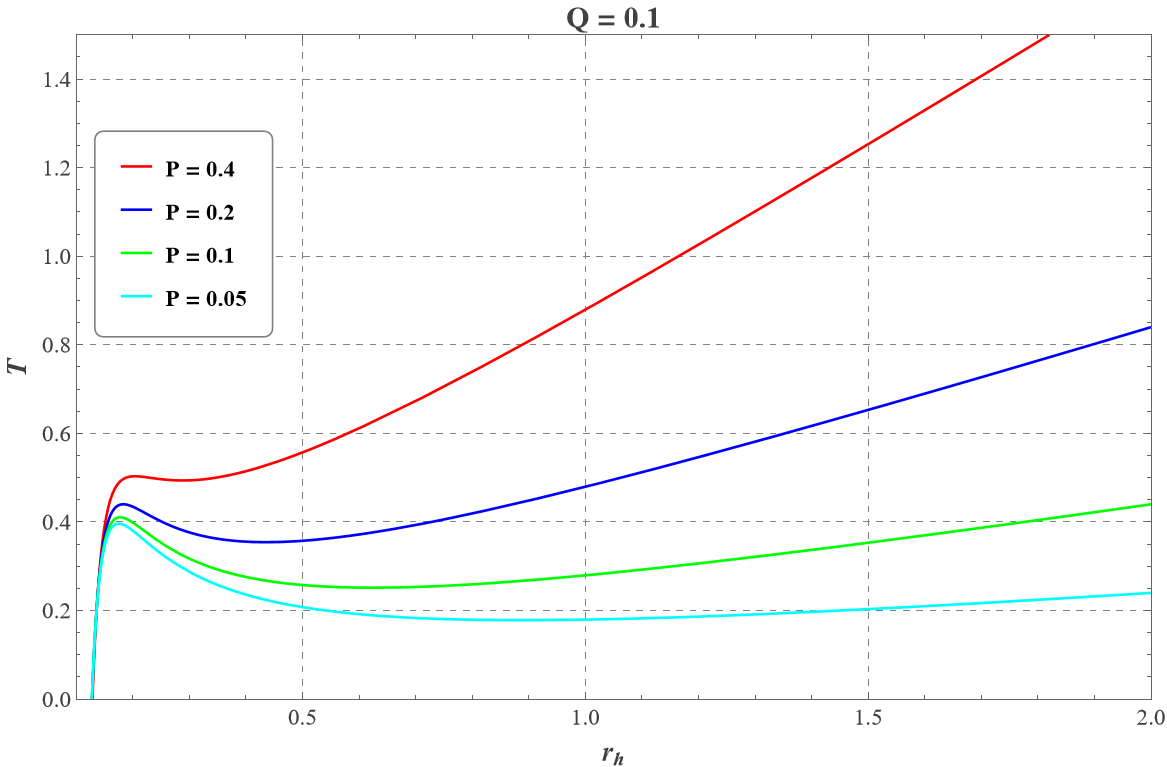}
		\caption{$Q=0.1$}
		\label{fig:temp_b}
	\end{subfigure}
	
	\vspace{0.5cm}
	
	\begin{subfigure}[b]{0.4\textwidth}
		\centering
		\includegraphics[width=\textwidth]{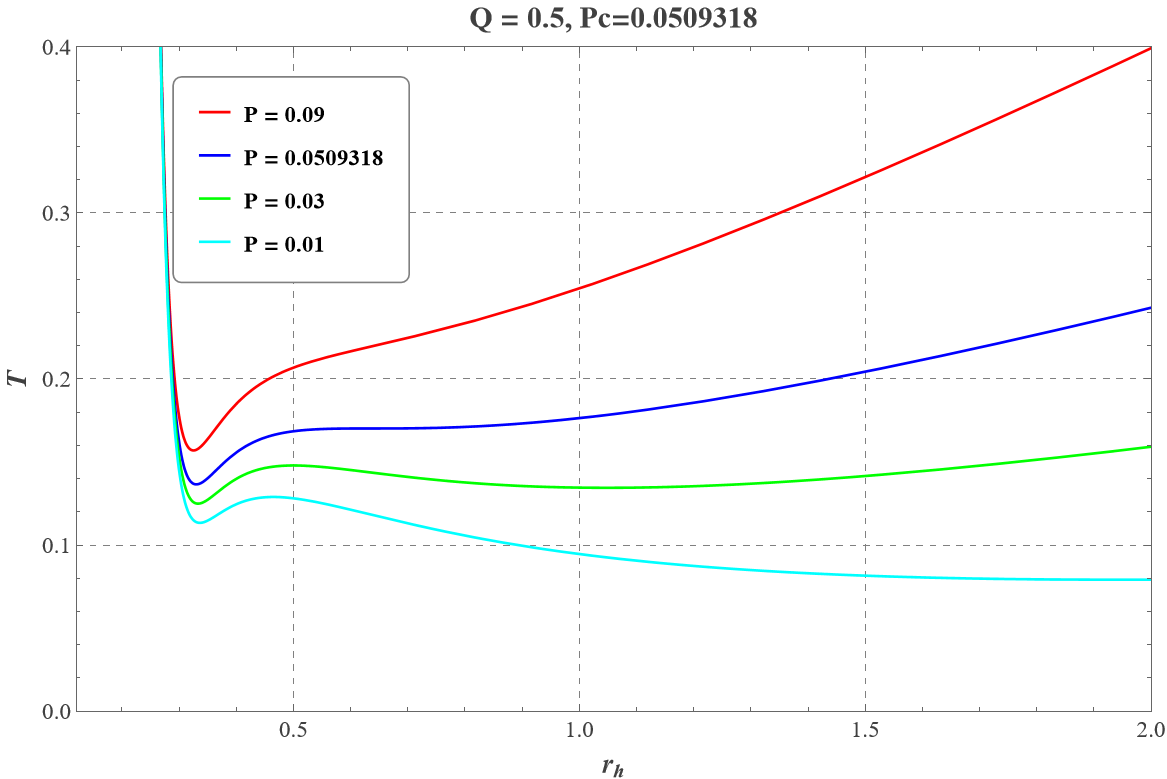}
		\caption{$Q=0.5$}
		\label{fig:temp_c}
	\end{subfigure}
	\quad
	\begin{subfigure}[b]{0.4\textwidth}
		\centering
		\includegraphics[width=\textwidth]{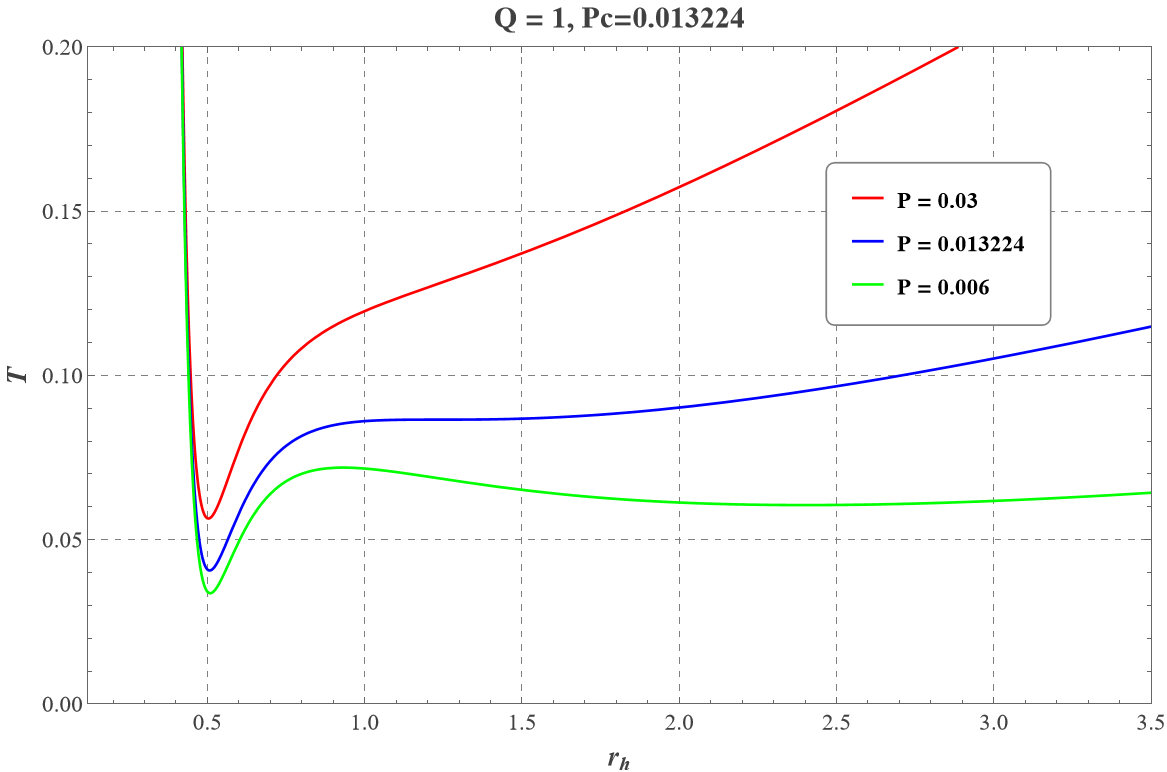}
		\caption{$Q=1$}
		\label{fig:temp_d}
	\end{subfigure}
	\caption{Isobars of the temperature \(T\) versus the horizon radius \(r_h\) for several values of the charge \(Q\) and pressure \(P\). Panels (a) and (b) correspond to the NC region, where no real critical point exists, while panels (c) and (d) correspond to the critical region, where the critical-point conditions admit solutions. In the latter region, the critical points occur at intermediate horizon radii, whereas the temperature diverges as \(r_h\to0\).}	\label{fig:temp1}
\end{figure}

Figure~\ref{fig:temp1} shows the temperature as a function of the horizon radius \(r_h\). Panels (a)--(d) correspond to different values of the black hole charge. In panels (a) and (b), corresponding to \(Q=0.05\) and \(Q=0.1\), the temperature is plotted for several values of the pressure; in these cases, no critical point exists. Panels (c) and (d) correspond to charges for which critical points exist. These critical points occur at intermediate horizon radii, while the temperature diverges as \(r_h\to0\), in contrast to the usual van der Waals-like behavior. This feature is examined in detail in Sec.~\ref{sec:4}.

In summary, within the canonical ensemble, the system exhibits two distinct regimes: a critical region at sufficiently large charge (\(Q\gtrsim0.15\) for \(\epsilon=10^{-4})\), where the criticality conditions admit solutions, and an NC region at smaller charge, where no such solutions exist.

\subsection{Criticality in the grand canonical ensemble}

We now apply the same criticality condition, Eq.~(\ref{crit_t}), in the
grand canonical ensemble, in which the electric potential \(\psi\)
rather than the charge \(Q\) is held fixed. For \(\epsilon=10^{-4}\),
the resulting critical-point coordinates are collected in
Table~\ref{table2}.

\begin{table}[htbp]
	\centering
	\caption{Critical-point coordinates in the grand canonical ensemble for
		$\epsilon=10^{-4}$. The shaded rows indicate the non-critical (NC)
		regions. The critical regime occurs in the intermediate interval
		$0.111\leq\psi\leq0.441$.}
	\label{table2}
	\begingroup
	\scriptsize
	\setlength{\tabcolsep}{2.5pt}
	\renewcommand{\arraystretch}{0.88}
	
	\begin{minipage}[t]{0.49\textwidth}
		\centering
		\begin{tabular}{ccccc}
			\toprule
			$\psi$ & $r_c$ & $P_c$ & $T_c$ & Status \\
			\midrule
			\rowcolor{gray!25} 0.001 & -- & -- & -- & NC \\
			\rowcolor{gray!25} 0.011 & -- & -- & -- & NC \\
			\rowcolor{gray!25} 0.021 & -- & -- & -- & NC \\
			\rowcolor{gray!25} 0.031 & -- & -- & -- & NC \\
			\rowcolor{gray!25} 0.041 & -- & -- & -- & NC \\
			\rowcolor{gray!25} 0.051 & -- & -- & -- & NC \\
			\rowcolor{gray!25} 0.061 & -- & -- & -- & NC \\
			\rowcolor{gray!25} 0.071 & -- & -- & -- & NC \\
			\rowcolor{gray!25} 0.081 & -- & -- & -- & NC \\
			\rowcolor{gray!25} 0.091 & -- & -- & -- & NC \\
			\rowcolor{gray!25} 0.101 & -- & -- & -- & NC \\
			\midrule
			0.111 & 0.123966 & 2.333794 & 1.437044 & C \\
			0.121 & 0.136501 & 1.671933 & 1.125905 & C \\
			0.131 & 0.147686 & 1.342258 & 0.973832 & C \\
			0.141 & 0.158626 & 1.113186 & 0.864629 & C \\
			0.151 & 0.169455 & 0.941162 & 0.778865 & C \\
			0.161 & 0.180215 & 0.806621 & 0.708368 & C \\
			0.171 & 0.190921 & 0.698553 & 0.648722 & C \\
			0.181 & 0.201574 & 0.610034 & 0.597208 & C \\
			0.191 & 0.212175 & 0.536406 & 0.552015 & C \\
			0.201 & 0.222718 & 0.474389 & 0.511870 & C \\
			0.211 & 0.233197 & 0.421595 & 0.475838 & C \\
			0.221 & 0.243603 & 0.376243 & 0.443218 & C \\
			0.231 & 0.253926 & 0.336971 & 0.413464 & C \\
			0.241 & 0.264153 & 0.302727 & 0.386148 & C \\
			\bottomrule
		\end{tabular}
	\end{minipage}
	\hfill
	\begin{minipage}[t]{0.49\textwidth}
		\centering
		\begin{tabular}{ccccc}
			\toprule
			$\psi$ & $r_c$ & $P_c$ & $T_c$ & Status \\
			\midrule
			0.251 & 0.274271 & 0.272678 & 0.360928 & C \\
			0.261 & 0.284262 & 0.246165 & 0.337525 & C \\
			0.271 & 0.294106 & 0.222652 & 0.315709 & C \\
			0.281 & 0.303779 & 0.201708 & 0.295292 & C \\
			0.291 & 0.313251 & 0.182976 & 0.276115 & C \\
			0.301 & 0.322485 & 0.166162 & 0.258046 & C \\
			0.311 & 0.331436 & 0.151025 & 0.240971 & C \\
			0.321 & 0.340047 & 0.137360 & 0.224798 & C \\
			0.331 & 0.348244 & 0.125000 & 0.209447 & C \\
			0.341 & 0.355931 & 0.113806 & 0.194853 & C \\
			0.351 & 0.362981 & 0.103665 & 0.180966 & C \\
			0.361 & 0.369216 & 0.094488 & 0.167747 & C \\
			0.371 & 0.374392 & 0.086216 & 0.155177 & C \\
			0.381 & 0.378145 & 0.078817 & 0.143259 & C \\
			0.391 & 0.379923 & 0.072312 & 0.132032 & C \\
			0.401 & 0.378823 & 0.066805 & 0.121606 & C \\
			0.411 & 0.373242 & 0.062591 & 0.112243 & C \\
			0.421 & 0.359932 & 0.060509 & 0.104630 & C \\
			0.431 & 0.330549 & 0.063711 & 0.101166 & C \\
			0.441 & 0.243609 & 0.106954 & 0.125200 & C \\
			\midrule
			\rowcolor{gray!25} 0.451 & -- & -- & -- & NC \\
			\rowcolor{gray!25} 0.461 & -- & -- & -- & NC \\
			\rowcolor{gray!25} 0.471 & -- & -- & -- & NC \\
			\bottomrule
		\end{tabular}
	\end{minipage}
	
	\vspace{2mm}
	\begin{minipage}{0.95\textwidth}
		\centering
		\scriptsize
		\textit{For the remaining values $\psi=0.481,\ldots,1.991,2.001$,
			no critical point is found. The scan was extended up to $\psi=2.001$.}
	\end{minipage}
	
	\endgroup
\end{table}

Table~\ref{table2} is organized into three regions
according to the value of the electric potential. For
\(0<\psi\leq0.101\), the criticality conditions admit no
real solution; this is the low-potential NC region. A critical point
appears in the intermediate interval \(0.111\leq\psi\leq0.441\),
where the critical coordinates are reported explicitly. For
\(\psi\geq0.451\), no critical point is found again; this
is the high-potential NC region. As in the canonical ensemble, the
boundaries of these regions are estimated by scanning discrete
values of \(\psi\), rather than by determining the corresponding
threshold values analytically.

\begin{figure}[H]
	\centering
	
	\begin{subfigure}[b]{0.4\textwidth}
		\centering
		\includegraphics[width=\textwidth]{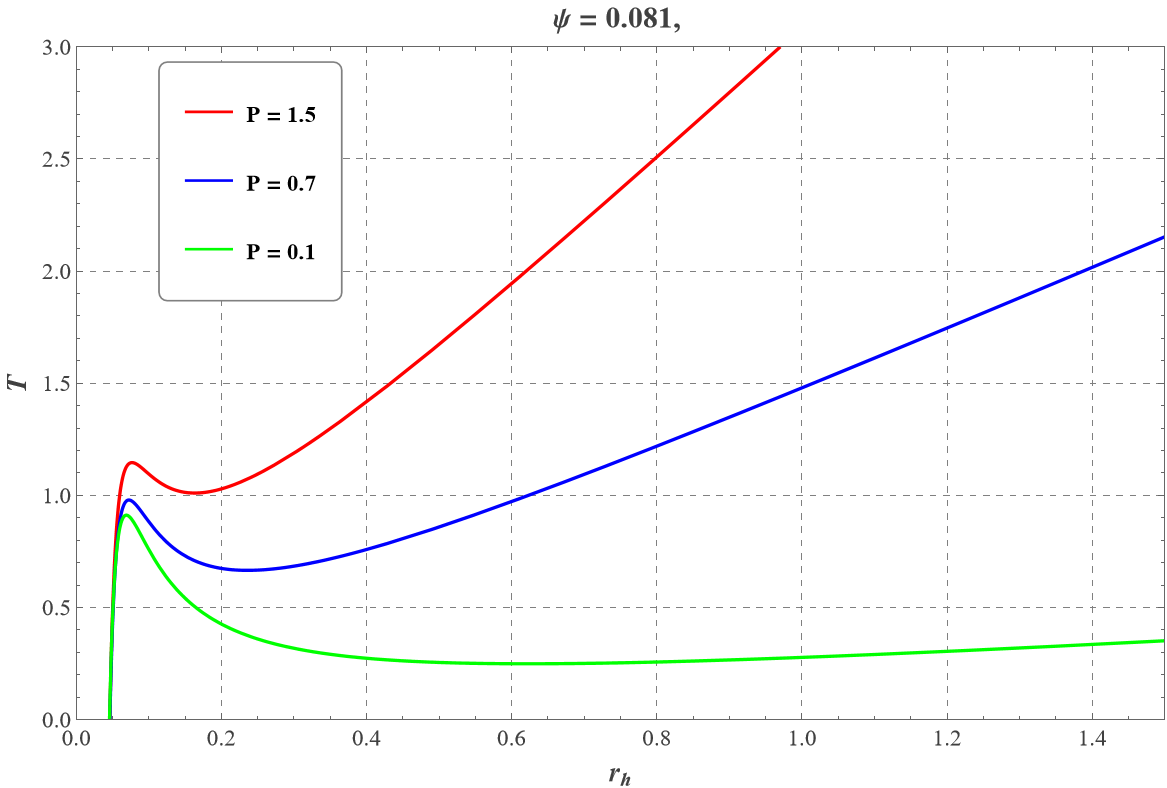}
		\caption{$\psi=0.081$}
		\label{fig:temp_a}
	\end{subfigure}
	\hfill
	\begin{subfigure}[b]{0.4\textwidth}
		\centering
		\includegraphics[width=\textwidth]{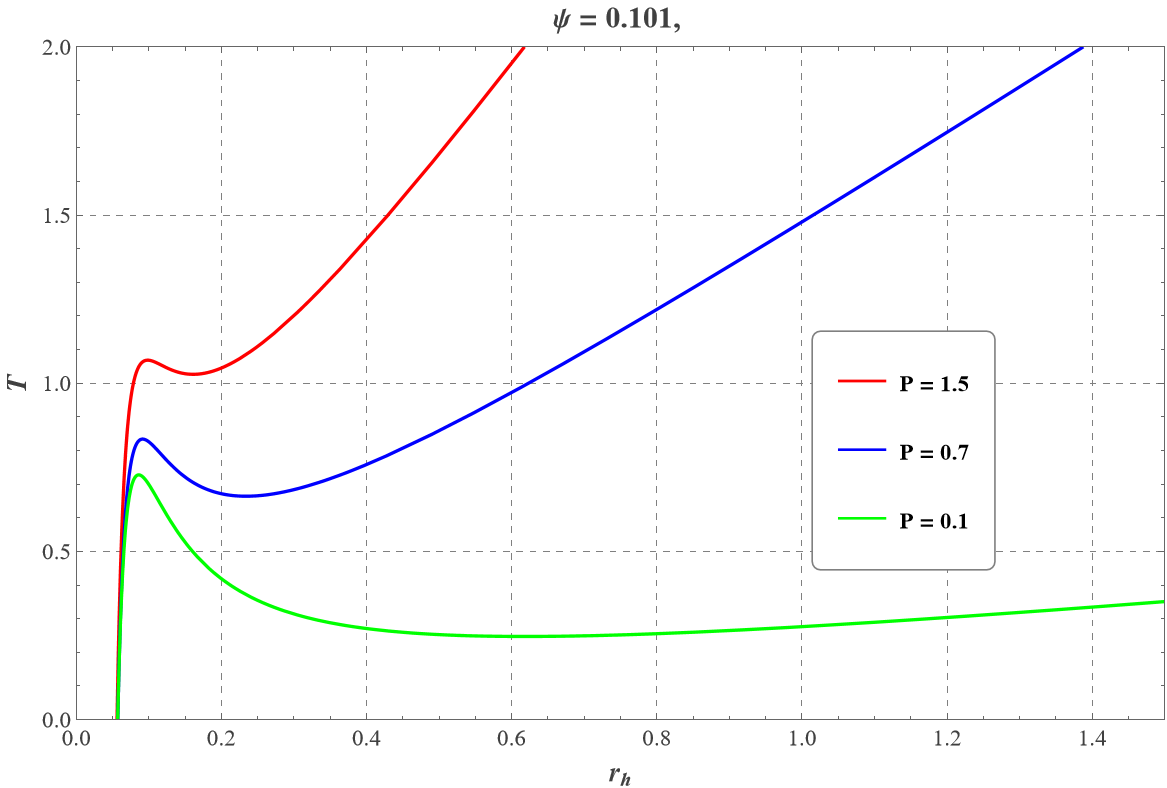}
		\caption{$\psi=0.101$}
		\label{fig:temp_b}
	\end{subfigure}
	\\[0.5cm]
	
	\begin{subfigure}[b]{0.4\textwidth}
		\centering
		\includegraphics[width=\textwidth]{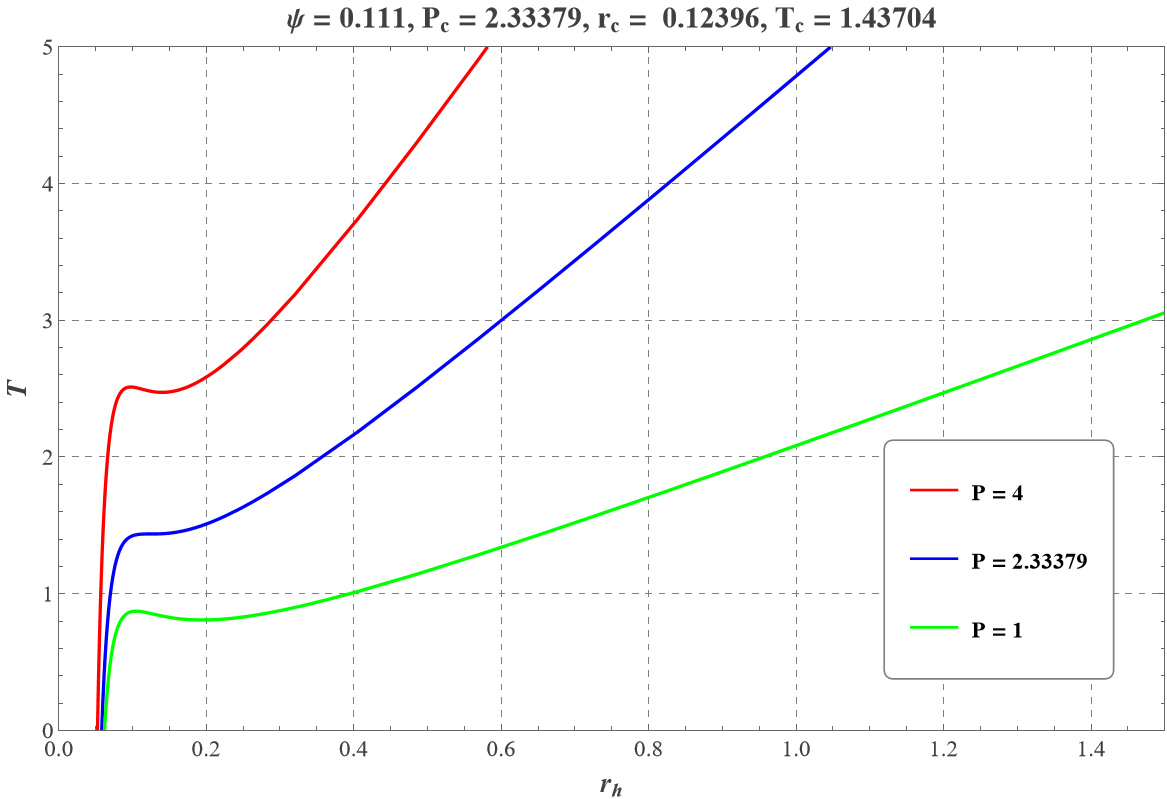}
		\caption{$\psi=0.111$}
		\label{fig:temp_c}
	\end{subfigure}
	\hfill
	\begin{subfigure}[b]{0.4\textwidth}
		\centering
		\includegraphics[width=\textwidth]{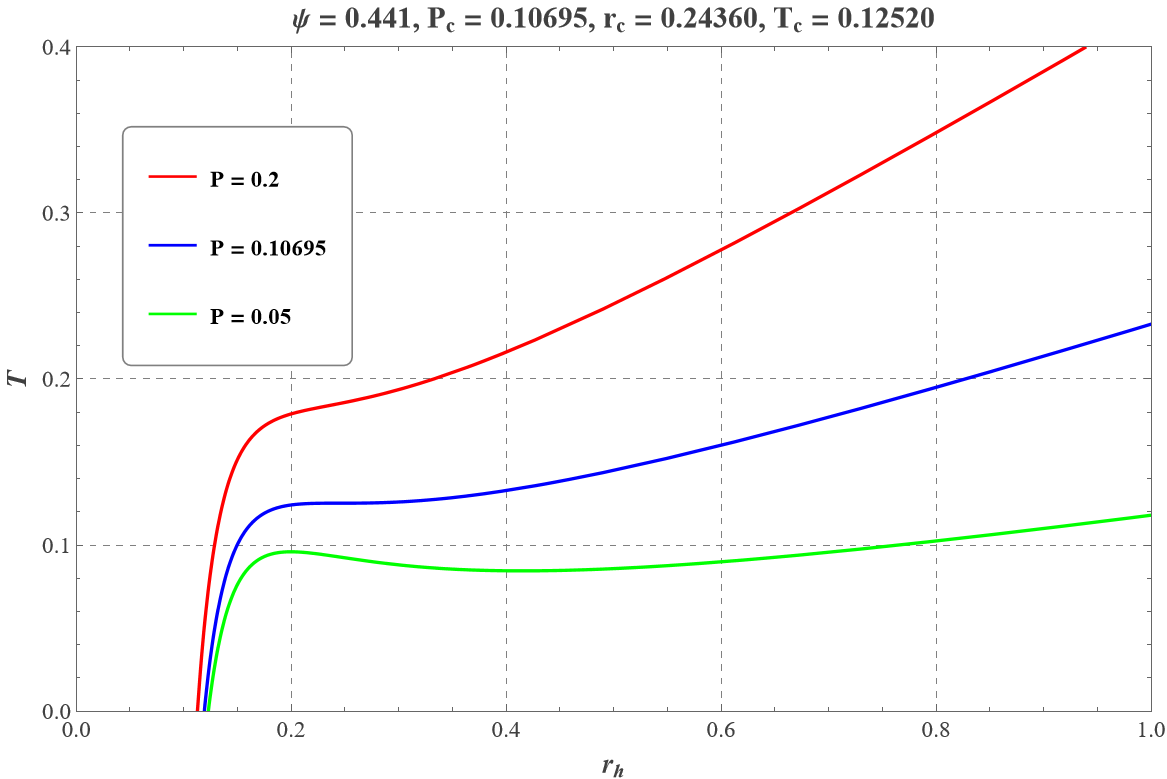}
		\caption{$\psi=0.441$}
		\label{fig:temp_d}
	\end{subfigure}
	\\[0.5cm]
	
	\begin{subfigure}[b]{0.4\textwidth}
		\centering
		\includegraphics[width=\textwidth]{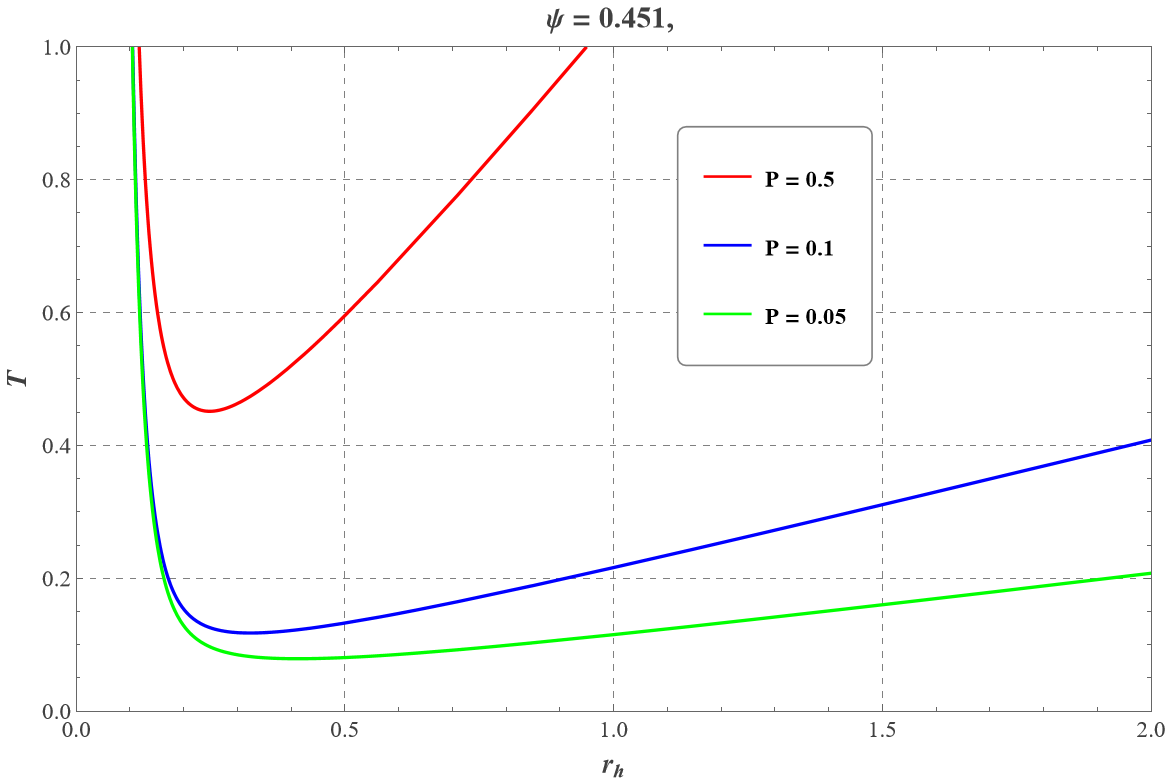}
		\caption{$\psi=0.451$}
		\label{fig:temp_e}
	\end{subfigure}
	\hfill
	\begin{subfigure}[b]{0.4\textwidth}
		\centering
		\includegraphics[width=\textwidth]{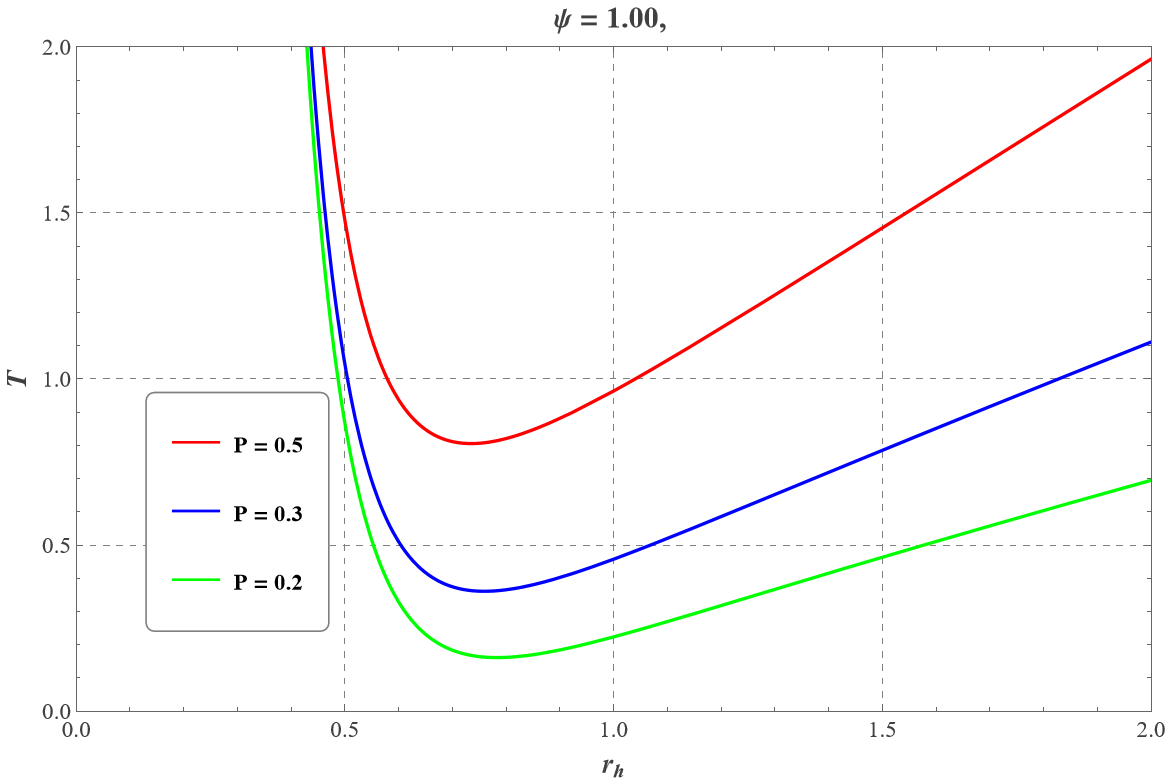}
		\caption{$\psi=1.00$}
		\label{fig:temp_f}
	\end{subfigure}
	
	\caption{Isobars of the temperature \(T\) as a function of the
		horizon radius \(r_h\) for several values of the electric potential \(\psi\) and pressure \(P\). Panels (a) and (b) correspond to the low-potential NC region, where the criticality conditions admit no	real solution. Panels (c) and (d) correspond to the intermediate	critical region, while panels (e) and (f) correspond to the high-potential NC region.}
	\label{fig:temp2}
\end{figure}

Figure~\ref{fig:temp2} shows the isobars for several values of
the electric potential \(\psi\), corresponding to the three regions
identified in Table~\ref{table2}. For \(\psi=0.081\) and
\(\psi=0.101\), which lie in the low-potential NC region, no
critical point exists. For \(\psi=0.111\) and \(\psi=0.441\),
which belong to the intermediate region, critical points are
present. Similarly, for \(\psi=0.451\) and \(\psi=1.00\), which
lie in the high-potential NC region, no critical point is found.

In summary, within the grand canonical ensemble, the parameter
space is divided into three regions according to the electric
potential: a low-potential NC region, an intermediate critical
region, and a high-potential NC region. This threefold structure
differs from the twofold structure found in the canonical ensemble.
As will be shown in Sec.~\ref{sec:4}, these three regions correspond
to the topological classes \(W=+1\), \(W=+1\), and \(W=0\),
respectively.

Before turning to the topological analysis, we assess the validity
of the perturbative expansion underlying the results obtained so
far. This is particularly important because the first-order
corrections contain inverse powers of \(r_h\), which may become
large at small horizon radii.


\section{A Brief Review of the Topological Approach to Black Hole Phase Transitions}
\label{sec:4}

The fundamental idea of the topological approach is to regard black hole
solutions as the zero points of a suitable vector field in an extended
parameter space and to assign an integer topological number, namely the
winding number, to the corresponding black hole branches. This number
remains invariant under continuous deformations of the system's
parameters, provided that no zero crosses the boundary of the parameter
space. It therefore provides a global topological characterization of
the phase structure.

The formal framework of this method is based on the definition of an
off-shell generalized free energy. In the Euclidean approach, the black
hole is characterized by an inverse temperature \(\tau\) associated
with the Euclidean time period. The on-shell configuration is recovered
when \(\tau\) equals the inverse Hawking temperature of the black hole
\cite{York:1986it}. The generalized free energy is defined as
\begin{equation}
	\mathcal{F} = \mathcal{H} - \frac{S}{\tau},
\end{equation}
where \(\mathcal{H}\) is the enthalpy (identified with the black hole
mass in the extended phase space) and \(S\) is the black hole entropy.
A two-component vector field is then introduced as
\begin{equation}
	\phi^{r_h} = \frac{\partial \mathcal{F}}{\partial r_h},
	\qquad
	\phi^{\Theta} = -\cot\Theta\,\csc\Theta,
\end{equation}
where \(\Theta\in(0,\pi)\) is an auxiliary angular coordinate whose
role is to ensure a consistent orientation of the vector field on the
boundaries of the \((r_h,\Theta)\) plane. The zeros of the radial
component, \(\partial\mathcal{F}/\partial r_h=0\), correspond to the
on-shell black hole configurations, for which the Hawking temperature
matches the inverse temperature \(1/\tau\).

Normalizing this field yields the unit vector field
\(n^a=\phi^a/\|\phi\|\), and Duan's conserved topological current
\cite{Duan1979,Duan1984} is then defined as
\begin{equation}
	j^\mu = \frac{1}{2\pi}
	\epsilon^{\mu\nu\rho}\epsilon_{ab}\,
	\partial_\nu n^a\,\partial_\rho n^b,
	\qquad
	\partial_\mu j^\mu=0.
\end{equation}
This current is localized, in the distributional sense, at the zero
points of \(\phi\), where it can be expressed as
\begin{equation}
	j^\mu = \delta^2(\phi)\,
	J^\mu\!\left(\frac{\phi}{x}\right),
\end{equation}
with \(J^\mu\) denoting the Jacobian vector of the mapping. The
topological charge associated with each isolated zero point, namely
the winding number \(w_i\), is obtained by integrating the time
component of the current over a two-dimensional surface surrounding
that point:
\begin{equation}
	w_i = \int j^0\,d^2x = \beta_i\,\eta_i,
\end{equation}
where
\begin{equation}
	j^0 = \beta_i\eta_i\,
	\delta^2(\vec{x}-\vec{z}_i).
\end{equation}
Here, \(\beta_i\) is the Hopf index, a positive integer that counts
the number of coverings of the internal space by the mapping around
the zero point \(\vec{z}_i\), while
\begin{equation}
	\eta_i =
	\mathrm{sign}\left[
	J^0\!\left(\frac{\phi}{x}\right)_{\vec{z}_i}
	\right] = \pm 1
\end{equation}
is the Brouwer degree, which determines the local orientation of the
vector field around the zero.

For isolated non-degenerate zero points of our vector field, where
\(\phi^{r_h}\) depends only on \(r_h\) and \(\phi^\Theta\) depends
only on \(\Theta\), the Hopf index is unity, \(\beta_i=1\), and the
winding number reduces to
\begin{equation}
	w_i = \eta_i
	= \mathrm{sign}\left(
	\frac{\partial^2\mathcal{F}}{\partial r_h^2}
	\right)\Bigg|_{r_h=\mathrm{ZP}_i},
\end{equation}
where \(\mathrm{ZP}_i\) denotes the corresponding zero point.

The winding number is directly related to the local thermodynamic
stability of an on-shell branch. Indeed, at an on-shell configuration,
\begin{equation}
	\left.
	\frac{\partial^2\mathcal{F}}{\partial r_h^2}
	\right|_{\mathrm{on-shell}}
	=
	\frac{\partial S}{\partial r_h}
	\frac{\partial T}{\partial r_h},
\end{equation}
where the derivatives are evaluated at fixed values of the natural
thermodynamic variables of the ensemble under consideration. The
heat capacity at fixed values of these variables is given by
\begin{equation}
	C=T\left(\frac{\partial S}{\partial T}\right)
	=T\frac{\partial S/\partial r_h}
	{\partial T/\partial r_h}.
\end{equation}
Thus, for \(T>0\) and \(\partial S/\partial r_h>0\), the signs
of the heat capacity and the on-shell second derivative of
\(\mathcal{F}\) coincide. Consequently, \(w_i=+1\) corresponds
to a locally thermodynamically stable branch with positive heat
capacity, whereas \(w_i=-1\) corresponds to an unstable branch
with negative heat capacity. This correspondence relies on
the positivity of \(\partial S/\partial r_h\) and does not
necessarily hold outside the physical domain.

At a critical point, where \(\partial T/\partial r_h=0\),
the Jacobian vanishes and the zero becomes degenerate;
therefore, the above local expression for the winding
number of a nondegenerate zero cannot be applied directly.

The winding number can also be computed from the deflection angle
of the unit vector field along a closed contour \(C\) surrounding
the zero point:
\begin{equation}
	w = \frac{1}{2\pi}
	\int_0^{2\pi}
	\epsilon_{ab}\,n^a
	\frac{\partial n^b}{\partial\vartheta}\,d\vartheta
	= \frac{\Omega(2\pi)}{2\pi}.
\end{equation}

In practice, we choose the contour to be traversed counterclockwise
by convention, with the parametrization
\begin{equation}
	\label{eq:contour_param}
	\begin{aligned}
		r_h(\vartheta)&=a\cos\vartheta+z_0,\\
		\Theta(\vartheta)&=b\sin\vartheta+\frac{\pi}{2},
	\end{aligned}
	\qquad
	0\leq\vartheta\leq2\pi,
\end{equation}
where the parameters \(a\) and \(b\) are chosen such that the contour
encloses the desired zero point \(z_0\) without enclosing other
zero points.

The sum of the winding numbers of all isolated zero points defines
the global topological invariant of the system:
\begin{equation}
	W = \sum_{i=1}^{N} w_i.
\end{equation}
This quantity remains invariant under continuous deformations as long
as no zero crosses the boundary of the parameter domain. It therefore
provides a topological classification of the thermodynamic phase
structure. A fundamental classification is based on the asymptotic
behavior of the inverse temperature
\(\beta(r_h)=1/T(r_h)\) near the minimum physical horizon radius
\(r_m\) and in the limit \(r_h\to\infty\), leading to four principal
classes:
\begin{align}
	W^{1-}: &\quad \beta(r_m)=0,\quad
	\beta(\infty)=\infty,\quad W=-1, \nonumber\\
	W^{0+}: &\quad \beta(r_m)=\infty,\quad
	\beta(\infty)=\infty,\quad W=0, \nonumber\\
	W^{0-}: &\quad \beta(r_m)=0,\quad
	\beta(\infty)=0,\quad W=0, \nonumber\\
	W^{1+}: &\quad \beta(r_m)=\infty,\quad
	\beta(\infty)=0,\quad W=+1.
\end{align}
These classes can be identified from the orientation of the unit
vector field on the boundaries of the parameter region, namely on
the segments \(I_1\) (at \(r_h\to\infty\)), \(I_3\) (at
\(r_h\to r_m\)), and \(I_2\) and \(I_4\) (at
\(\Theta=\pi\) and \(\Theta=0\), respectively). The sign of the
radial component of the vector field on these boundaries is determined
by
\begin{equation}
	\phi^{r_h} =
	\left(\frac{\partial S}{\partial r_h}\right)
	\left(\frac{1}{\beta}-\frac{1}{\tau}\right).
\end{equation}
On the physical domain, where
\(\partial S/\partial r_h>0\), the sign of
\(\phi^{r_h}\) is determined solely by the difference
\(1/\beta-1/\tau\). Consequently, the asymptotic behavior of
\(\beta(r_h)\) determines the direction of the radial component
of the vector field on the corresponding boundaries.

More recently, an extended classification has been proposed that
includes temperature-dependent classes such as
\(W^{0\leftrightarrow 1+}\) and the subclasses
\(\overline{W}^{1+}\), \(\widehat{W}^{1+}\),
\(\widetilde{W}^{1+}\), and \(\ddot{W}^{1-}\). These classes are
distinguished not only by the global topological number \(W\), but
also by the stability of the innermost and outermost branches and
by the arrangement of stable and unstable states in the low- and
high-temperature limits
\cite{Wu:2024asq,Ai:2025vno,Chen:2025vsz}.

In the present work, the generalized free energy is constructed
according to the thermodynamic ensemble under consideration. In the
canonical ensemble, the electric charge is held fixed and
\(\mathcal{F}=\mathcal{H}-S(Q,r_h,\epsilon)/\tau\), whereas in the grand canonical ensemble, the electric potential is fixed and the appropriate generalized
free energy is \(\mathcal{G}=\mathcal{M}-\psi Q-S(\psi,r_h,\epsilon)/\tau\). The corresponding vector fields are then used to investigate the topological properties of the black hole branches in both ensembles.

\subsection{Thermodynamic topology in the canonical ensemble}

We now apply the general framework of Sec.~\ref{sec:4} to the canonical
ensemble, in which the black hole charge \(Q\) is held fixed. To study
topological phase transitions in this ensemble, we first construct the
off-shell Helmholtz free energy:
\begin{equation}\label{off-F}
	\begin{split}
		\mathcal{F} &= \mathcal{H} - \frac{S}{\tau} = \frac{4\pi P r_h^3}{3} + \frac{Q^2}{8r_h} + \frac{r_h}{2} \\
		&\quad + \epsilon\left(
		\frac{256\pi^2 P^2 Q^2}{r_h}
		+ \frac{5\pi P Q^4}{r_h^5}
		+ \frac{16\pi P Q^2}{3r_h^3}
		- \frac{9Q^6}{32r_h^9}
		+ \frac{53Q^4}{56r_h^7}
		\right) - \frac{1}{\tau}\left(
		\pi r_h^2 + \frac{5\pi Q^4 \epsilon}{3r_h^6}
		\right).
	\end{split}
\end{equation}

The corresponding vector field is defined as
\begin{equation}\label{canovec}
	\begin{split}
		\phi^{r_h} &= \frac{\partial \mathcal{F}}{\partial r_h} = 4\pi P r_h^2 - \frac{Q^2}{8r_h^2} + \frac{1}{2} \\
		&\quad + \epsilon\left(
		-\frac{256\pi^2 P^2 Q^2}{r_h^2}
		- \frac{25\pi P Q^4}{r_h^6}
		- \frac{16\pi P Q^2}{r_h^4}
		+ \frac{81Q^6}{32r_h^{10}}
		- \frac{53Q^4}{8r_h^8}
		\right) \\
		&\quad - \frac{1}{\tau}\left(
		2\pi r_h - \frac{10\pi Q^4 \epsilon}{r_h^7}
		\right).
	\end{split}
\end{equation}

Setting the radial component of the vector field to zero,
\(\phi^{r_h}=0\), we obtain the inverse temperature of the heat bath:
\begin{equation}
	\begin{aligned}
		\tau &= \frac{N}{D}, \\[4pt]
		N &= 64\pi r_h^3\left(-r_h^8 + 5Q^4\epsilon\right), \\[4pt]
		D &= 8192\pi^2 P^2 Q^2 \epsilon\, r_h^8 - 128\pi P r_h^{12}
		+ 800\pi P Q^4 \epsilon\, r_h^4 \\
		&\quad + 512\pi P Q^2 \epsilon\, r_h^6
		+ 4Q^2 r_h^8 - 16r_h^{10}
		- 81Q^6\epsilon + 212Q^4\epsilon\, r_h^2.
	\end{aligned}
\end{equation}

\begin{figure}[H]
	\centering
	
	\begin{subfigure}[b]{0.4\textwidth}
		\centering
		\includegraphics[width=\textwidth]{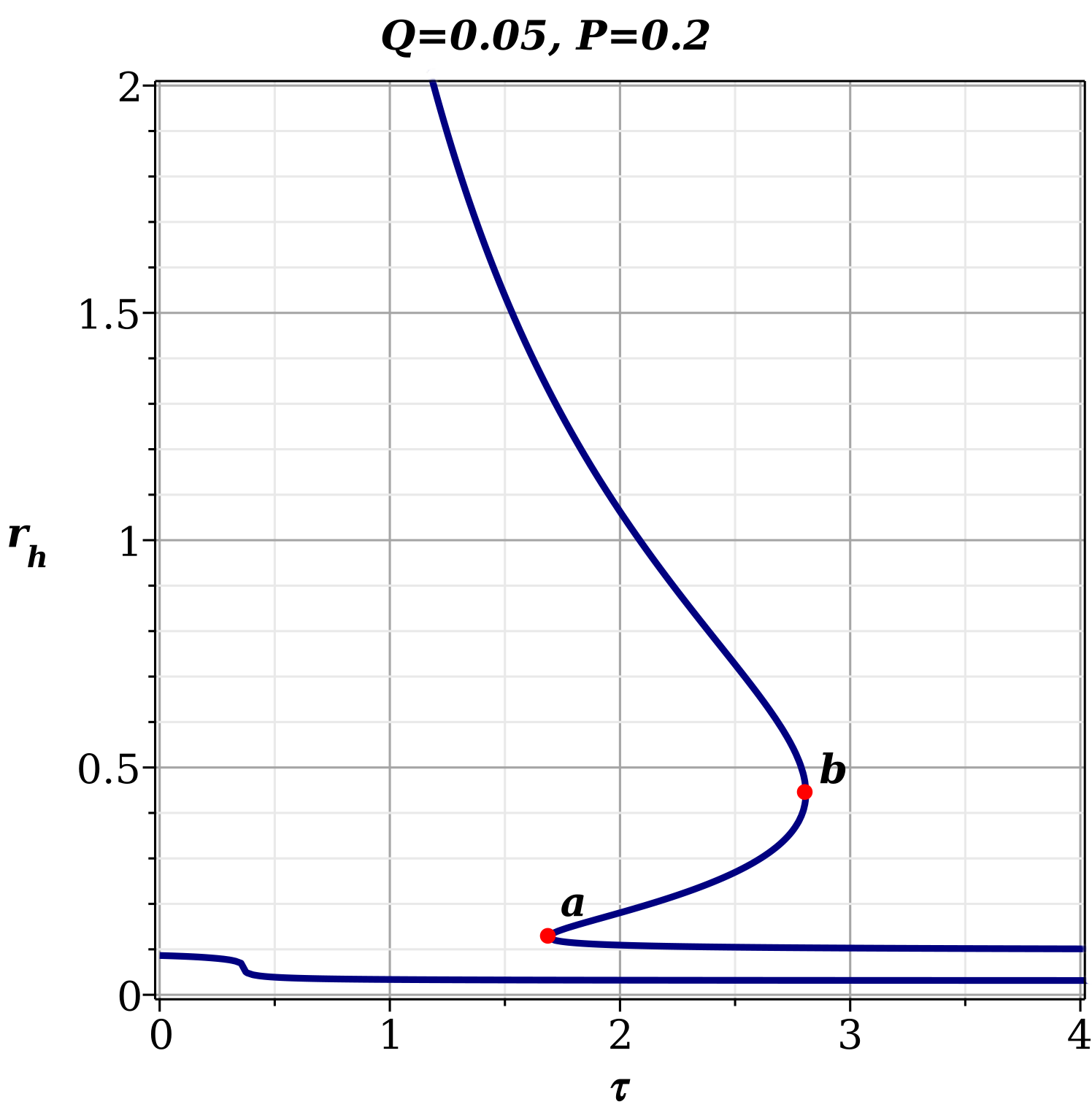}
		\caption{\(Q=0.05\), \(P=0.2\), with a lower unphysical branch;
			\(W=+1\).}
		\label{fig:tau1}
	\end{subfigure}
	\hfill
	\begin{subfigure}[b]{0.4\textwidth}
		\centering
		\includegraphics[width=\textwidth]{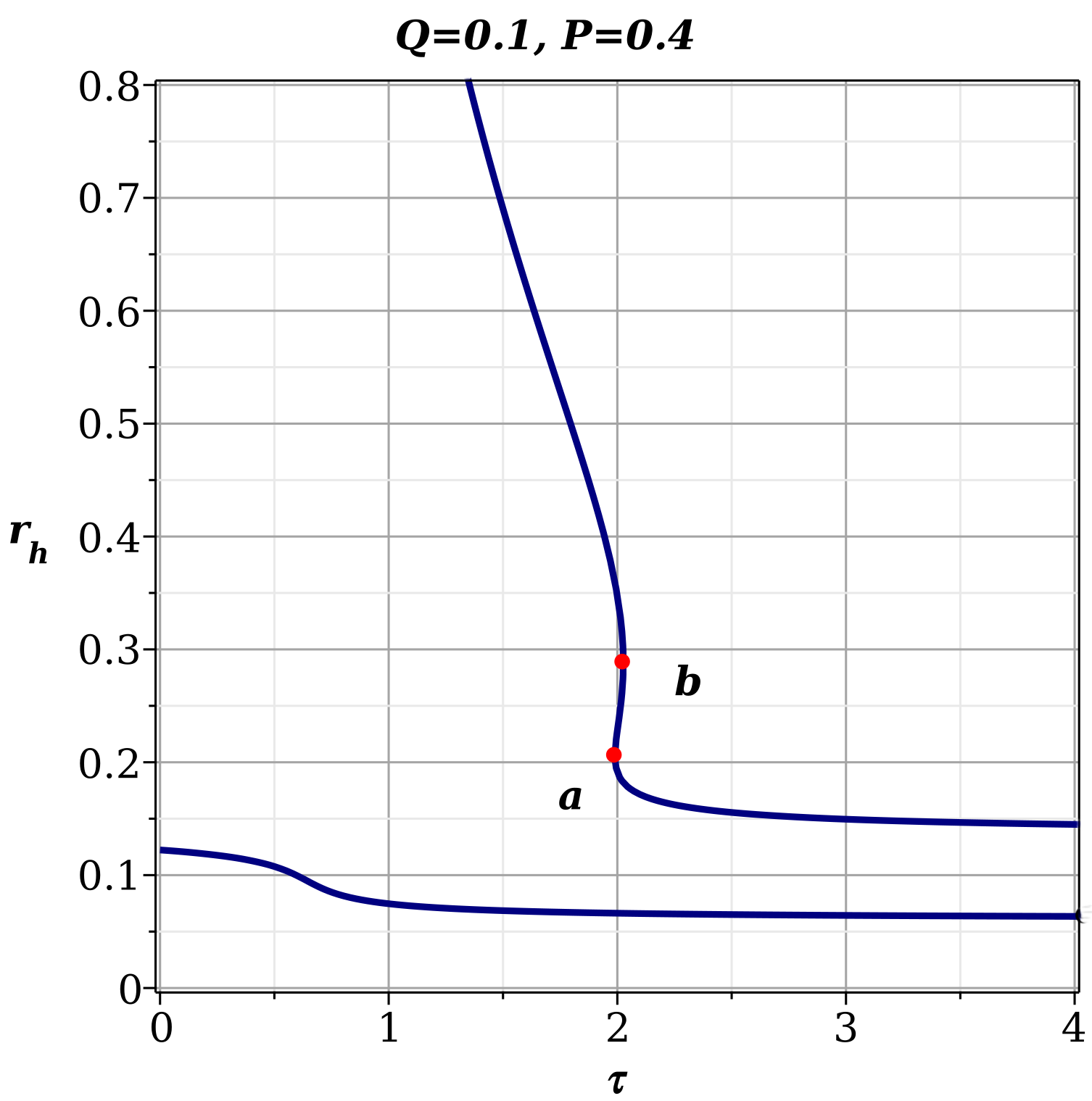}
		\caption{\(Q=0.1\), \(P=0.4\), with a lower unphysical branch;
			\(W=+1\).}
		\label{fig:tau2}
	\end{subfigure}
	
	\vspace{0.3cm}

	\begin{subfigure}[b]{0.4\textwidth}
		\centering
		\includegraphics[width=\textwidth]{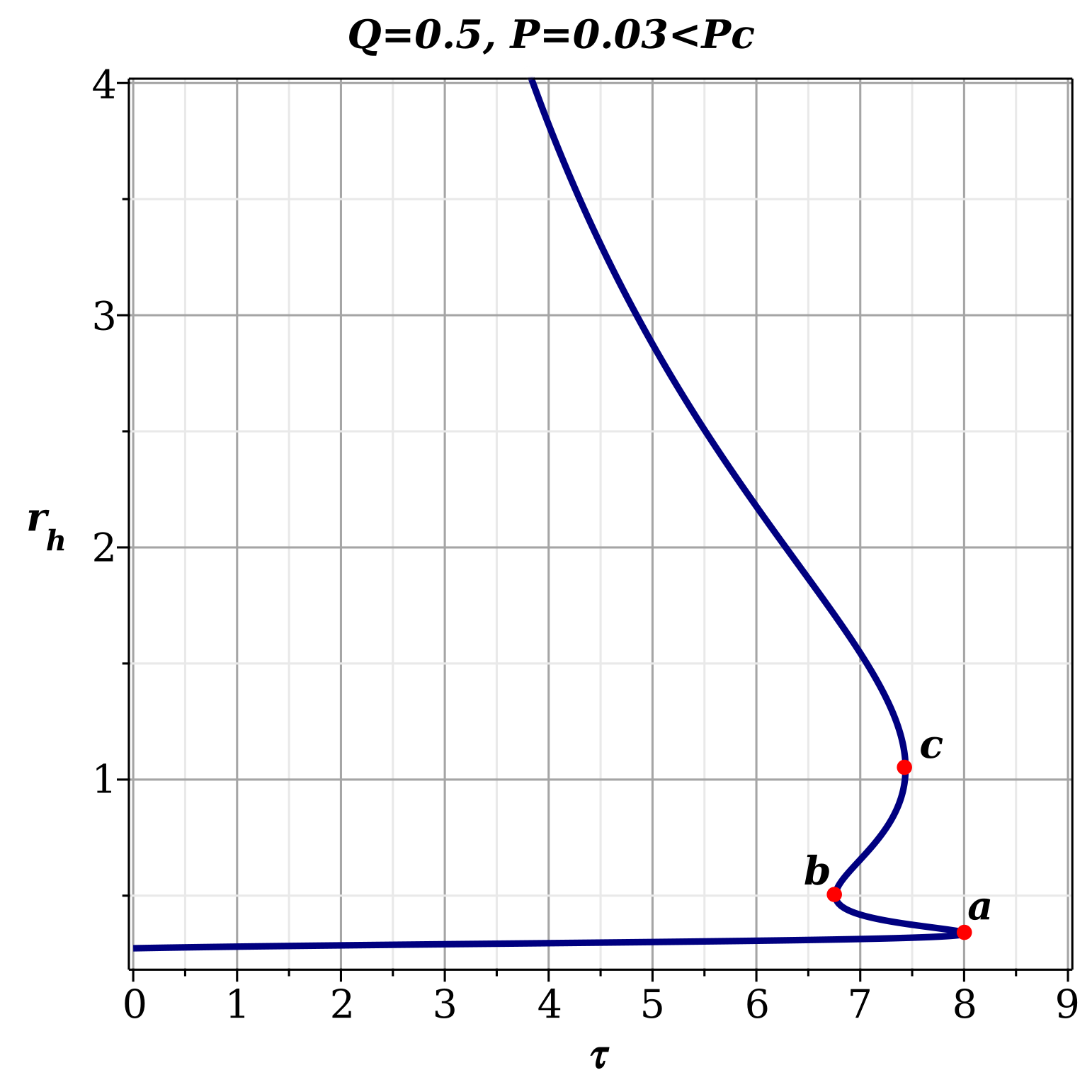}
		\caption{\(Q=0.5\), \(P=0.03<P_c\).}
		\label{fig:tau3}
	\end{subfigure}
	\hfill
	\begin{subfigure}[b]{0.4\textwidth}
		\centering
		\includegraphics[width=\textwidth]{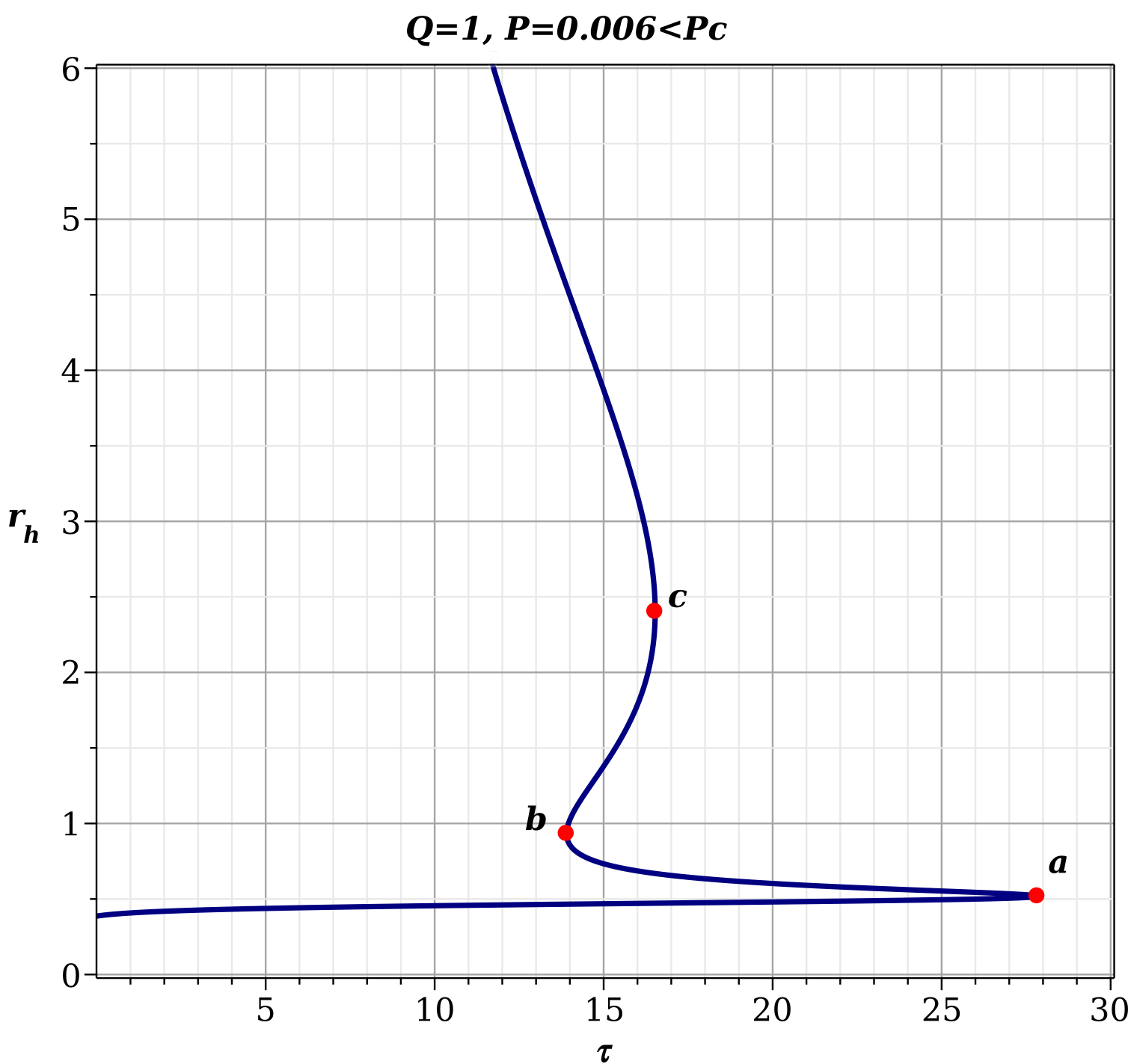}
		\caption{\(Q=1\), \(P=0.006<P_c\).}
		\label{fig:tau4}
	\end{subfigure}
	
	\caption{The \(r_h\)--\(\tau\) diagrams for different values of the
		charge \(Q\) and pressure \(P\). Panels~(a) and~(b) correspond to
		the NC region, where the physical equilibrium structure carries
		\(W=+1\), while panels~(c) and~(d) illustrate the critical region
		at pressures below the critical pressure, including the
		intermediate equilibrium branches whose winding numbers cancel in
		pairs, giving \(W=0\).}
	\label{fig:canotau}
\end{figure}

Figure~\ref{fig:canotau} illustrates the thermodynamic phase
structure through the \(r_h\)--\(\tau\) diagrams. Panel~(a)
corresponds to \(Q=0.05\) and \(P=0.2\), for which the temperature
exhibits no critical behavior; this case belongs to the NC region.
A distinct branch appears at small horizon radii in the lower part
of the plot. Although the winding number associated with each
equilibrium point \((\tau,r_h)\) on this branch could in principle
be evaluated, the branch lies in the region where
\(\partial S/\partial r_h<0\) and is therefore unphysical. Indeed,
for \(r_h<r_*\), where
\(r_*\equiv(5Q^4\epsilon)^{1/8}\), the entropy derivative is
negative. As \(\tau\) increases, this branch moves toward smaller
horizon radii and remains entirely within the unphysical region
for all \(\tau>0\). Consequently, it does not contribute to the
physical topological count. For the parameters of panel~(a),
\(r_*\approx0.086\), which coincides with the horizon radius at
which the branch intersects the \(r_h\) axis at \(\tau=0\).

We therefore consider the other family of connected branches.
In the intermediate interval \(\tau_a<\tau<\tau_b\), three
branches emerge, alternating between unstable and stable.
For \(\tau>\tau_b\), only one stable branch remains. The total
winding number \(W=+1\) is conserved throughout the range of
\(\tau\) considered. It is worth noting that when
\(\partial S/\partial r_h<0\), the relation between the sign
of the Jacobian and thermodynamic stability is reversed.
Thus, the usual identification \(w=+1\) with a stable branch
does not apply in this region. Including such roots in the
topological count would therefore assign winding numbers to
configurations outside the physical domain.

As discussed above, the stability of equilibrium branches
at non-degenerate zeros is determined by the sign of the
Jacobian, \(J=\partial^2\mathcal{F}/\partial r_h^2\).
Accordingly, the stability of each branch can be identified
directly from the sign of \(J\).

Panel~(b) shows the thermodynamic phase structure for \(Q=0.1\)
and \(P=0.4\). As in panel~(a), the total winding number is
\(W=+1\). Although no conventional critical point exists in the
NC region, the value \(W=+1\) indicates that a first-order
transition between small and large black holes occurs in this
region, as reflected in the change in sign of
\(\partial^2\mathcal{F}/\partial r_h^2\) along the equilibrium
branches. 
We have verified this behavior by computing and plotting the
corresponding curves; the two representative panels~(a) and~(b)
are sufficient to illustrate it.

For \(Q=0.5\), within the critical region, the system exhibits
a nontrivial equilibrium structure at pressures below the
critical pressure. However, because the temperature behavior
at small horizon radii differs from that of a conventional
van der Waals-like system, the topological structure is also
different, and the total winding number remains zero throughout
the regions considered. Consequently, the creation and
annihilation of equilibrium branches in the intermediate
regions occur in pairs with opposite topological charges,
preserving the total winding number \(W=0\). This behavior
originates from the nonminimal coupling, which modifies the
equilibrium structure compared with that of the conventional
Reissner--Nordstr\"om--AdS black hole. Panel~(c) illustrates
this behavior for \(Q=0.5\) and \(P=0.03<P_c\), while
panel~(d) shows the corresponding behavior for \(Q=1\) and
\(P=0.006<P_c\). In both panels, all branches satisfy
\(\partial S/\partial r_h>0\) and can therefore be regarded
as physical. The small-horizon branch emerges smoothly from
the \(\tau=0\) axis, where it intersects the \(r_h\) axis.
This intersection is determined by
\(\partial S/\partial r_h=0\), and the lower branch lies
above \(r_*\), where the entropy derivative is positive.
For the parameters of panel~(c), \(r_*\approx0.2734\),
consistent with the value of \(r_h\) at the \(\tau=0\)
intersection. Based on the qualitative boundary behavior
of the curves in this region, we expect the same pattern
to persist over a wide range of charges and pressures;
panels~(c) and~(d) provide two representative examples.
These plots also show that, at fixed charge and pressure,
increasing \(\tau\) (decreasing the heat-bath temperature)
leads to the emergence of additional branches, and hence
additional equilibrium zeros and intermediate topological
transitions. Conversely, decreasing \(\tau\) (increasing
the temperature) results in a more direct evolution of
the equilibrium structure, without intermediate pair
creation and annihilation processes.

To further confirm the topological classification, we examine the
unit vector field and the deflection-angle profiles. The orientation
of the unit vector field along the boundaries \(I_1\) and \(I_3\)
determines the topological class, while the deflection-angle method
provides an independent check of the winding numbers associated
with individual zeros.

\begin{figure}[H]
	\centering
	
	\begin{subfigure}[b]{0.45\textwidth}
		\centering
		\includegraphics[width=\textwidth]{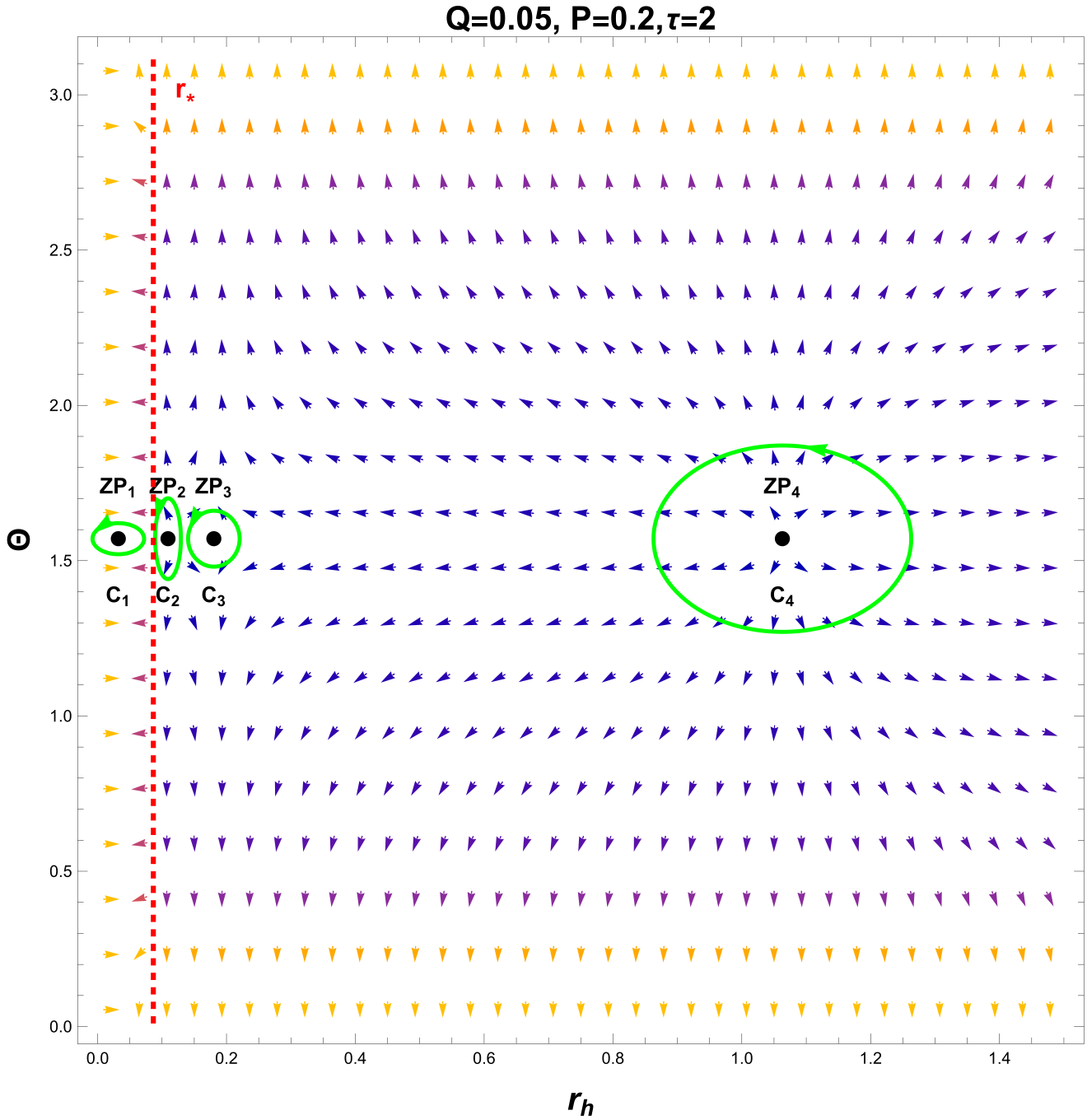}
		\caption{\(Q=0.05\), \(P=0.2\), \(\tau=2\), \(r_{*}=0.086\), $ZP_1<r_{*}$;
			\(W=+1-1+1=+1\).}
		\label{fig:v1}
	\end{subfigure}
	\hfill
	\begin{subfigure}[b]{0.45\textwidth}
		\centering
		\includegraphics[width=\textwidth]{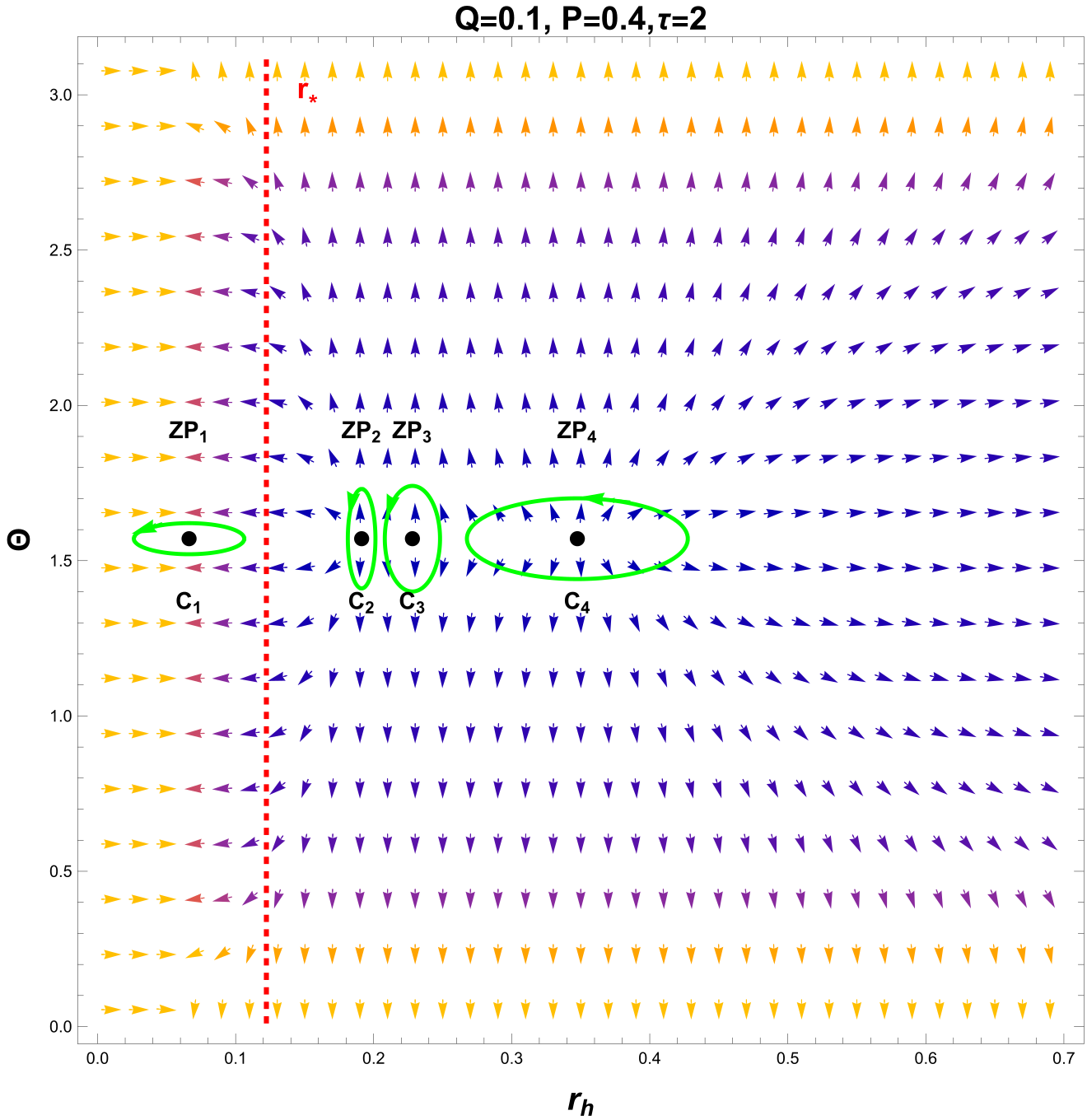}
		\caption{\(Q=0.1\), \(P=0.4\), \(\tau=2\), \(r_{*}=0.122\), $ZP_1<r_{*}$;
			\(W=+1-1+1=+1\).}
		\label{fig:v2}
	\end{subfigure}
	
	\vspace{0.3cm}

	\begin{subfigure}[b]{0.45\textwidth}
		\centering
		\includegraphics[width=\textwidth]{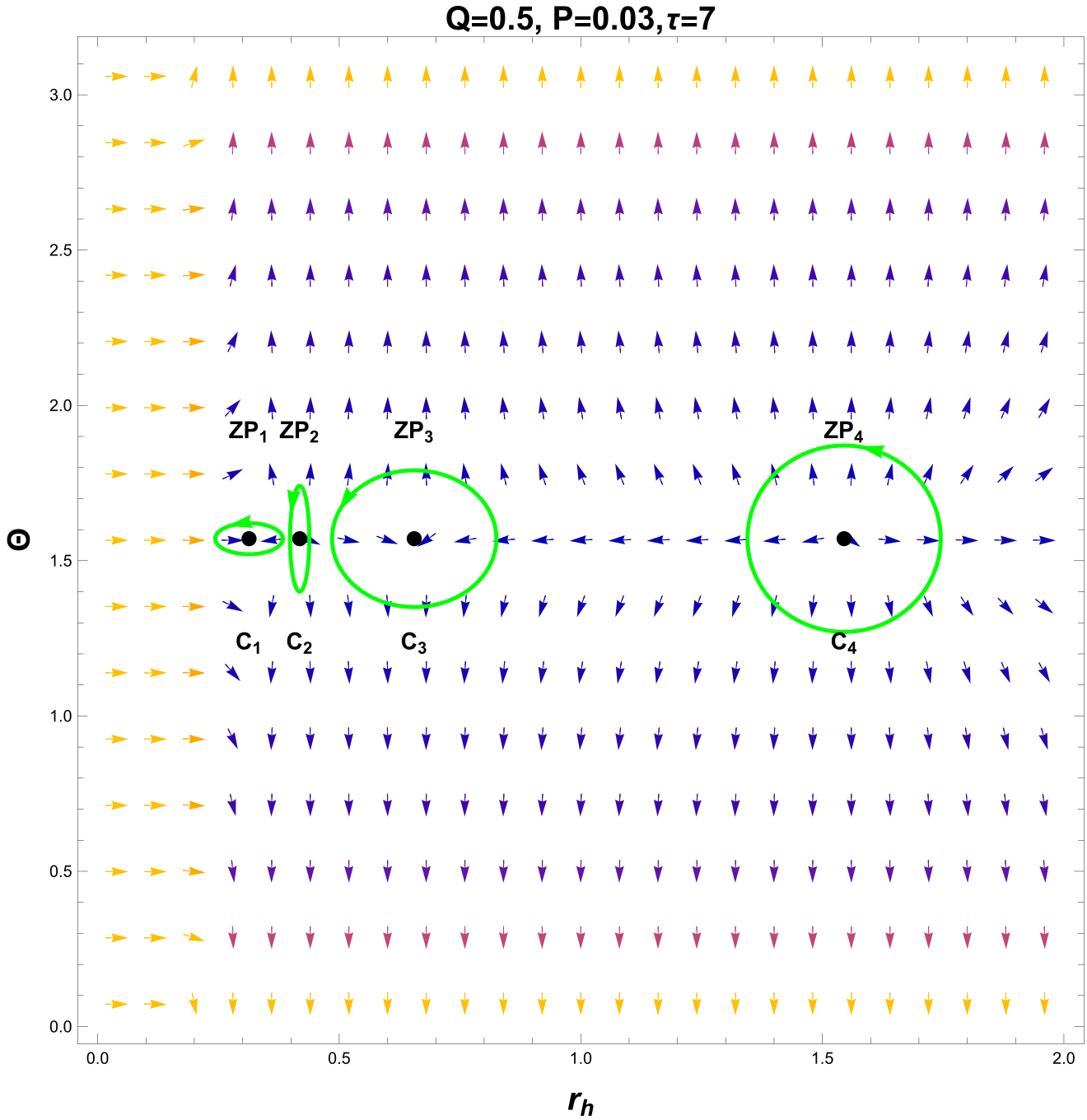}
		\caption{\(Q=0.5\), \(P=0.03<P_c\), \(\tau=7\);
			\(W=-1+1-1+1=0\). Here \(r_{*}=0.273\), and all zeros satisfy
			\(ZP_i>r_{*}\).}
		\label{fig:v3}
	\end{subfigure}
	\hfill
	\begin{subfigure}[b]{0.45\textwidth}
		\centering
		\includegraphics[width=\textwidth]{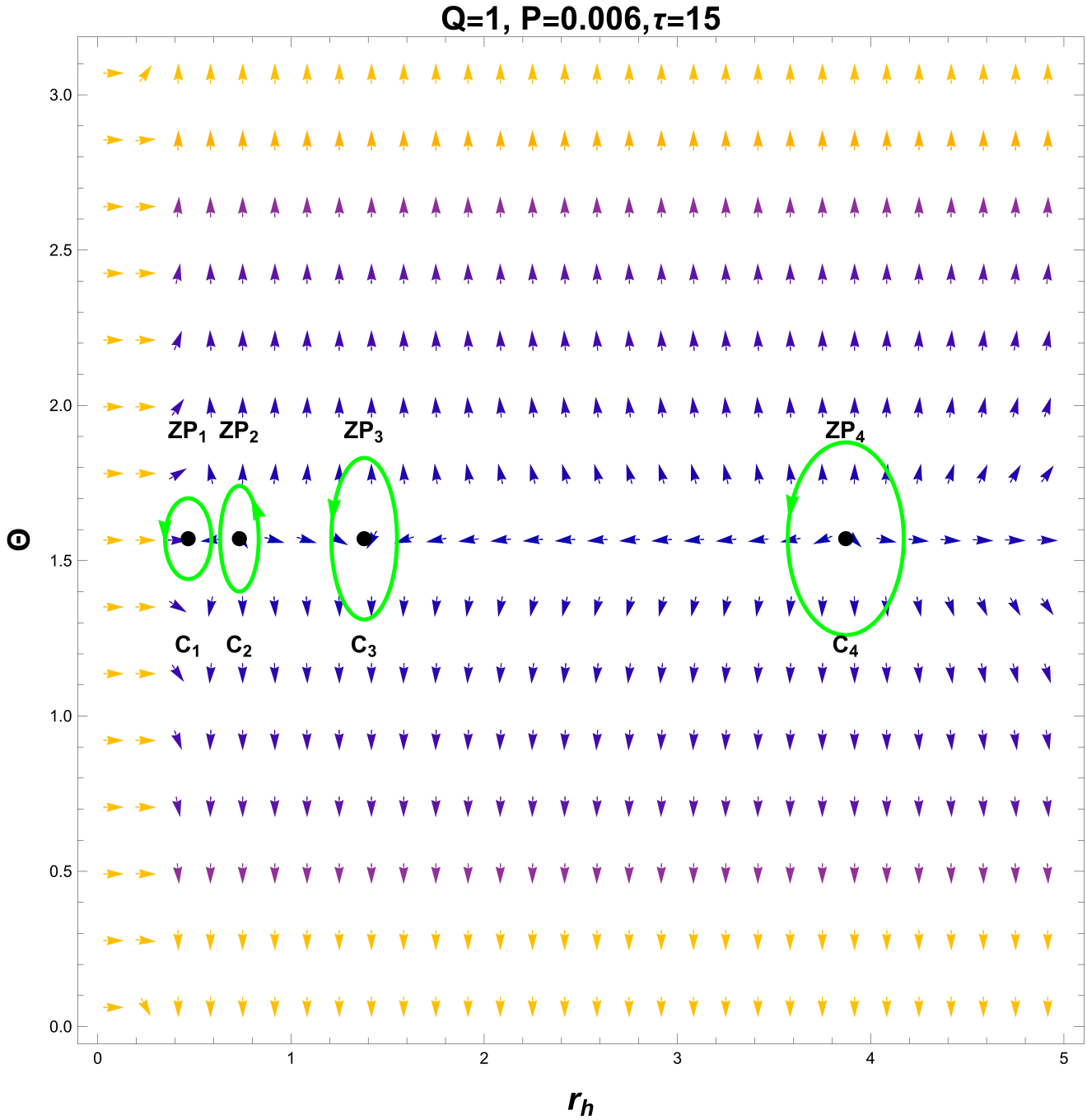}
		\caption{\(Q=1\), \(P=0.006<P_c\), \(\tau=15\);
			\(W=-1+1-1+1=0\). Here \(r_{*}=0.3866\), and all zeros satisfy
			\(ZP_i>r_{*}\).}
		\label{fig:v4}
	\end{subfigure}
	
	\caption{Configurations of the unit vector field for different values
		of the charge \(Q\), pressure \(P\), and inverse bath temperature
		\(\tau\). The boundary orientations confirm the \(W=+1\)
		topological classification in the NC region and \(W=0\) in the
		critical region. The contours enclosing the zeros are traversed
		counterclockwise. Although the winding numbers can be calculated
		from these contours or from the corresponding deflection angles,
		for non-degenerate zeros their signs can also be determined from
		the sign of the Jacobian determinant.}
	\label{fig:v1-v4}
\end{figure}

Figure~\ref{fig:v1-v4} presents four configurations of the unit
vector field for different choices of the charge, pressure, and
inverse bath temperature. Based on the discussion above concerning
the physical and unphysical regions, which are distinguished by the
sign of the entropy derivative with respect to the horizon radius,
we conclude that the smallest zero in panels~(a) and~(b) lies in
the region where \(\partial S/\partial r_h<0\) and is therefore
unphysical, whereas the remaining zeros lie in the physical region.
Taking the boundary \(r_h=r_*\), rather than the full range of
\(r_h\), as the inner boundary \(I_3\), we exclude the unphysical
region \(r_h<r_*\) from the topological count. The orientation of
the unit vector field along \(I_3\) is then leftward. Thus, the
system in the NC region belongs to the class \(W=+1\), or more
precisely, \(W^{1+}\).

Panels~(c) and~(d) correspond to charge values within the critical
region. All zeros in these panels lie in the physical region.
This behavior is also found for all other charge values and
pressures considered, as expected from the boundary behavior of
the temperature as a function of the horizon radius shown in
Fig.~\ref{fig:temp1}. The orientations of the unit vector field
along the boundaries in this region indicate that the system
belongs to the class \(W=0\).

\begin{figure}[H]
	\centering
	
	\begin{subfigure}[b]{0.45\textwidth}
		\centering
		\includegraphics[width=\textwidth]{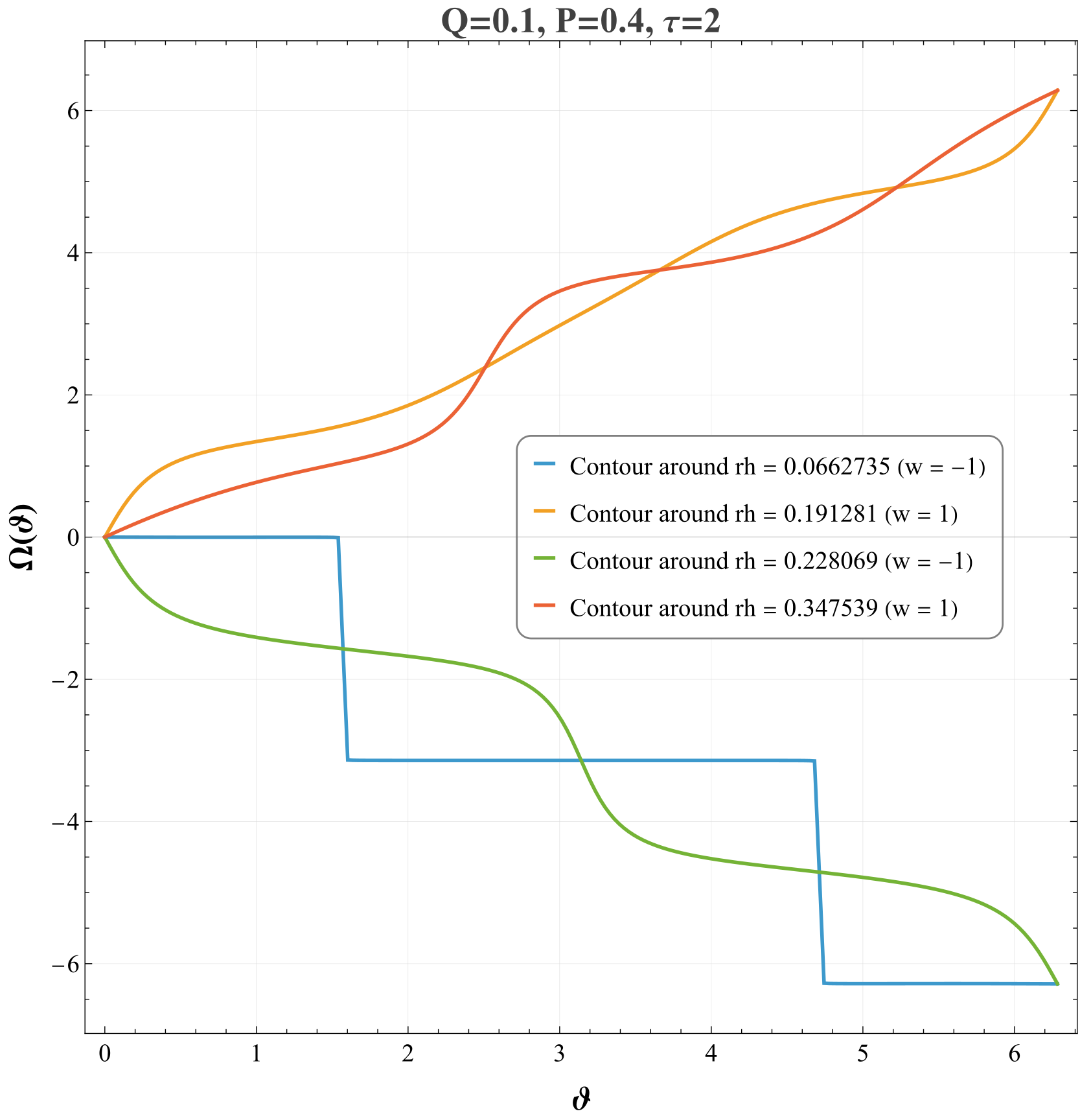}
		\caption{\(Q=0.1\), \(P=0.4\), \(\tau=2\); the winding number
			of the smallest zero is excluded from the count. The
			winding number sequence is \((+,-,+)\).}
		\label{fig:d1}
	\end{subfigure}
	\hfill
	\begin{subfigure}[b]{0.45\textwidth}
		\centering
		\includegraphics[width=\textwidth]{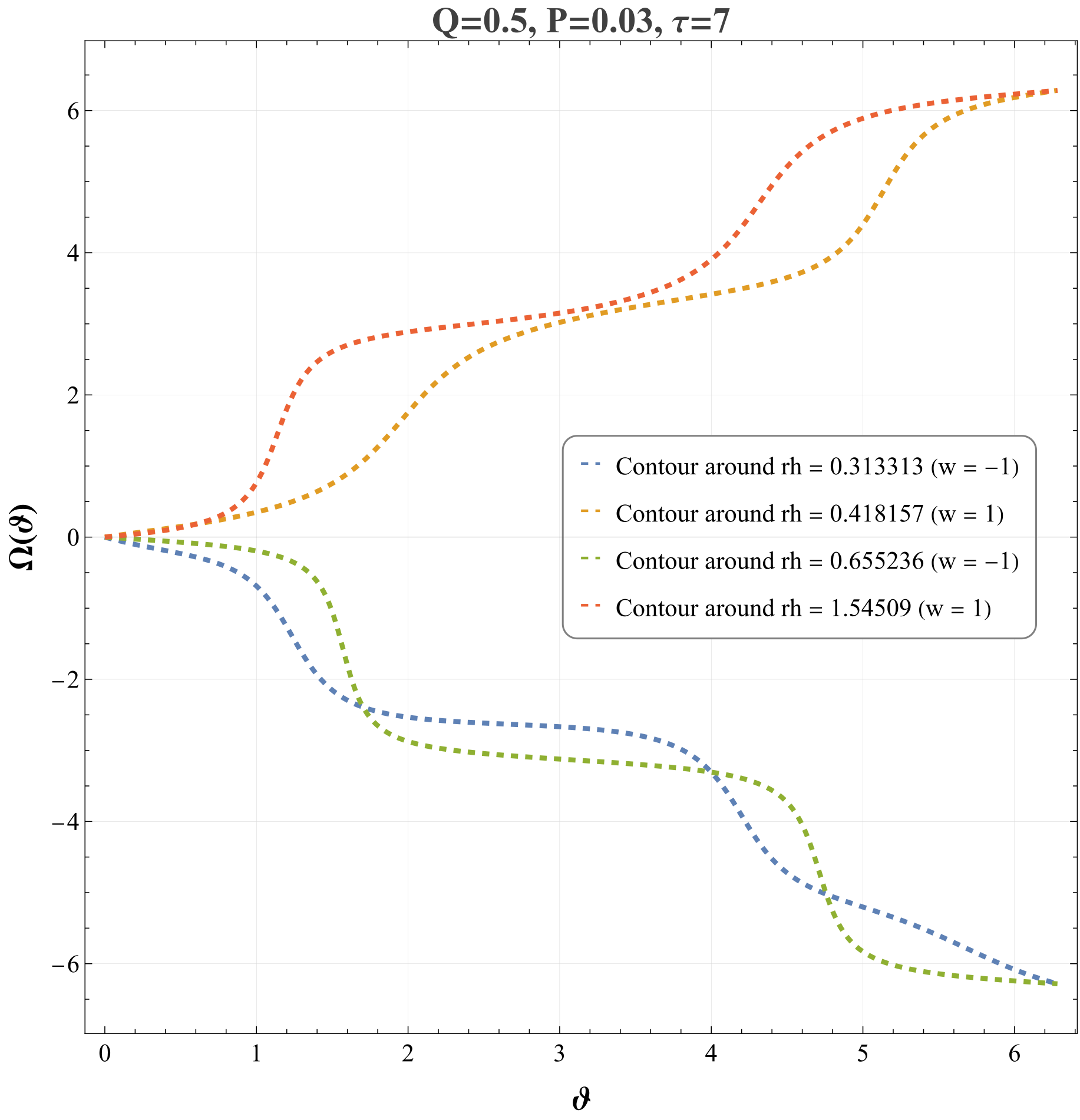}
		\caption{\(Q=0.5\), \(P=0.03\), \(\tau=7\); the total winding
			number is \(W=0\) and the
			winding number sequence is \((-,+,-,+)\)}.
		\label{fig:d2}
	\end{subfigure}
	
	\caption{Deflection-angle profiles illustrating the winding numbers
		associated with the equilibrium zeros. For each configuration,
		the contours used to calculate the winding numbers are chosen
		independently and need not coincide with those shown in
		Fig.~\ref{fig:v1-v4}.}
	\label{fig:d1-d2}
\end{figure}

Figure~\ref{fig:d1-d2} presents the deflection-angle profiles for
\(Q=0.1\) and \(Q=0.5\), at different values of pressure and inverse
bath temperature, representing the NC and critical regions,
respectively. These profiles provide an additional check of the
winding numbers associated with the individual equilibrium zeros
and confirm that their sum gives the total topological charge
\(W=+1\) in the NC region, since the winding number of the smallest
zero is excluded from the count. Panel~(b) shows the corresponding
deflection-angle profile for the zeros in the critical region,
yielding a total winding number of \(W=0\).

The results show that, in the canonical ensemble, the system belongs
to the \(W=+1\) topological class in the NC region and to the
\(W=0\) class in the critical region. A notable feature is that
the vector field can possess up to four zeros. In the critical
region, all four zeros are physical, and their winding numbers
cancel in pairs. In contrast, in the NC region, the smallest zero
always lies in the unphysical domain \(r_h<r_*\) and is excluded
from the count, leaving at most three physical zeros with a net
winding number of \(W=+1\).


\subsection{Thermodynamic topology in the grand canonical ensemble}

We now extend the topological analysis to the grand canonical
ensemble, in which the electric potential \(\psi\) is held fixed,
while the electric charge is allowed to vary. The off-shell Gibbs
free energy is  given by
\begin{equation}\label{offG}
	\begin{split}
		\mathcal{G} = \mathcal{M} - \psi Q - \frac{S}{\tau}
		&= \frac{4 r_h^3 \pi P}{3} - 2\psi^2 r_h + \frac{r_h}{2} - \frac{r_h^2 \pi}{\tau} \\
		&\quad + \Bigg( 4096\,\psi^2 r_h \pi^2 P^2 + \frac{1280\,\psi^4 \pi P}{r_h} + \frac{256\,\psi^2 \pi P}{3\,r_h} \\
		&\qquad - \frac{1152\,\psi^6}{r_h^3} + \frac{1696\,\psi^4}{7\,r_h^3} - \frac{1280\,\psi^4 \pi}{3\,r_h^2 \tau} \Bigg)\epsilon,
	\end{split}
\end{equation}
and the corresponding vector field is
\begin{equation}\label{vecG}
	\begin{split}
		\phi^{r_h}
		&= 4 r_h^2 \pi P - 2\psi^2 + \frac{1}{2} - \frac{2 r_h \pi}{\tau} \\
		&\quad + \Bigg( 4096\,\psi^2 \pi^2 P^2 - \frac{1280\,\psi^4 \pi P}{r_h^2} - \frac{256\,\psi^2 \pi P}{3\,r_h^2} \\
		&\qquad + \frac{3456\,\psi^6}{r_h^4} - \frac{5088\,\psi^4}{7\,r_h^4} + \frac{2560\,\psi^4 \pi}{3\,r_h^3 \tau} \Bigg)\epsilon.
	\end{split}
\end{equation}

Imposing the equilibrium condition \(\phi^{r_h}=0\), we obtain
the inverse temperature of the heat bath as a function of the
horizon radius:
\begin{equation}
	\begin{split}
		\tau &= -28\,r_h \pi \cdot \frac{N}{D}, \\[6pt]
		N &= 1280\,\epsilon\,\psi^4 - 3\,r_h^4, \\[6pt]
		D &= 172032\,P^2 \pi^2 \epsilon\,\psi^2 r_h^4 - 53760\,P \pi \epsilon\,\psi^4 r_h^2 + 168\,P \pi r_h^6 \\
		&\quad - 3584\,P \pi \epsilon\,\psi^2 r_h^2 + 145152\,\epsilon\,\psi^6 - 84\,\psi^2 r_h^4 \\
		&\quad - 30528\,\epsilon\,\psi^4 + 21\,r_h^4.
	\end{split}
\end{equation}

\begin{figure}[H]
	\centering
	\begin{subfigure}[b]{0.31\textwidth}
		\centering
		\includegraphics[width=\textwidth]{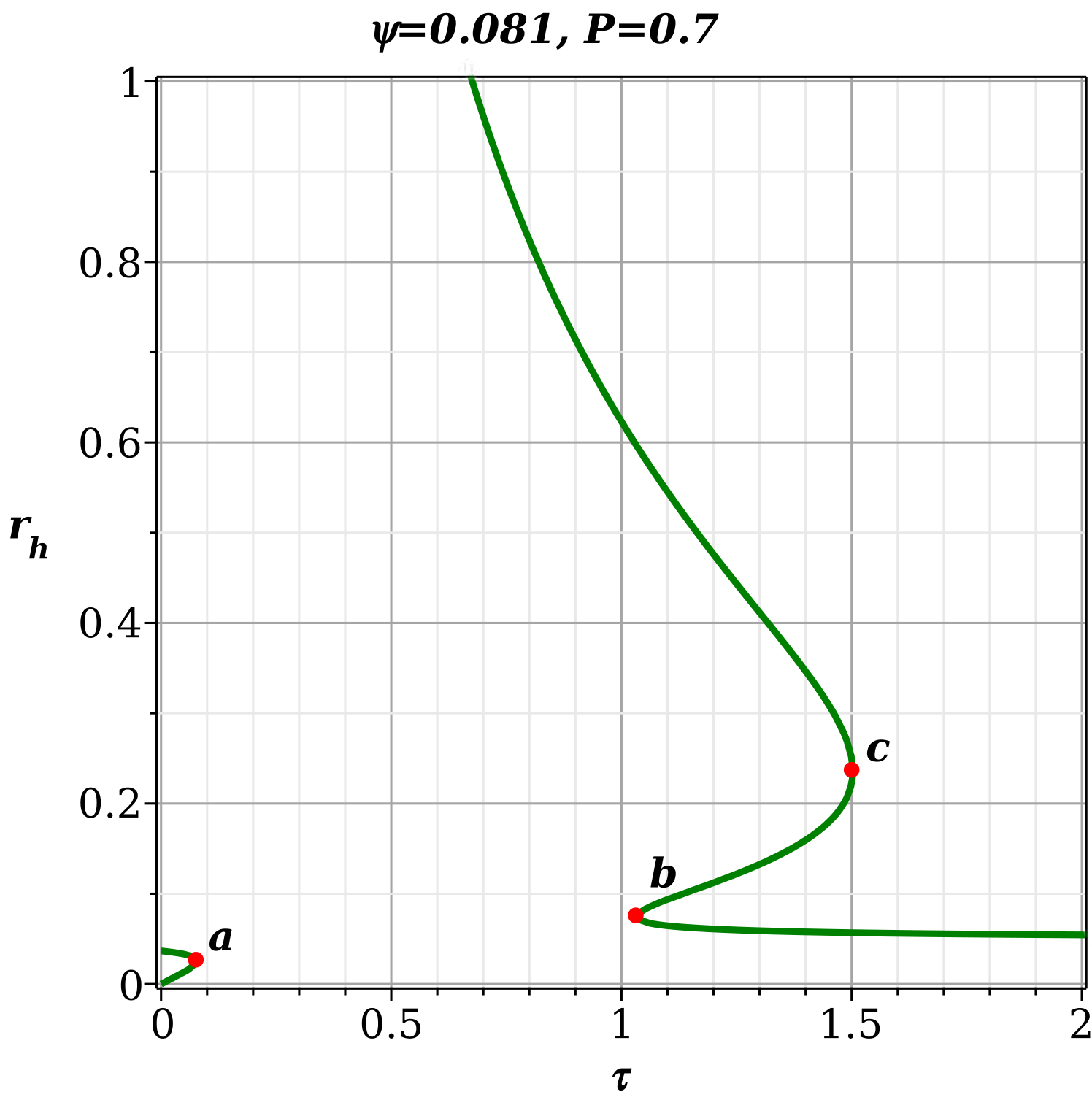}
		\caption{\(\psi=0.081\), \(P=0.7\).}
		\label{fig:tau5}
	\end{subfigure}
	\hfill
	\begin{subfigure}[b]{0.31\textwidth}
		\centering
		\includegraphics[width=\textwidth]{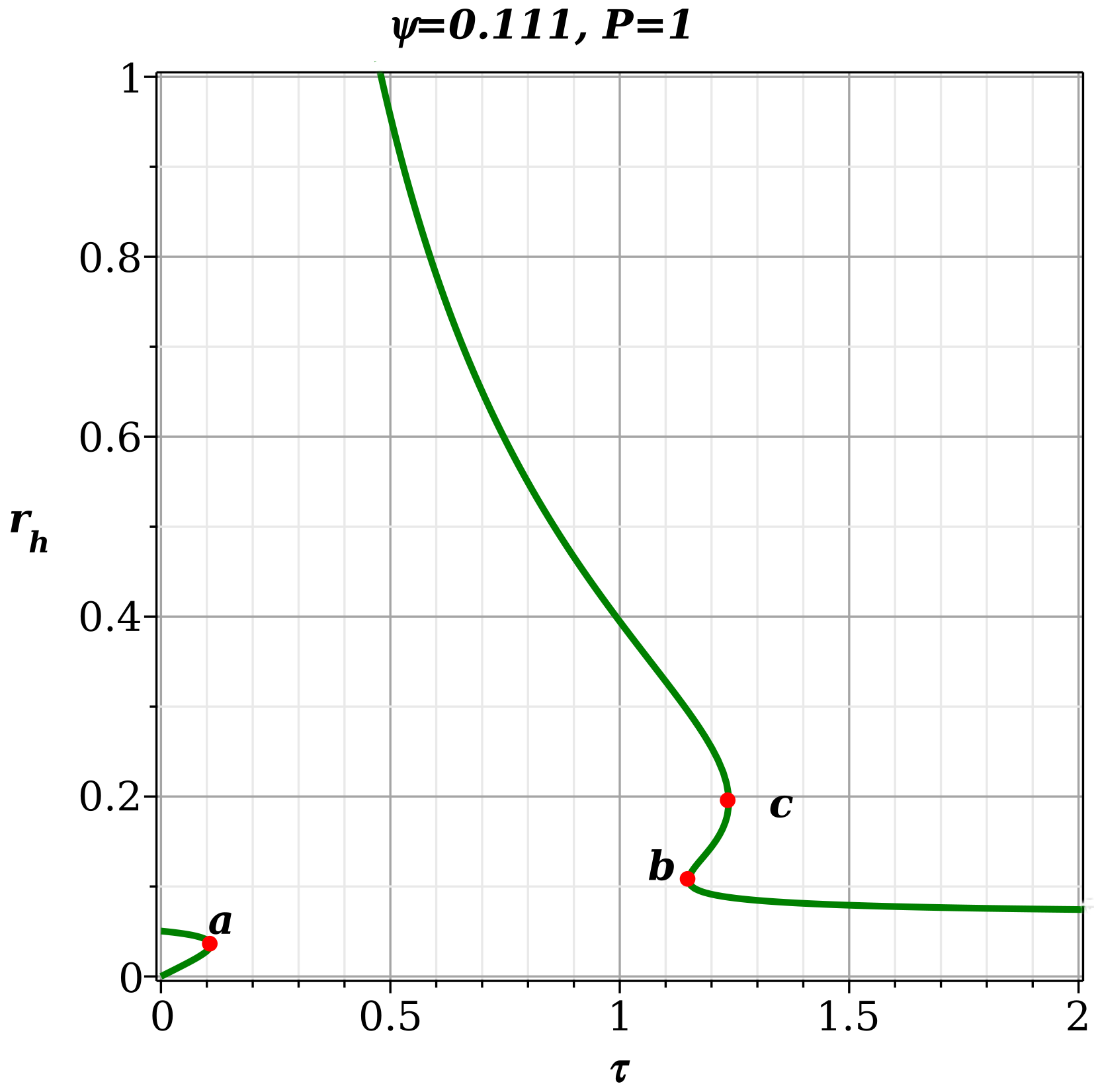}
		\caption{\(\psi=0.111\), \(P=1\).}
		\label{fig:tau6}
	\end{subfigure}
	\hfill
	\begin{subfigure}[b]{0.31\textwidth}
		\centering
		\includegraphics[width=\textwidth]{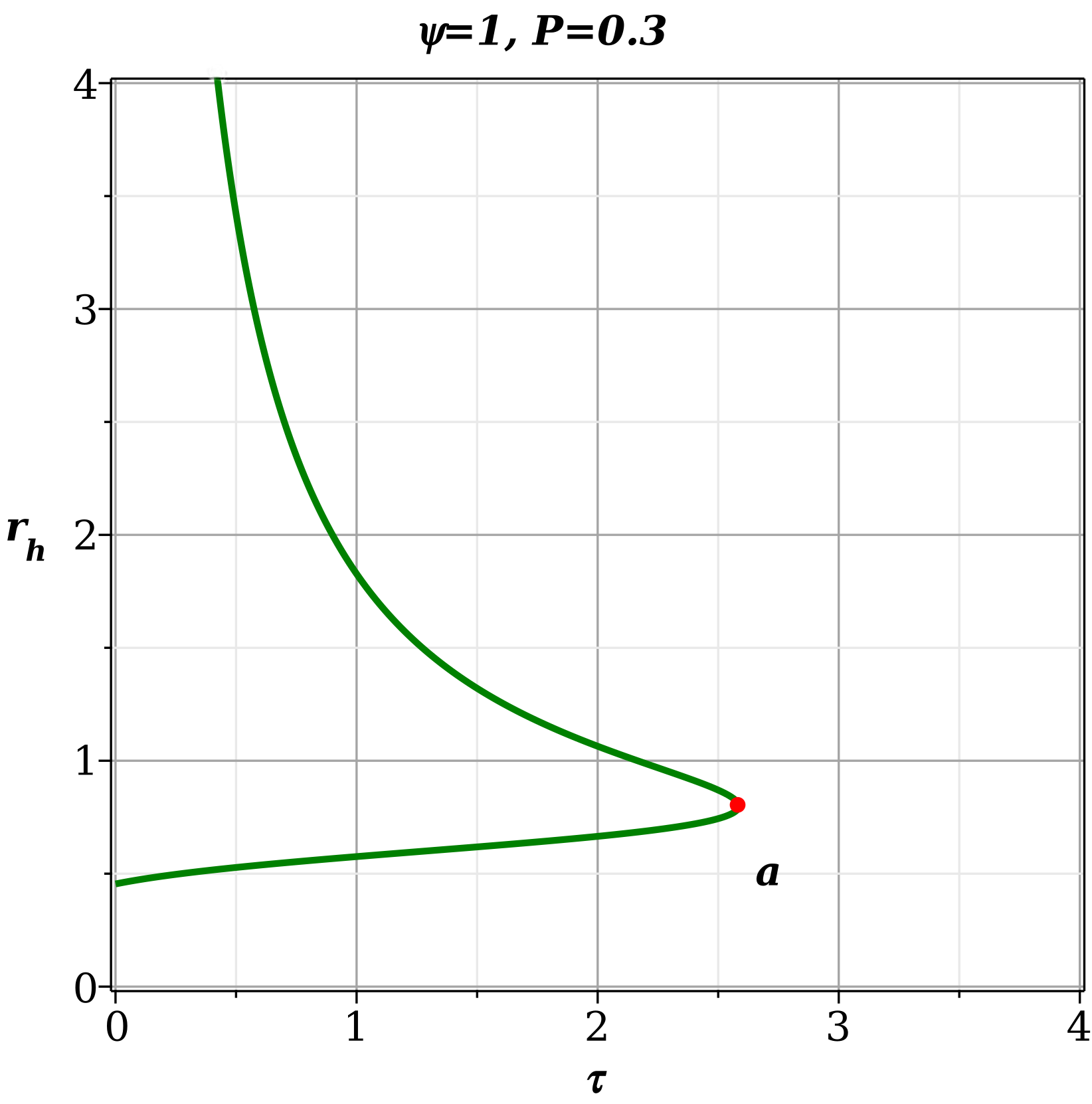}
		\caption{\(\psi=1\), \(P=0.3\).}
		\label{fig:tau7}
	\end{subfigure}
	
	\caption{The \(r_h\)--\(\tau\) diagrams for different values of the
		electric potential \(\psi\) and pressure \(P\), corresponding to the
		first NC, critical, and second NC regions of the grand
		canonical ensemble, respectively. The left and middle panels have
		\(W=+1\), whereas the right panel has \(W=0\). For all branches,
		\(dS/dr_h>0\).}
	\label{fig:tau5-tau7}
\end{figure}

Figure~\ref{fig:tau5-tau7} displays the corresponding \(r_h\)--\(\tau\)
diagrams for different values of the electric potential and pressure.

We consider three representative values of the electric potential:
\(\psi=0.081\) from the first NC region, \(\psi=0.111\) from the
critical region, and \(\psi=1\) from the second NC region. Although
we have examined a wide range of potential values, only these three
representative cases are presented to avoid redundant plots.
The corresponding \(r_h\)--\(\tau\) diagrams are shown in
Fig.~\ref{fig:tau5-tau7}.

In panel~(a), corresponding to \(\psi=0.081\) and \(P=0.7\), three
equilibrium branches exist for \(\tau<\tau_a\). From smaller to larger
horizon radii, these branches are ordered as stable, unstable, and
stable. Their total winding number is \(W=+1\). For
\(\tau_a<\tau<\tau_b\), only one stable branch remains. As \(\tau\)
increases further, three branches reappear in the interval
\(\tau_b<\tau<\tau_c\), with the same ordering: stable, unstable,
and stable. For \(\tau>\tau_c\), only one stable branch,
corresponding to a small black hole, remains. Along all these
branches, the entropy derivative with respect to the horizon radius
is positive, so they lie within the physical domain.

A similar equilibrium structure is observed in panel~(b), which
corresponds to \(\psi=0.111\) and \(P=1\). However, the local
thermodynamic analysis identifies this case as belonging to the
critical region, whereas panel~(a) corresponds to the first NC
region. Thus, despite their similar topological structures, the two
cases exhibit different local thermodynamic behavior: criticality
is present in panel~(b) but absent in panel~(a). Nevertheless, both
configurations have the same total winding number, \(W=+1\). This
demonstrates that the global topological classification alone
cannot distinguish between the presence and absence of
thermodynamic criticality.

Panel~(c), corresponding to \(\psi=1\) and \(P=0.3\), represents
the second NC region. In contrast to the first NC region, this case
has a total winding number of \(W=0\), indicating that the two NC
regions belong to different topological classes.

\begin{figure}[H]
	\centering
	\begin{subfigure}[b]{0.31\textwidth}
		\centering
		\includegraphics[width=\textwidth]{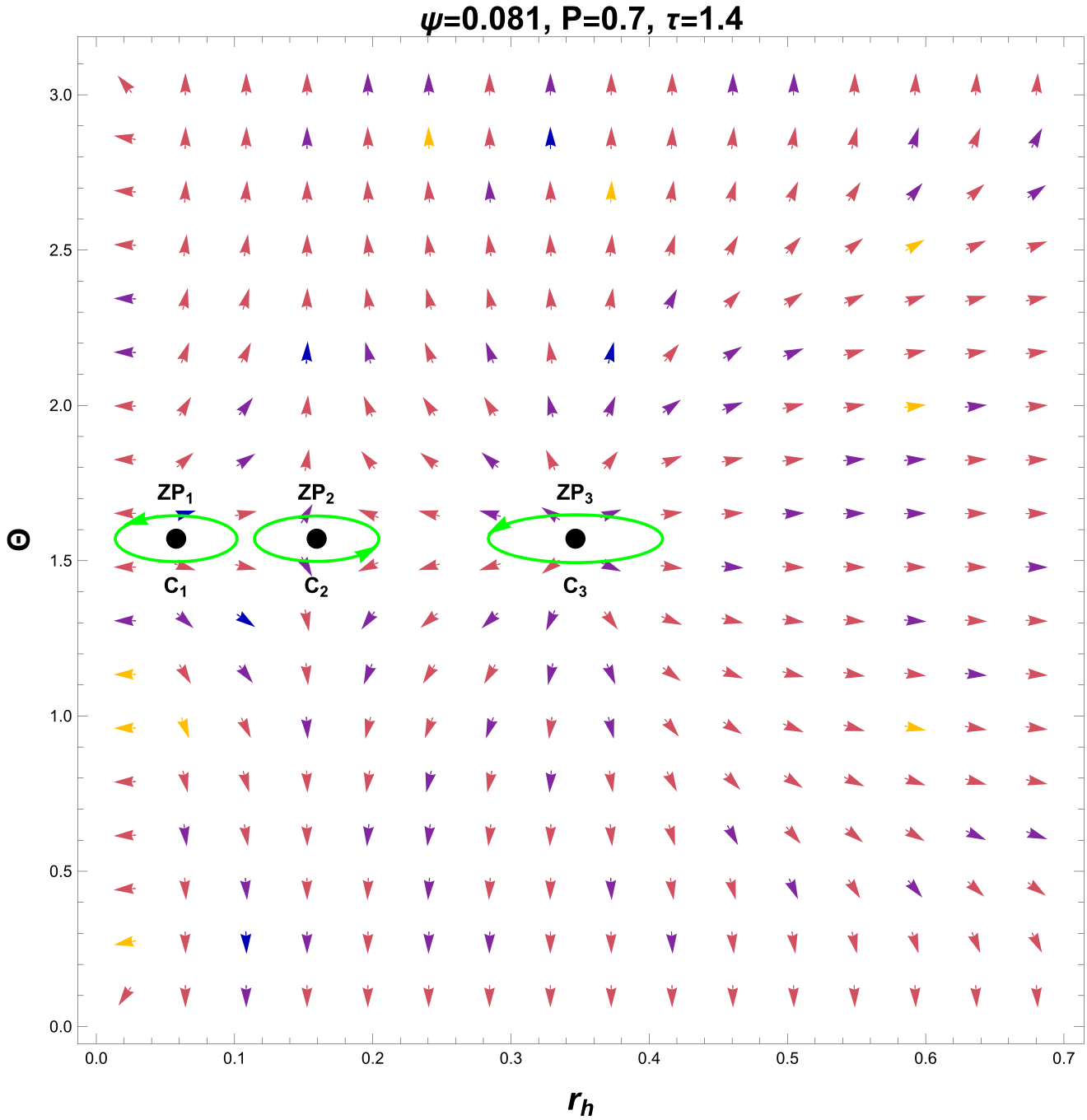}
		\caption{\(\psi=0.081\), \(P=0.7\), \(\tau=1.4\); the zeros are at
			\(0.0577\), \(0.1594\), and \(0.3466\).}
		\label{fig:v5}
	\end{subfigure}
	\hfill
	\begin{subfigure}[b]{0.31\textwidth}
		\centering
		\includegraphics[width=\textwidth]{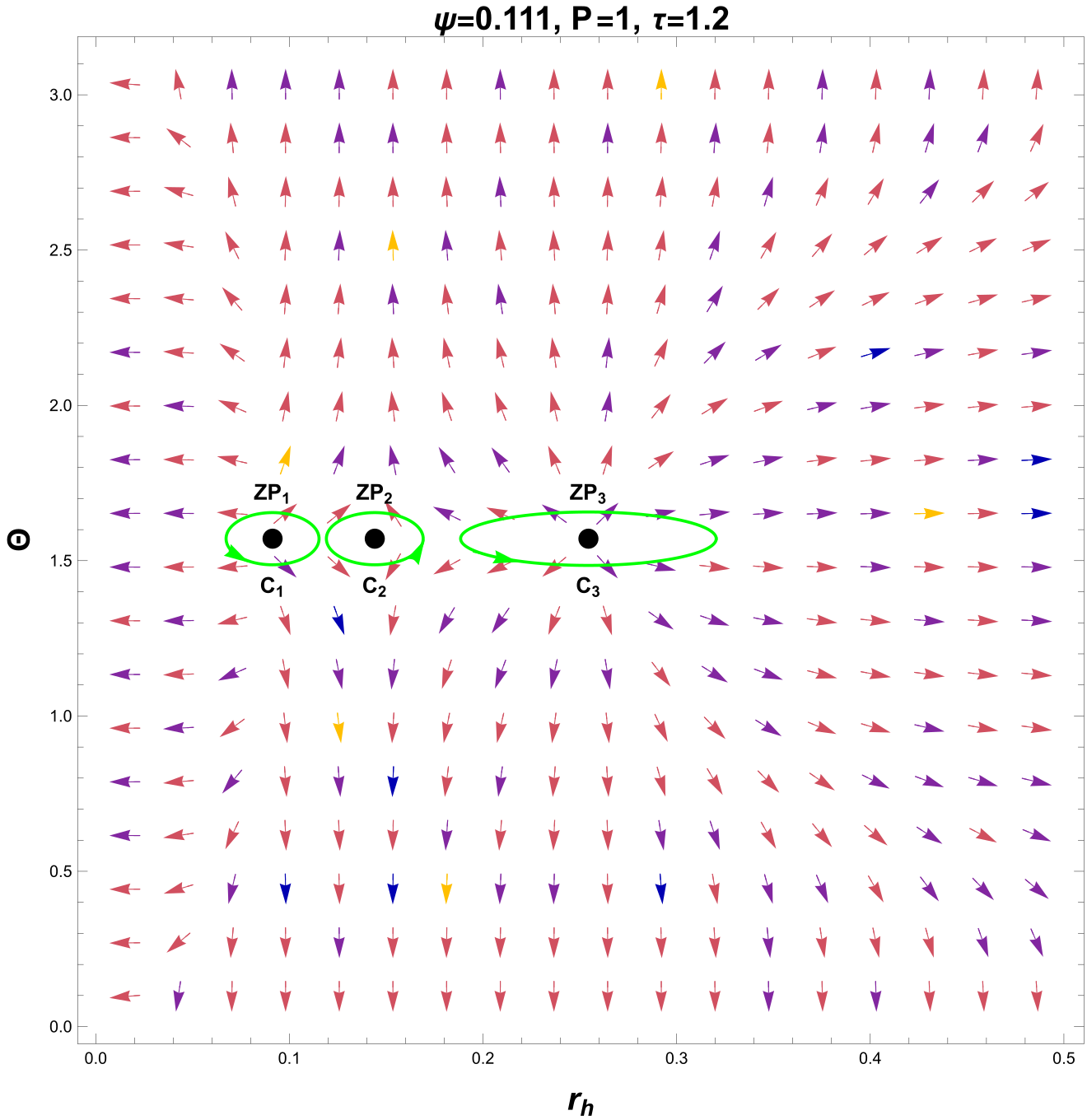}
		\caption{\(\psi=0.111\), \(P=1\), \(\tau=1.2\); the zeros are at
			\(0.0913\), \(0.1440\), and \(0.2544\).}
		\label{fig:v6}
	\end{subfigure}
	\hfill
	\begin{subfigure}[b]{0.33\textwidth}
		\centering
		\includegraphics[width=\textwidth]{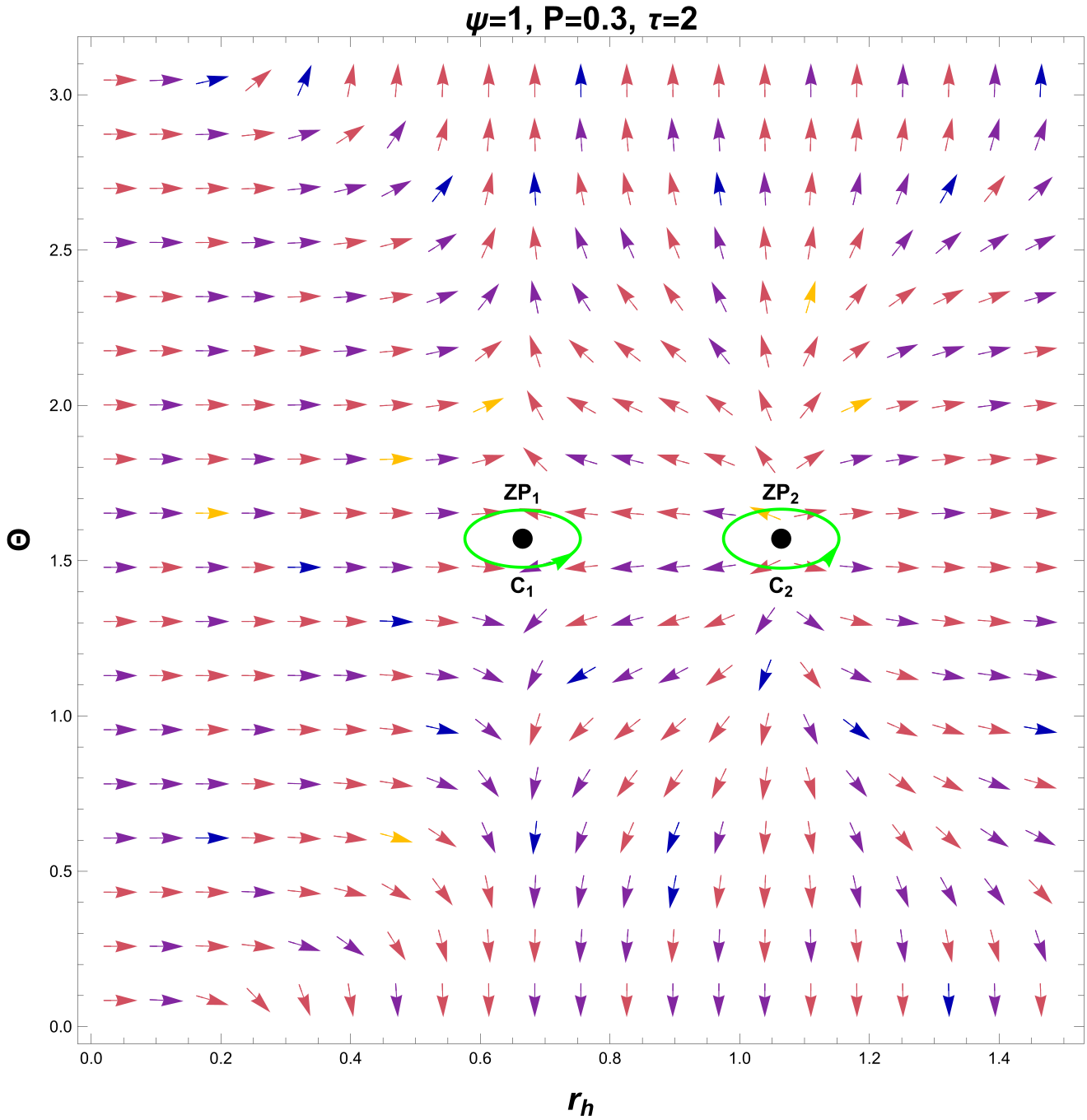}
		\caption{\(\psi=1\), \(P=0.3\), \(\tau=2\); the zeros are at
			\(0.6652\) and \(1.0640\).}
		\label{fig:v7}
	\end{subfigure}
	\caption{Unit-vector field configurations for three representative
		parameter sets belonging to the different regions discussed
		above. Panels~(a) and~(b) correspond to the \(W=+1\)
		classification, whereas panel~(c) corresponds to \(W=0\).}
	\label{fig:v5-v7}
\end{figure}

The unit-vector field configurations shown in Fig.~\ref{fig:v5-v7}
provide a direct visualization of the topological classification in the
\((r_h,\Theta)\) plane for the three representative parameter sets
discussed above. The orientation of the field along the boundaries of
the parameter region determines the global topological class, while
the zeros of its radial component correspond to the on-shell black hole
solutions.

In panels~(a) and~(b), representing the low- and intermediate-potential
regions, respectively, the unit vector field exhibits three zeros,
and the boundary orientation yields a global winding number of
\(W=+1\). Although these panels belong to different conventional
thermodynamic regimes---noncritical and critical, respectively---they
share the same topological classification. This further confirms that
the global winding number alone cannot distinguish between the
presence and absence of conventional criticality. In panel~(c),
which represents the high-potential region, the vector field has two
zeros, and the boundary orientation gives \(W=0\). In all three
panels, the zeros lie in the physical region where
\(\partial S/\partial r_h>0\); hence, none of the roots needs to be
excluded from the topological count.

\begin{figure}[H]
	\centering
	\begin{subfigure}[b]{0.31\textwidth}
		\centering
		\includegraphics[width=\textwidth]{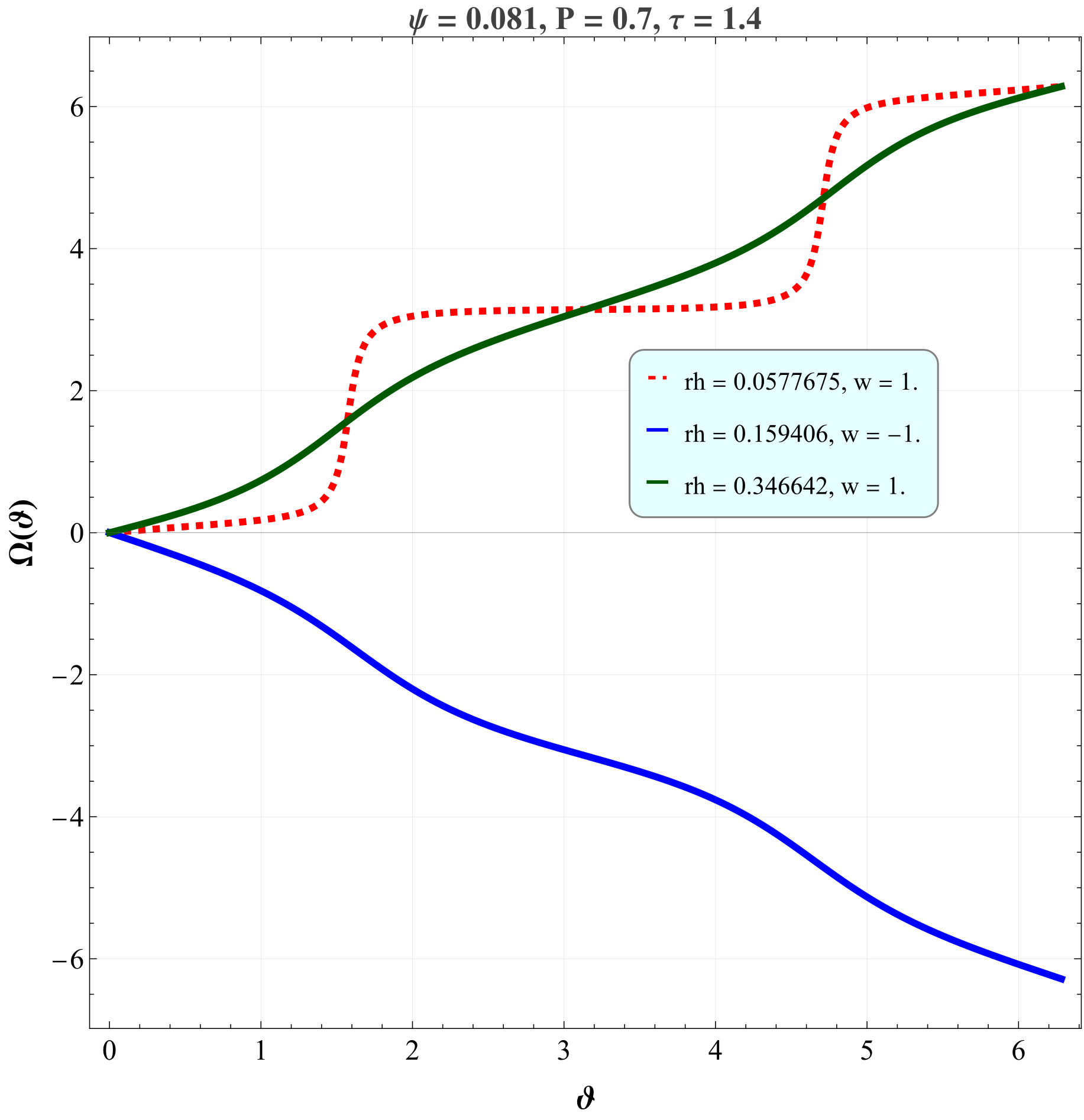}
		\caption{\(\psi=0.081\), \(P=0.7\), \(\tau=1.4\); \(W=+1\).}
		\label{fig:d3}
	\end{subfigure}
	\hfill
	\begin{subfigure}[b]{0.31\textwidth}
		\centering
		\includegraphics[width=\textwidth]{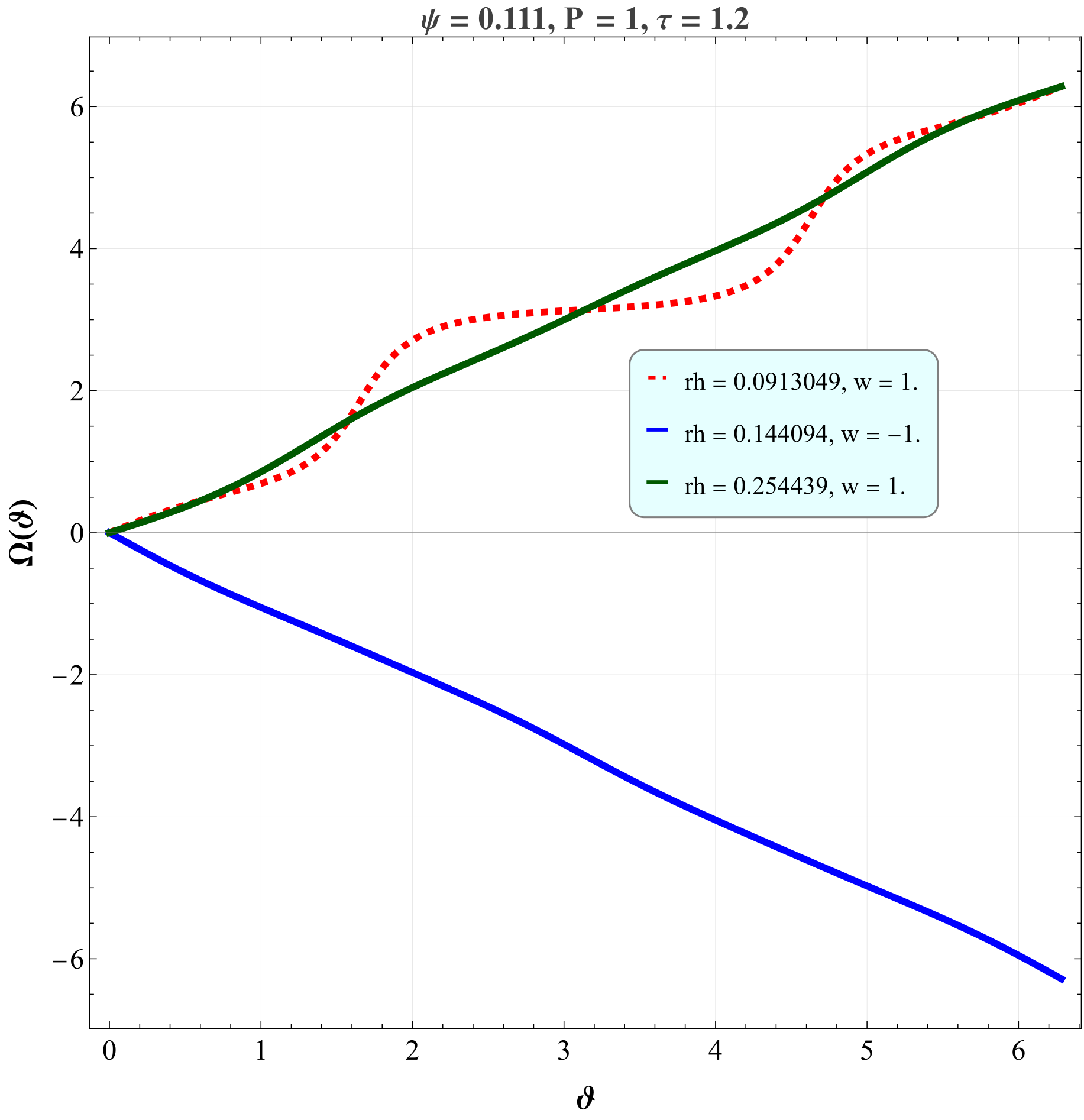}
		\caption{\(\psi=0.111\), \(P=1\), \(\tau=1.2\); \(W=+1\).}
		\label{fig:d4}
	\end{subfigure}
	\hfill
	\begin{subfigure}[b]{0.31\textwidth}
		\centering
		\includegraphics[width=\textwidth]{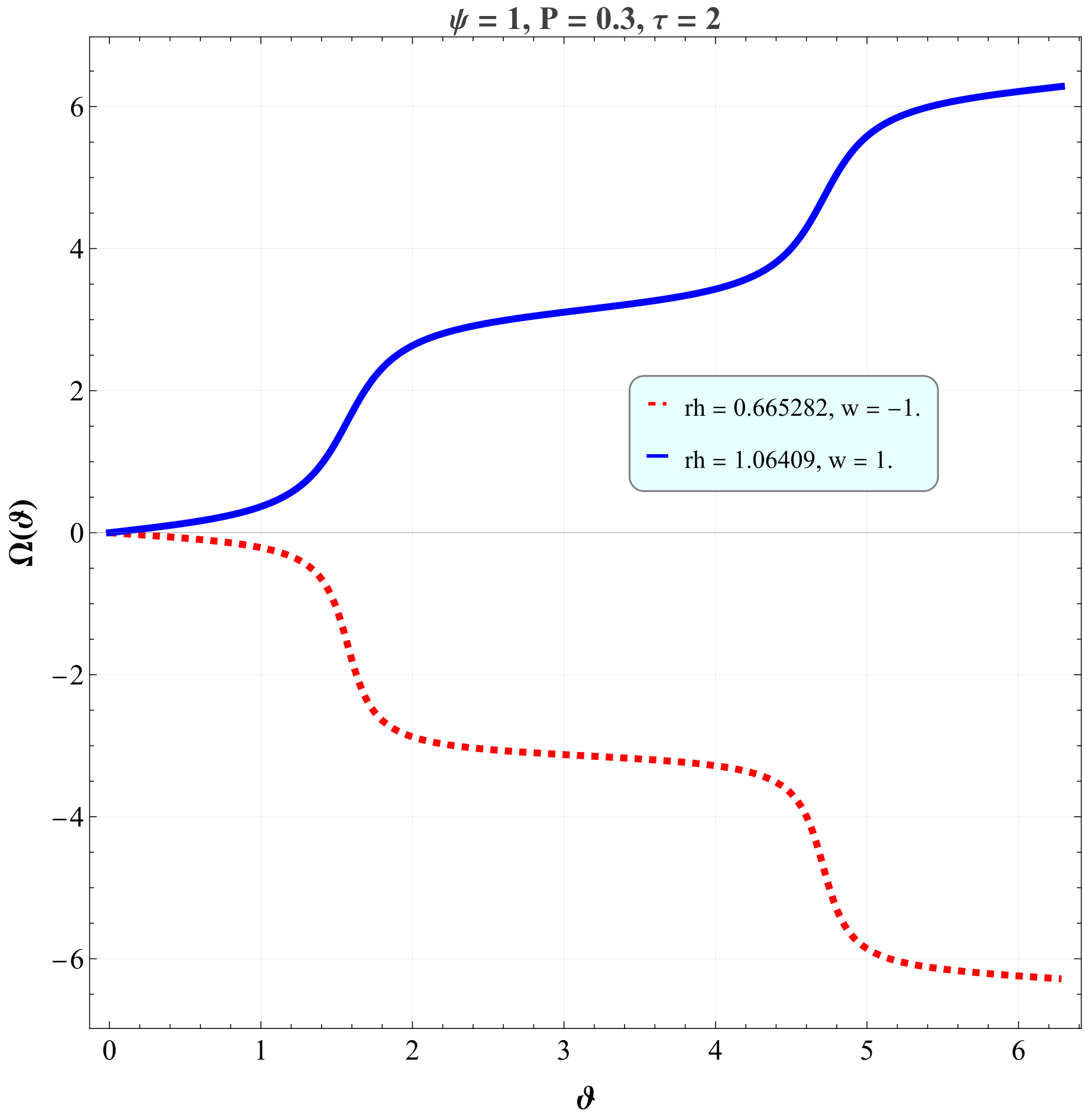}
		\caption{\(\psi=1\), \(P=0.3\), \(\tau=2\); \(W=0\).}
		\label{fig:d5}
	\end{subfigure}
	\caption{Deflection-angle profiles for three representative
		parameter sets, providing an independent check of the
		corresponding winding numbers. The first NC region and the
		intermediate critical region belong to the \(W=+1\)
		(\(W^{1+}\)) class, whereas the second (high-potential)
		region belongs to the \(W=0\) (\(W^{0-}\)) class.}
	\label{fig:d3-d5}
\end{figure}

The deflection-angle profiles for the same three representative
parameter sets are shown in Fig.~\ref{fig:d3-d5}. These profiles
provide an independent check of the winding numbers obtained from
the local analysis of the vector field.

In panels~(a) and~(b), the total deflection angle, divided by
\(2\pi\), yields \(W=+1\), in agreement with the classification
obtained from the unit-vector field. In panel~(c), the same
procedure gives \(W=0\). These results confirm that the
deflection-angle method and the local Jacobian analysis lead to
the same topological classification.

The results reveal an important distinction between conventional
criticality and thermodynamic topology in the grand canonical
ensemble. Although the low- and intermediate-potential regions
share the same global topological number, \(W=+1\), conventional
critical points occur only in the latter. By contrast, the
high-potential region is characterized by \(W=0\), despite also
lacking conventional criticality. Thus, the winding number
provides a global classification of the equilibrium structure,
whereas the conventional criticality conditions are needed to
distinguish the local thermodynamic behavior of different
regions. A summary of the classification in both ensembles is
provided in Table~\ref{table3}.

\begin{table}[H]
	\centering
	\caption{Summary of the topological classification in the
		canonical and grand canonical ensembles. The physical zeros
		are those lying in the region where \(dS/dr_h>0\).}
	\label{table3}
	\begin{tabular}{lccc}
		\hline
		Ensemble & Region & Physical zeros & Total winding number \(W\) \\
		\hline
		Canonical & NC region & Three & \(+1\) \\
		Canonical & Critical region & All zeros & \(0\) \\
		\hline
		Grand canonical & Low-potential region & All zeros & \(+1\) \\
		Grand canonical & Intermediate-potential region & All zeros & \(+1\) \\
		Grand canonical & High-potential region & All zeros & \(0\) \\
		\hline
	\end{tabular}
\end{table}

As summarized in Table~\ref{table3}, the canonical and grand
canonical ensembles exhibit different relationships between
conventional criticality and topological classification. In the
canonical ensemble, the transition from the NC region to the
critical region changes the global winding number from \(W=+1\)
to \(W=0\), making the two regimes topologically distinct. In
the grand canonical ensemble, however, the low- and
intermediate-potential regions share the same winding number,
\(W=+1\), despite differing in their conventional criticality.
The high-potential region, with \(W=0\), is topologically
distinct from both. These results demonstrate that the global
winding number and the conventional criticality conditions
encode complementary, rather than identical, information.
We return to this point in the conclusions.


\section{Conclusions}\label{sec:5}

In this work, we have investigated the thermodynamic criticality and
topological properties of a four-dimensional charged AdS black hole
with a nonminimal electromagnetic-curvature interaction of the
\(F^4R\) type. The black-hole solution and its thermodynamic
quantities were constructed perturbatively to first order in the
coupling parameter \(\epsilon\). The resulting corrections modify
the dependence of the Hawking temperature and entropy on the horizon
radius and, consequently, affect the equilibrium structure of the
system. The entropy was obtained consistently from the first law of
black-hole thermodynamics. Throughout the analysis, we have
restricted attention to the perturbative regime, in which the
dimensionless coupling \(\epsilon/\ell^4\) remains sufficiently
small. In particular, the choice \(\epsilon=10^{-4}\) considered
here lies within this regime for the parameter values investigated.

We first examined conventional thermodynamic criticality by imposing
the critical-point conditions on the Hawking temperature in the
canonical and grand canonical ensembles. This analysis identifies
distinct critical and noncritical regions in the corresponding
parameter spaces and provides the conventional thermodynamic
classification against which the topological results can be compared.

In the canonical ensemble, the critical and noncritical regions
belong to different topological classes. The critical region is
characterized by a global winding number \(W=0\), whereas the
noncritical region has \(W=+1\). An important aspect of this
classification is the physical admissibility of the equilibrium
solutions. Since the entropy contains a correction proportional to
\(r_h^{-6}\), its derivative with respect to the horizon radius is
not necessarily positive throughout the entire mathematical domain.
Thus, not every zero of the equilibrium equation represents a
physically admissible black-hole configuration. In the noncritical
region, the smallest zero in the representative configurations
satisfies \(dS/dr_h<0\) and must therefore be excluded from the
physical topological count. After imposing this restriction, the
remaining equilibrium points yield \(W=+1\). In the critical
region, the equilibrium points considered satisfy the entropy
admissibility condition, and their local winding numbers sum to
zero. We also find that up to four physically admissible
equilibrium points can occur, with the contributions of
intermediate branches canceling in the global winding number.

The grand canonical ensemble exhibits a different topological
structure, with three distinct regions determined by the electric
potential. The low-potential region is noncritical and has
\(W=+1\), while the intermediate-potential region contains
conventional critical points and has the same winding number.
In contrast, the high-potential region is noncritical and is
characterized by \(W=0\). All equilibrium points considered in
the grand canonical configurations satisfy \(dS/dr_h>0\) and
are therefore physically admissible. These results highlight an
important distinction: the low- and intermediate-potential
regions have the same global winding number despite their
different conventional criticality, whereas the two noncritical
regions at low and high potentials belong to different
topological classes.

We further examined the consistency of the topological
classification using complementary methods. In addition to
calculating the local winding numbers from the Jacobian of the
thermodynamic vector field, we investigated the orientation of
the unit vector field along the boundary of the parameter domain
and employed the deflection-angle method. These approaches
confirm the winding numbers obtained from the local analysis
and emphasize the importance of identifying the physical domain
when determining the global topological charge.

Overall, our findings demonstrate that thermodynamic topology
and conventional temperature-based criticality analysis provide
complementary, rather than identical, information about black-hole
equilibrium configurations. While the conventional criticality
conditions determine the existence of critical points, the
winding number characterizes the global structure of the
equilibrium solutions. The comparison between the two ensembles
also shows that the relationship between these descriptions
depends on the thermodynamic ensemble. Therefore, a complete
classification requires both the local criticality conditions
and the global topological analysis, together with an explicit
assessment of the physical admissibility of the equilibrium
configurations. These results provide a basis for further
investigations of the effects of higher-order electromagnetic-
curvature interactions on black-hole thermodynamics and topology.

\vspace{1cm}
\noindent {\large {\bf Acknowledgments}}

The authors acknowledge the use of AI-assisted tools for language
editing and the preparation of tables. All physical and mathematical
derivations, results, and scientific arguments are the authors' own.

\medskip
\noindent {\large {\bf Data availability statement}}

All data supporting the findings of this study are included in
this article.

\vspace{1cm}


\appendix
\section{Explicit form of the field equations}
\label{app:A}

In this appendix, we present the explicit forms of the field equations
used in the main text. We first provide the interaction contribution
to the energy-momentum tensor, followed by the \(tt\)- and
\(rr\)-components of the modified Einstein equations and the
nonvanishing component of the Maxwell equation.

The interaction term of the total energy-momentum tensor is given by
\begin{align}
	T^{(I)}_{\mu \nu }=	& - F_{\beta }{}^{\gamma } F^{\eta \lambda } F_{\mu }{}^{\alpha } F_{\nu}{}^{\beta } R_{\alpha \gamma \eta \lambda }
	- F_{\alpha }{}^{\gamma } F^{\eta \lambda } F_{\mu }{}^{\alpha } F_{\nu }{}^{\beta } R_{\beta \gamma \eta \lambda } \notag \\
	& + \tfrac{1}{2} F_{\alpha }{}^{\gamma } F^{\alpha \beta } F_{\beta }{}^{\eta } F^{\lambda \rho } g_{\mu \nu } R_{\gamma \eta \lambda \rho }
	- \tfrac{1}{2} F_{\beta }{}^{\eta } F^{\beta \gamma } F_{\gamma }{}^{\lambda } F_{\nu }{}^{\alpha } R_{\mu \alpha \eta \lambda } \notag \\
	& + F_{\alpha }{}^{\beta } F_{\beta }{}^{\gamma } F^{\eta \lambda } F_{\nu }{}^{\alpha } R_{\mu \gamma \eta \lambda }
	- \tfrac{1}{2} F_{\beta }{}^{\eta } F^{\beta \gamma } F_{\gamma }{}^{\lambda } F_{\mu }{}^{\alpha } R_{\nu \alpha \eta \lambda } \notag \\
	& + F_{\alpha }{}^{\beta } F_{\beta }{}^{\gamma } F^{\eta \lambda } F_{\mu }{}^{\alpha } R_{\nu \gamma \eta \lambda }
	+ F^{\gamma \eta } F_{\mu }{}^{\alpha } F_{\nu }{}^{\beta } \nabla_{\alpha }\nabla_{\eta }F_{\beta \gamma } \notag \\
	& + F_{\beta }{}^{\gamma } F_{\mu }{}^{\alpha } F_{\nu }{}^{\beta } \nabla_{\alpha }\nabla_{\eta }F_{\gamma }{}^{\eta }
	- F_{\beta }{}^{\eta } F^{\beta \gamma } F_{\nu }{}^{\alpha } \nabla_{\alpha }\nabla_{\eta }F_{\mu \gamma } \notag \\
	& - F_{\beta }{}^{\eta } F^{\beta \gamma } F_{\mu }{}^{\alpha } \nabla_{\alpha }\nabla_{\eta }F_{\nu \gamma }
	+ F^{\beta \gamma } F_{\nu }{}^{\alpha } \nabla_{\alpha }F_{\gamma }{}^{\eta } \nabla_{\beta }F_{\mu \eta } \notag \\
	& + F^{\beta \gamma } F_{\mu }{}^{\alpha } \nabla_{\alpha }F_{\gamma }{}^{\eta } \nabla_{\beta }F_{\nu \eta }
	+ F^{\gamma \eta } F_{\mu }{}^{\alpha } F_{\nu }{}^{\beta } \nabla_{\beta }\nabla_{\eta }F_{\alpha \gamma } \notag \\
	& + F_{\alpha }{}^{\gamma } F_{\mu }{}^{\alpha } F_{\nu }{}^{\beta } \nabla_{\beta }\nabla_{\eta }F_{\gamma }{}^{\eta }
	- F^{\beta \gamma } F_{\nu }{}^{\alpha } \nabla_{\alpha }F_{\mu }{}^{\eta } \nabla_{\gamma }F_{\beta \eta } \notag \\
	& - F^{\beta \gamma } F_{\mu }{}^{\alpha } \nabla_{\alpha }F_{\nu }{}^{\eta } \nabla_{\gamma }F_{\beta \eta }
	+ F_{\mu }{}^{\alpha } F_{\nu }{}^{\beta } \nabla_{\beta }F^{\gamma \eta } \nabla_{\eta }F_{\alpha \gamma } \notag \\
	& - F^{\beta \gamma } F_{\nu }{}^{\alpha } \nabla_{\beta }F_{\mu }{}^{\eta } \nabla_{\eta }F_{\alpha \gamma }
	- F^{\beta \gamma } F_{\mu }{}^{\alpha } \nabla_{\beta }F_{\nu }{}^{\eta } \nabla_{\eta }F_{\alpha \gamma } \notag \\
	& + F_{\mu }{}^{\alpha } F_{\nu }{}^{\beta } \nabla_{\alpha }F^{\gamma \eta } \nabla_{\eta }F_{\beta \gamma }
	+ F_{\alpha }{}^{\beta } F_{\nu }{}^{\alpha } \nabla_{\gamma }F_{\mu }{}^{\gamma } \nabla_{\eta }F_{\beta }{}^{\eta } \notag \\
	& + F_{\alpha }{}^{\beta } F_{\mu }{}^{\alpha } \nabla_{\gamma }F_{\nu }{}^{\gamma } \nabla_{\eta }F_{\beta }{}^{\eta }
	+ F_{\mu }{}^{\alpha } F_{\nu }{}^{\beta } \nabla_{\alpha }F_{\beta }{}^{\gamma } \nabla_{\eta }F_{\gamma }{}^{\eta } \notag \\
	& + F^{\beta \gamma } F_{\nu }{}^{\alpha } \nabla_{\alpha }F_{\mu \beta } \nabla_{\eta }F_{\gamma }{}^{\eta }
	+ F^{\beta \gamma } F_{\mu }{}^{\alpha } \nabla_{\alpha }F_{\nu \beta } \nabla_{\eta }F_{\gamma }{}^{\eta } \notag \\
	& + F_{\mu }{}^{\alpha } F_{\nu }{}^{\beta } \nabla_{\beta }F_{\alpha }{}^{\gamma } \nabla_{\eta }F_{\gamma }{}^{\eta }
	+ F^{\beta \gamma } F_{\nu }{}^{\alpha } \nabla_{\alpha }F_{\gamma }{}^{\eta } \nabla_{\eta }F_{\mu \beta } \notag \\
	& - F_{\alpha }{}^{\gamma } F^{\alpha \beta } \nabla_{\gamma }F_{\nu }{}^{\eta } \nabla_{\eta }F_{\mu \beta }
	+ F_{\alpha }{}^{\beta } F_{\nu }{}^{\alpha } \nabla_{\gamma }F_{\beta }{}^{\eta } \nabla_{\eta }F_{\mu }{}^{\gamma } \notag \\
	& + F^{\beta \gamma } F_{\nu }{}^{\alpha } \nabla_{\gamma }F_{\alpha \beta } \nabla_{\eta }F_{\mu }{}^{\eta }
	- F_{\alpha }{}^{\gamma } F^{\alpha \beta } \nabla_{\gamma }F_{\nu \beta } \nabla_{\eta }F_{\mu }{}^{\eta } \notag \\
	& + F^{\beta \gamma } F_{\mu }{}^{\alpha } \nabla_{\alpha }F_{\gamma }{}^{\eta } \nabla_{\eta }F_{\nu \beta }
	- F_{\alpha }{}^{\gamma } F^{\alpha \beta } \nabla_{\beta }F_{\mu }{}^{\eta } \nabla_{\eta }F_{\nu \gamma } \notag \\
	& + F_{\alpha }{}^{\beta } F_{\mu }{}^{\alpha } \nabla_{\gamma }F_{\beta }{}^{\eta } \nabla_{\eta }F_{\nu }{}^{\gamma }
	+ F^{\beta \gamma } F_{\mu }{}^{\alpha } \nabla_{\gamma }F_{\alpha \beta } \nabla_{\eta }F_{\nu }{}^{\eta } \notag \\
	& - F_{\alpha }{}^{\gamma } F^{\alpha \beta } \nabla_{\gamma }F_{\mu \beta } \nabla_{\eta }F_{\nu }{}^{\eta }
	+ F_{\alpha }{}^{\beta } F_{\beta }{}^{\gamma } F_{\nu }{}^{\alpha } \nabla_{\eta }\nabla_{\gamma }F_{\mu }{}^{\eta } \notag \\
	& + F_{\alpha }{}^{\beta } F_{\beta }{}^{\gamma } F_{\mu }{}^{\alpha } \nabla_{\eta }\nabla_{\gamma }F_{\nu }{}^{\eta }.
\end{align}

The $tt$- and $rr$-components of Einstein's equations are
\begin{equation}\label{tt-comp}
	\begin{split}
		&    f(r)-1 + \Lambda r^2 + r f'(r) + \frac{1}{4} e^{2 H(r)}\kappa r^2 \alpha h'(r)^2+2e^{4H(r)}\kappa \epsilon h'(r)^2  B_1(r)=0,
	\end{split}
\end{equation}
where
\begin{equation}
	\begin{split}
		B_1(r)=	& 2 f(r) h'(r)^2
		+ r f'(r) h'(r)^2  + 16 f(r) r h'(r)^2 H'(r)
		\\
		&+ 8 r^2 f'(r) h'(r)^2 H'(r) + 12 f(r) r^2 h'(r)^2 H'(r)^2
		\\
		&- 2 r^2 h'(r)^2 f''(r)  + 16 f(r) r h'(r) h''(r)
		+ 2 r^2 f'(r) h'(r) h''(r) \\
		& + 32 f(r) r^2 h'(r) H'(r) h''(r)
		+ 12 f(r) r^2 h''(r)^2 \\
		& + 8 f(r) r^2 h'(r)^2 H''(r)
		+ 4 f(r) r^2 h'(r) h'''(r)
	\end{split}
\end{equation}
and
\begin{equation}\label{rr-comp}
	f(r)-1 + \Lambda r^2 + r f'(r) - 2 f(r) r H'(r) +\frac{1}{4}e^{2H(r)}  \kappa r^2 \alpha h'(r)^2+\frac{1}{4}\kappa \epsilon e^{4H(r)} B_2(r)=0,
\end{equation}
where
\begin{equation}
	\begin{split}
		B_2(r)= & 8 r f'(r) h'(r)^4  - 16 f(r) r h'(r)^4 H'(r) \\
		&+ 64 r^2 f'(r) h'(r)^4 H'(r)  - 64 f(r) r^2 h'(r)^4 H'(r)^2 \\
		&- 16 r^2 h'(r)^4 f''(r) + 16 r^2 f'(r) h'(r)^3 h''(r) \\
		& - 32 f(r) r^2 h'(r)^3 H'(r) h''(r) + 32 f(r) r^2 h'(r)^4 H''(r) ,
	\end{split}
\end{equation}
respectively.

The non-zero component of the Maxwell equation of Eq.~(\ref{EOM-Maxwell}) is
\begin{eqnarray}
	r^3 \, \alpha \Big( 2 h'(r) + r h'(r) H'(r) + r h''(r) \Big)+\epsilon B_4(r)=0,
\end{eqnarray}
where $B_4(r)$ is as follows:
\begin{equation}
	\begin{split}
		B_4(r)=& 8 f(r) f'(r) h'(r) - 8 r f'(r)^2 h'(r) - 8 f(r)^2 h'(r) H'(r) \\
		& + 8 f(r) r f'(r) h'(r) H'(r)  + 64 r^2 f'(r)^2 h'(r) H'(r)  + 8 f(r)^2 r h'(r) H'(r)^2 \\
		& - 64 f(r) r^2 f'(r) h'(r) H'(r)^2  - 60 r^3 f'(r)^2 h'(r) H'(r)^2 + 36 f(r) r^3 f'(r) h'(r) H'(r)^3 \\
		& + 6 r^4 f'(r)^2 h'(r) H'(r)^3 + 8 f(r)^2 r^3 h'(r) H'(r)^4  + 8 f(r) r^4 f'(r) h'(r) H'(r)^4 \\
		& - 8 f(r)^2 r^4 h'(r) H'(r)^5  - 8 f(r) r h'(r) f''(r) - 48 r^2 f'(r) h'(r) f''(r) \\
		& + 52 f(r) r^2 h'(r) H'(r) f''(r) + 96 r^3 f'(r) h'(r) H'(r) f''(r) \\
		& - 64 f(r) r^3 h'(r) H'(r)^2 f''(r) - 32 r^4 f'(r) h'(r) H'(r)^2 f''(r)\\
		&+ 16 f(r) r^4 h'(r) H'(r)^3 f''(r)  - 16 r^3 h'(r) f''(r)^2  + 10 r^4 h'(r) H'(r) f''(r)^2  \\
		&- 8 f(r) r f'(r) h''(r) - 16 r^2 f'(r)^2 h''(r)  + 8 f(r)^2 r H'(r) h''(r)\\
		& + 44 f(r) r^2 f'(r) H'(r) h''(r)  + 36 r^3 f'(r)^2 H'(r) h''(r)  - 24 f(r)^2 r^2 H'(r)^2 h''(r)  \\
		& - 60 f(r) r^3 f'(r) H'(r)^2 h''(r)  - 18 r^4 f'(r)^2 H'(r)^2 h''(r)  + 24 f(r)^2 r^3 H'(r)^3 h''(r) \\
		& + 24 f(r) r^4 f'(r) H'(r)^3 h''(r) - 8 f(r)^2 r^4 H'(r)^4 h''(r) \\
		& - 4 f(r) r^2 f''(r) h''(r)  - 12 r^3 f'(r) f''(r) h''(r)+ 12 f(r) r^3 H'(r) f''(r) h''(r) \\
		& + 12 r^4 f'(r) H'(r) f''(r) h''(r)- 8 f(r) r^4 H'(r)^2 f''(r) h''(r) - 2 r^4 f''(r)^2 h''(r) \\
		& + 8 f(r)^2 r h'(r) H''(r) + 84 f(r) r^2 f'(r) h'(r) H''(r) + 60 r^3 f'(r)^2 h'(r) H''(r) \\
		& - 64 f(r)^2 r^2 h'(r) H'(r) H''(r) - 192 f(r) r^3 f'(r) h'(r) H'(r) H''(r) \\
		& - 60 r^4 f'(r)^2 h'(r) H'(r) H''(r)  + 80 f(r)^2 r^3 h'(r) H'(r)^2 H''(r) \\
		& + 80 f(r) r^4 f'(r) h'(r) H'(r)^2 H''(r)  - 16 f(r)^2 r^4 h'(r) H'(r)^3 H''(r) \\
		& + 52 f(r) r^3 h'(r) f''(r) H''(r)  + 20 r^4 f'(r) h'(r) f''(r) H''(r) \\
		& - 32 f(r) r^4 h'(r) H'(r) f''(r) H''(r) + 8 f(r)^2 r^2 h''(r) H''(r) \\
		& + 24 f(r) r^3 f'(r) h''(r) H''(r)  - 24 f(r)^2 r^3 H'(r) h''(r) H''(r) \\
		& - 24 f(r) r^4 f'(r) H'(r) h''(r) H''(r)  + 16 f(r)^2 r^4 H'(r)^2 h''(r) H''(r) \\
		& + 8 f(r) r^4 f''(r) h''(r) H''(r) - 40 f(r)^2 r^3 h'(r) H''(r)^2 \\
		& - 40 f(r) r^4 f'(r) h'(r) H''(r)^2  + 24 f(r)^2 r^4 h'(r) H'(r) H''(r)^2 \\
		&- 8 f(r)^2 r^4 h''(r) H''(r)^2 - 4 f(r) r^2 h'(r) f^{(3)}(r)  - 12 r^3 f'(r) h'(r) f^{(3)}(r) \\
		& + 12 f(r) r^3 h'(r) H'(r) f^{(3)}(r)  + 12 r^4 f'(r) h'(r) H'(r) f^{(3)}(r)\\
		&- 8 f(r) r^4 h'(r) H'(r)^2 f^{(3)}(r)  - 4 r^4 h'(r) f''(r) f^{(3)}(r) + 8 f(r) r^4 h'(r) H''(r) f^{(3)}(r) \\
		& + 8 f(r)^2 r^2 h'(r) H^{(3)}(r) + 24 f(r) r^3 f'(r) h'(r) H^{(3)}(r) - 24 f(r)^2 r^3 h'(r) H'(r) H^{(3)}(r) \\
		& - 24 f(r) r^4 f'(r) h'(r) H'(r) H^{(3)}(r)  + 16 f(r)^2 r^4 h'(r) H'(r)^2 H^{(3)}(r) \\
		& + 8 f(r) r^4 h'(r) f''(r) H^{(3)}(r) - 16 f(r)^2 r^4 h'(r) H''(r) H^{(3)}(r).
	\end{split}
\end{equation}
This completes the explicit form of all the field equations required in the main text.

\section{Perturbative regime and range of validity}
\label{app:B}

The first-order correction functions \(f_1\), \(h_1\), and \(H_1\)
contain inverse powers of the radial coordinate up to \(r^{-10}\).
Consequently, the smallness of the dimensionless coupling
\(\epsilon/\ell^4\) alone does not ensure that the perturbative
corrections remain small at all horizon radii. We therefore estimate
the relative size of the first-order contributions along the
physical equilibrium branches. For this purpose, we introduce
\begin{equation}
	\delta_f(r)\equiv\left|\frac{\epsilon f_1(r)}{f_0(r)}\right|,
	\qquad
	\delta_h(r)\equiv\left|\frac{\epsilon h_1(r)}{h_0(r)}\right|,
	\qquad
	\delta_H(r)\equiv\left|\epsilon H_1(r)\right|.
\end{equation}
The leading inverse-power contributions to the first two ratios
come from the terms \(2\kappa^2Q^6/r^{10}\) in \(f_1\) and
\(7\kappa Q^5/(12r^9)\) in \(h_1\). Comparing these terms with
the corresponding leading contributions \(Q^2/(4r^2)\) in
\(f_0\) and \(Q/r\) in \(h_0\), respectively, gives the common
estimate
\begin{equation}
	\delta(r)\equiv\frac{\epsilon Q^4}{r^8}.
	\label{eq:delta}
\end{equation}
The correction to \(H\) is related to this quantity by
\(\delta_H(r)=\frac{21}{4}\delta(r)\). Thus, a conservative
estimate of the relative size of the first-order contributions is
\begin{equation}
	\eta(r)\equiv
	\max\{\delta_f(r),\delta_h(r),\delta_H(r)\}
	\simeq \frac{21}{4}\delta(r).
	\label{eq:eta}
\end{equation}
The condition \(\eta(r)\ll1\) identifies the region in which
the first-order corrections are parametrically small. We use
\(\eta\) as an estimate of perturbative control, rather than as
a direct measure of the error due to neglected higher-order terms.

For \(\epsilon=10^{-4}\), we have evaluated this estimate at the
critical points reported in Tables~\ref{table1} and
\ref{table2}. In the canonical ensemble, the values of
\(\delta(r_c)\) are at most of order \(10^{-3}\) throughout
the critical region (\(Q\geq0.15\)), and decrease rapidly
with increasing \(Q\), reaching approximately \(2\times10^{-5}\)
at \(Q=1\). Accordingly, \(\eta(r_c)\) remains below
approximately \(5.3\times10^{-3}\). In the grand canonical
ensemble, \(\delta(r_c)\) ranges from about \(1.6\times10^{-2}\)
at \(\psi=0.111\) to about \(3\times10^{-2}\) for
\(\psi\lesssim0.40\). The corresponding values of \(\eta(r_c)\)
are approximately \(0.08\)--\(0.16\). These estimates indicate
strong perturbative control at the canonical critical points
and moderate control over most of the grand canonical critical
region. Near its upper-potential boundary
(\(\psi\approx0.44\)), where \(\delta\) reaches about \(0.1\),
the first-order corrections are no longer uniformly small.
We therefore regard this part of the parameter space as less
well controlled quantitatively.

In the noncritical (NC) regions, the relevant physical branches
can extend to smaller horizon radii. Their lower boundary is
determined by the condition \(dS/dr_h=0\), which, using the
corrected entropy~(\ref{entropy1}), gives
\begin{equation}
	r_*^8=5Q^4\epsilon.
	\label{eq:rstar}
\end{equation}
At this boundary, the estimates introduced above become
\begin{equation}
	\delta(r_*)=\frac{1}{5},
	\qquad
	\eta(r_*)\simeq\frac{21}{20}.
	\label{eq:deltarstar}
\end{equation}
Thus, the perturbative expansion is not uniformly controlled
arbitrarily close to the lower boundary of the physical domain.
This limitation should be kept in mind when interpreting results
associated with very small horizon radii. Away from this boundary,
however, the corrections decrease rapidly as \(r_h^{-8}\).
For example, for \(Q=0.05\) and \(r_h=0.15\), one obtains
\(\delta\simeq2\times10^{-3}\), corresponding to
\(\eta\simeq1.1\times10^{-2}\). Hence, the less controlled
part of the branch is confined to the vicinity of \(r_*\),
rather than extending over the entire physical branch. The
relative extent of this region depends on the charge: for
\(Q\geq0.15\), the critical points lie well away from
\(r_*\), whereas for \(Q\lesssim0.14\), the lower-radius
part of the physical branches is more relevant.

Finally, we have checked that the qualitative features of the
thermodynamic phase structure and the associated global winding
numbers, \(W=+1\) and \(W=0\), remain unchanged when the
coupling is varied to \(\epsilon=10^{-3}\) and
\(\epsilon=10^{-5}\). This comparison provides a consistency
check on the qualitative conclusions, although it does not by
itself quantify the effects of higher-order terms. Overall, the
estimates above show that the perturbative description is well
controlled at the canonical critical points and remains
reasonably controlled over most of the grand canonical critical
region. Greater caution is required near the upper-potential
boundary of that region and near the lower-radius boundary
\(r_*\) of the physical branches.


\end{document}